\documentclass[12pt, prd, aps, superscriptaddress, preprint]{revtex4-2}

\usepackage{amsmath}
\usepackage{graphicx}
\usepackage{mathrsfs}
\usepackage{hyperref}
\usepackage{xcolor}
\usepackage{subfigure}
\usepackage{slashed}
\usepackage{epsfig,color}
\usepackage{multirow}

\usepackage{float}

\definecolor{red}{rgb}{1.00, 0.00, 0.00}

\definecolor{blue}{rgb}{0.00, 0.00, 1.00}
\definecolor{green}{rgb}{0.10, 1.00, .10}

\newcommand{\Real}{\operatorname{Re}}
\newcommand{\Imag}{\operatorname{Im}}
\newcommand{\MS}{\overline{\mathrm{MS}}}
\newcommand{\BL}{\mathrm{BL}}
\newcommand{\Loc}{\mathrm{L}}
\newcommand{\RI}{\mathrm{RI}}
\newcommand{\Lat}{\mathrm{Lat}}
\newcommand{\LD}{\mathrm{LD}}
\newcommand{\SD}{\mathrm{SD}}

\begin{document}
\title{Long-distance contribution to $\epsilon_K$ from lattice QCD}

\newcommand\bnl{Brookhaven National Laboratory, Upton, NY 11973, USA}
\newcommand\cu{Physics Department, Columbia University, New York, NY 10027, USA}
\newcommand\jlab{Thomas Jefferson National Accelerator Facility, Newport News, Virginia, USA.}
\newcommand\kntky{Department of Physics and Astronomy, University of Kentucky, Lexington, KY 40506, USA}
\newcommand\riken{RIKEN-BNL Research Center, Brookhaven National
Laboratory, Upton, NY 11973, USA}
\newcommand\soton{School of Physics and Astronomy, University of
Southampton,  Southampton SO17 1BJ, UK}

\author{Ziyuan Bai}\affiliation{\cu} 
\author{Norman H. Christ}\affiliation{\cu}
\author{Joseph M. Karpie}\affiliation{\cu}\affiliation{\jlab}
\author{Christopher T.~Sachrajda}\affiliation{\soton}
\author{Amarjit Soni}\affiliation{\bnl}
\author{Bigeng Wang}\affiliation{\cu}\affiliation{\kntky}

\date{September 01, 2023}

\begin{abstract}
A lattice QCD approach to the calculation of the long-distance contributions to $\epsilon_K$ is presented. This parameter describes indirect CP violation in $K\to\pi\pi$ decay.  While the short-distance contribution to $\epsilon_K$ can be accurately calculated in terms of standard model parameters and a single hadronic matrix element, $B_K$, there is a long-distance part which is estimated to be approximately $5\%$ of the total and is more difficult to determine.  A method for determining this small but phenomenologically important contribution to $\epsilon_K$ using lattice QCD is proposed and a complete exploratory calculation of the contribution is presented. This exploratory calculation uses an unphysical light quark mass corresponding to a 339 MeV pion mass and an unphysical charm quark mass of 968 MeV, expressed in the $\MS$ scheme at 2 GeV. This calculation demonstrates that future work should be able to determine this long-distance contribution from first principles with a controlled error of 10\% or less.
\end{abstract}

\maketitle
%\tableofcontents

\section{Introduction}

The $K_L - K_S$ mass difference, $\Delta M_K$ and the measure of indirect CP violation in kaon decay, $\epsilon_K$, are two important quantities originating from highly-suppressed, second-order weak processes.  Both have precisely-measured experimental values, making them ideal tests of the standard model if the standard-model predictions for these quantities could be accurately computed.  As second-order weak processes both involve the exchange of two $W$ bosons and correspond to the CP conserving ($\Delta M_K$) and CP violating ($\epsilon_K$) components of $K^0$-$\overline{K}^0$ mixing.  However, the largest contributions to each of these quantities come from very different kinematic regions. 

With their larger Cabibbo-Kobayashi-Maskawa (CKM) matrix elements, the up and charm quarks are the dominant intermediate quarks in the calculation of $\Delta M_K$ giving a much larger contribution than the top quark.  Consequently $\Delta M_K$ is described as a long-distance quantity, coming predominately from the energy scale of the charm quark mass.  As a result the two $W$ boson exchanges that contribute to $\Delta M_K$ can be treated as two effective $\Delta S=1$ four-quark interactions and these two local operators are typically separated by a distance on the order of the Compton wavelength of the charm quark.  For shorter distances the difference between the up and charm quark masses can be neglected and the Glashow-Iliopoulos-Maiani  (GIM) mechanism implies that the integral over the spatial separation between these two operators will converge in this short-distance region where the two operators approach each other.

The dominance of the charm energy scale makes $\Delta M_K$ a difficult quantity to compute using QCD perturbation theory. As discovered by Brod and Gorbahn~\cite{Brod:2011ty}, the next-to-next-to-leading order (NNLO) contribution to $\Delta M_K$ is 36\% of the sum of the leading-order~(LO) and next-to-leading-order (NLO) contributions, making such a perturbative calculation unreliable.  Therefore, lattice QCD is at present the most promising approach to determine $\Delta M_K$ from the standard model, with all errors controlled.  Lattice methods to calculate $\Delta M_K$ have been introduced and demonstrated in Refs.~\cite{Christ:2012se} and \cite{Bai:2014cva}. However, these are difficult calculations at present~\cite{Wang:2022lfq} because they must be performed with a lattice spacing that is small compared to the Compton wavelength of the charm quark.

The situation is quite different for the standard model contribution to $\epsilon_K$.  Because this quantity is CP violating, the magnitudes of the relevant CKM matrix elements no longer suppress the top quark and the GIM mechanism no longer applies.  As a result the largest contribution to $\epsilon_K$ comes from energies on the order of the top quark mass and the two $W$ boson exchanges are well represented at the mass scale of the decaying kaon by a single local $\Delta S=2$ four-quark operator.  Thus,  $\epsilon_K$ can be described as short-distance dominated.  However, there is a subdominant part which comes from longer distances and can be described by the product of two distinct $\Delta S=1$ four-quark operators, separated by distances much larger than $1/M_W$.

Because of their small size, these long-distance contributions to $\epsilon_K$ are conventionally treated in an approximate way.  All charm quark contributions are treated as far above the QCD energy scale, $\Lambda_{\mathrm{QCD}}$ and represented by a local, $\Delta S=2$ operator while the component coming from the up quark is effectively neglected.  The errors associated with this approximate treatment of the long-distance contribution to  $\epsilon_K$ are estimated to be a few percent~\cite{Buras:2010pza}.  

With this approximate treatment of the long-distance part of $\epsilon_K$, both the short- and long-distance contributions to $\epsilon_K$ can be written as the product of a perturbatively computed Wilson coefficient and the matrix element of a local $\Delta S=2$ operator between $K^0$ and $\overline{K}^0$ states.  This matrix element can now be evaluated using lattice methods to 2\% accuracy (see for example Refs.~\cite{Durr:2011ap, Blum:2014tka, Jang:2015sla} and the recent compilation~\cite{FlavourLatticeAveragingGroupFLAG:2021npn}) with the largest uncertainty in the standard model prediction for $\epsilon_K$ coming from the CKM matrix element $V_{cb}$ which appears to the fourth power in the Wilson coefficient.  

As future experiments reduce the uncertainty in $V_{cb}$, it will become increasingly important to improve the accuracy of the calculation of these long-distance effects. In this paper we give a complete description of methods based on lattice QCD which directly evaluate these long-distance contributions with full control of systematic errors.  We present an exploratory calculation which demonstrates these methods and provides evidence that such a lattice calculation determining this long-distance contribution to 10\% accuracy should be practical as a large-scale project on the current generation of supercomputers.  As we will describe, the largest difficulty at present is the practical challenge of using a sufficiently small lattice spacing that the charm quark can be treated accurately and at the same time a sufficiently large lattice volume that physical-mass pions can be included without large finite-volume distortions.

We briefly outline the new difficulties that such a calculation must overcome beyond those found in a calculation of $\Delta M_K$ .  First, because we are evaluating the imaginary part of the kaon mixing matrix element $M_{\overline{0}0}$, the top quark contribution can no longer be neglected. We therefore must include all the QCD penguin operators in a calculation of $\epsilon_K$ while in principle a result for $\Delta M_K$ that is accurate to 1\% could be obtained from only the current-current operators $Q_1$ and $Q_2$, defined below Eq.~\eqref{Eq:HW_mix2}. The inclusion of QCD penguin operators requires that significantly more matrix elements be determined as well as diagrams with a new topology, not present in the calculation of $\Delta M_K$.

Second, the absence of the GIM mechanism for these top quark contributions implies that many diagrams of interest will contain a logarithmic divergence arising when the positions of the pair of local, four-quark operators collide.  In the complete theory such divergences would be absent, regulated by the non-local structure at short distances coming from the $W$ boson and top quark propagators.  Of course, in our lattice calculation such short-distance structure is absent and these singular position-space sums are cut off at the lattice scale, proportional to the inverse lattice spacing $a^{-1}$. For a proper continuum limit to be taken, these divergences must be removed.

The appearance of such unphysical singularities is of course a standard occurrence when the operators that define a first-order effective field theory are used in a second order calculation.  Additional low energy constants must be specified before the second-order theory is well defined.  The ambiguities that appear when products of these lowest order operators are evaluated are resolved by these new second-order low-energy constants.  We refer to these products of pairs of local operators as bilocal operators.  The singularity that results when the two operators collide can be removed by the subtraction of a local operator.  Such a subtraction renormalizes the bilocal operator and can be specified in the $\MS$ or a generalized ``regularization invariant symmetric momentum'' (RI/SMOM) scheme suitable for continuum or lattice regularization.  Once the bilocal operators have been renormalized, additional local operators corresponding to the necessary subtractions can be added.  Their coefficients are these new, well-defined low-energy constants (LECs).  These LECs must then be determined by comparison with the underlying short-distance theory.  This procedure is standard in perturbative calculations~\cite{Buchalla:1995vs} and has been  previously discussed and implemented for lattice calculations as well~\cite{Christ:2012se, Christ:2016eae, Bai:2017fkh}.

Since the necessary low energy constants have already been determined when the bilocal operators are renormalized in the $\MS$ scheme~\cite{Buras:1990fn, Buchalla:1995vs, Brod:2010mj, Brod:2011ty}, we use QCD perturbation theory to relate the $\MS$ and RI/SMOM schemes and a lattice calculation to provide a non-perturbative relation between the lattice-regulated bilocal operator and a bilocal operator renormalized in the RI/SMOM scheme.

In this paper we will refer to the distance scales at which perturbation theory can be safely applied as ``short-distance'', expecting that these correspond to energies at or above a lower limit of 2-3~GeV.  We will describe lower energy scales than these as corresponding to ``long-distance''.  This language is intended to distinguish the regions in which perturbative and lattice methods can be used.  The fact that lattice methods may now be viable up to this 2-3 GeV energy scale and below is the main motivation for this paper.  This 2-3 GeV boundary between the short- and long-distance regions may be appropriate for a target accuracy on the order of a few percent.  Given the presumed asymptotic nature of the QCD perturbation series, it is likely that to achieve significantly higher accuracy, the perturbative, short-distance region will need to be revised upward and an even larger long-distance region treated using lattice methods. Achieving this improvement will require future calculations with an even finer lattice spacing in order to continue to control discretization errors.

In Section~\ref{section:Calculation} we review the theoretical framework for the calculation of $\epsilon_K$ in the standard model and present the combinations of four-quark effective operators that must be used in a lattice calculation of the long-distance contribution to $\epsilon_K$.  We then describe the techniques used to evaluate the needed matrix elements using lattice QCD.  In Section~\ref{section:SD_correction} we describe in detail the method used to subtract the unphysical, short-distance part of these lattice-regulated matrix elements so that the resulting renormalized operators obey regularization-independent conditions allowing these operators to be related to the more conventional bilocal operators appearing in perturbation theory, renormalized in the $\MS$ scheme.  In Section~\ref{section:Results} we present and discuss our numerical results for $\epsilon_K$ while Section~\ref{section:conclusion} contains our conclusions.  The Appendix contains details of the more conventional renormalization procedure used to renormalize the local, four-quark operators.

\section{Theoretical aspects of the calculation of $\epsilon_K$}
\label{section:Calculation}

In this section we will review the usual formulae which determine the indirect CP-violation parameter $\epsilon_K$ in the standard model and provide a connection between the perturbative treatment of $\epsilon_K$, now partialy carried out to NNLO~\cite{Brod:2010mj, Brod:2011ty}, and the lattice calculation which is the subject of this paper.  The theoretical framework for the standard model determination of $\epsilon_K$ and the usual electroweak and QCD perturbative approach to the calculation is reviewed in Section~\ref{sec:basic-formulae}.  Section~\ref{sec:lattice-modify} describes the modifications to this conventional perburbative approach which is proposed here, while in Section~\ref{sec:lattice-method} we present the new lattice QCD methods that we employ to calculate the long-distance contribution to $\epsilon_K$ with controlled errors.

\subsection{Basic standard model formulae}
\label{sec:basic-formulae}

The standard analysis of neutral kaon decay expresses $\epsilon_K$ as
\begin{eqnarray}\label{Eq:epsK1}
	\epsilon_K &=& e^{i\phi_{\epsilon}} \sin{\phi_{\epsilon}} \left( 
\frac{-\Imag{M_{\overline{0}0}}}{\Delta M_K} +\frac{\Imag{A_0} } {\Real{A_0}} \right)
\end{eqnarray}
where $\Delta M_K$ is the mass difference between the long- and short-lived neutral $K$ mesons and $\Gamma_S - \Gamma_L$ (which appears below) is the difference between their decay widths.  Here $A_0$ is the complex $I = 0$ amplitude for $K \rightarrow \pi \pi$ decay after its strong interaction ``Watson'' phase, $e^{i\delta_0}$, has been removed, where $\delta_0$ is the $I=0$, $s$-wave $\pi\pi$ scattering phase shift evaluated for a $\pi\pi$ center-of-mass energy equal to the kaon mass.  The angle $\phi_\epsilon$ is defined by
\begin{equation}\label{Eq:epsK2}
\phi_\epsilon = \tan^{-1}\left(\frac{2\Delta M_K}{\Gamma_S - \Gamma_L}\right) = 43.51(5)^\circ.
\end{equation}

The quantity $M_{\overline{0}0}$ is the dispersive part of the $K^0$-$\overline{K}^0$ mixing matrix and is conventionally written in the non-covariant form:
\begin{eqnarray}
M_{\overline{0} 0} = \langle \overline{K}^0 | H_W^{\Delta S=2} | K^0 \rangle
	 + \mathcal{P}\sum_n \frac{\langle \overline{K}^0 | H_W^{\Delta S=1} | n \rangle
     \langle n |H_W^{\Delta S=1}|  K^0 \rangle}{M_K - E_n} 
\label{Eq:M00},
\end{eqnarray}
where the $\mathcal{P}$ indicates that the principal part should be taken to resolve the singularity when $E_n = M_K$ in the generalized sum over intermediate states labeled by the index $n$.  Here and later in this paper we will use unit normalization for the finite-volume kaon states $|K(\vec p_{\vec n})\rangle$: $\langle K(\vec p_{\vec n}) | K(\vec p_{\vec n'}) \rangle = \delta_{\vec n,\vec n'}$.  We adopt the usual phase conventions in which $CP|K^0\rangle = -|\overline{K}^0\rangle$, time reversal symmetry requires $A_0$ to be real and $T$ or $CP$ symmetry requires $M_{\overline{0} 0}$ to be real.

The standard formula given in Eq.~\eqref{Eq:M00} involves two local, effective four-quark operators.  The first, $H_W^{\Delta S=2}$ describes a second-order-weak, $\Delta S=2$ transition arising from the short-distance part of the exchange of two $W$ bosons while $H_W^{\Delta S=1}$ describes a first-order process in which a single $W$ boson has been exchanged.  The second term, containing two insertions of $H_W^{\Delta S=1}$, includes both short- and long-distance effects depending on whether the intermediate state $|n\rangle$ carries a large or a small energy.  Of course, this term is appropriate only when the intermediate-state energy $E_n$ is sufficiently small that the transition amplitude $\langle n |H_W^{\Delta S=1}|  K^0 \rangle$ can be accurately described by a point-like treatment of the $W$ boson exchange. The contributions of intermediate states of higher energy must be represented by the first, $H_W^{\Delta S=2}$ matrix element.

\begin{figure}[ht]
	\centering
	\begin{tabular}{cc}
		\includegraphics[width=0.4\textwidth]{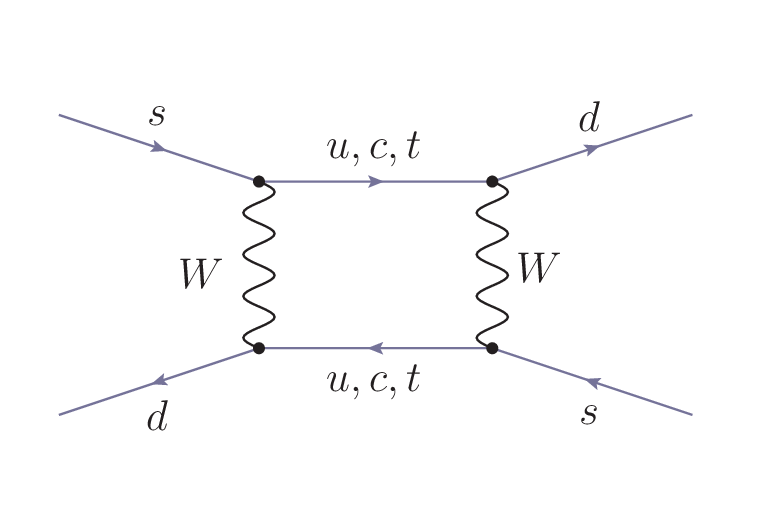} & 
		\includegraphics[width=0.4\textwidth]{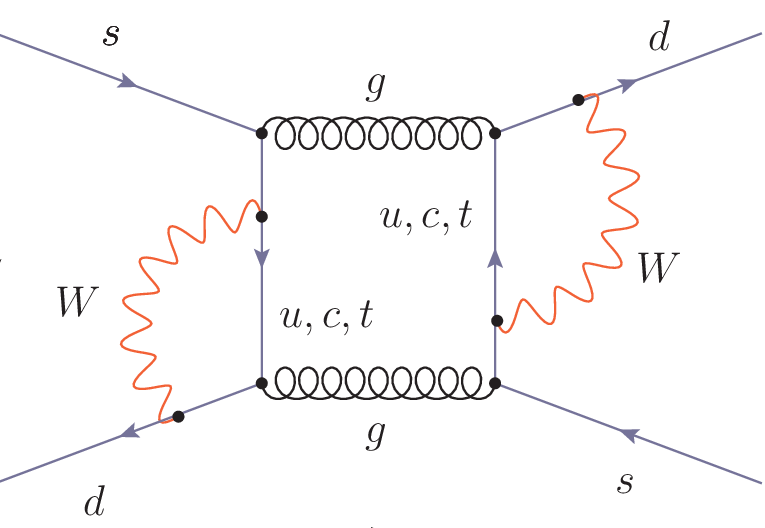} \\
		Connected topology & Disconnected topology\\
	\end{tabular}
	\caption{Two types of $\Delta S =2$ diagram contributing to $\epsilon_K$.}
	\label{figure:two_diagrams}
\end{figure}

A more explicit and covariant standard-model description of $K^0-\overline{K}^0$ mixing is represented by Feynman diagrams of the sort shown in Fig.~\ref{figure:two_diagrams}.  However, the expression for  $M_{\overline{0}0}$ given in Eq.~\eqref{Eq:M00} is more appropriate for a lattice QCD calculation in which $W$ boson exchange must always be described by a local effective four-quark coupling.  A well-defined division of the short- and long-distance effects between the two terms in  Eq.~\eqref{Eq:M00} is required for a meaningful lattice calculation and is an important part of this paper.

The two diagrams shown in Fig.~\ref{figure:two_diagrams} each represent one of the two types of diagram which contribute to $K^0-\overline{K}^0$ , mixing.  The left-hand diagram  shows the connected topology in which the $K^0$ and $\overline{K}^0$ sources cannot be separated unless two quark lines are cut.  The right-hand diagram shows an example of the remaining diagrams which have what we label as a disconnected topology.  Here the  $K^0$ and $\overline{K}^0$ sources can be separated by cutting only gluon lines.  In the left diagram we do not show the gluon lines that would be present in both a perturbative or lattice QCD calculation.  We do include two curly gluon lines in the right-hand diagram both to make it clear that it is only the set of quark lines that is disconnected and to show a topology which contributes to gluonic penguin amplitudes.

In both cases we can identify two quark lines that pass through the diagram and convert a strange quark into a down quark.  Each of these quark lines has two weak vertices: one converts the strange quark to an up-type quark (up, charm or top) and the second converts that up-type quark into a down quark.  Thus, each line will introduce a factor of $\lambda_i = V_{id} V^*_{is}$ where $V_{qq'}$ is the CKM matrix element connecting a $q'$ down-type quark  to a $q$ up-type quark and $i=u$, $c$ and $t$.  Because of the flavor symmetry of QCD each of these three terms are identical except for the factor of $\lambda_i$ and the mass which enters the up-type quark line joining the two weak vertices.

The orthogonality of the first and second columns of the unitary CKM matrix implies
\begin{eqnarray}
	\lambda_u + \lambda_c + \lambda_t = 0,
\label{eq:CKM-unitarity}
\end{eqnarray}
which can be used to combine the sum over the three up-type quarks that appear in Fig.~\ref{figure:two_diagrams} into a sum over two terms. The usual choice is to eliminate $\lambda_u$ by subtracting a term with the sum over the three CKM products in Eq.~\eqref{eq:CKM-unitarity} but with an up-quark line connecting the two weak vertices, a term which vanishes because of Eq.~\eqref{eq:CKM-unitarity}. This subtraction removes the up-quark contribution proportional to $\lambda_u$ and replaces the original terms proportional to $\lambda_c$ and $\lambda_t$ with terms containing the difference between the charm- and up-quark line and top- and up-quark line, respectively.

This can be illustrated for the lowest-order connected diagram in Fig.~\ref{figure:two_diagrams} by the free field propagator formula:
\begin{equation}
	\sum_{i = u,c,t} \frac{\lambda_i\, \slashed{p} } {p^2 +  m_i^2} = \lambda_c\left\{ \frac{
	\slashed{p} }{p^2 + m_c^2} - \frac{ \slashed{p} }{p^2 + m_u^2} \right\}
	+ \lambda_t \left\{ \frac{\slashed{p}} {p^2 + m_t^2} - \frac{\slashed{p}} {p^2 - m_u^2} \right\} .
\label{eq:up-subtraction}
\end{equation}
This use of CKM unitarity to eliminate one of the three products, $\lambda_u$, $\lambda_c$ or $\lambda_t$, as in Eq.~\eqref{eq:up-subtraction}, is easiest to illustrate for the case in which the up-type quark propagators directly connect the two weak vertices.  However, since the three flavors of quarks have the same gluon couplings, this simplification applies generally when an arbitrary number of gluon vertices are inserted on the intermediate up-type-quark line, including the case of one gluon vertex insertion as in the disconnected graph shown in Fig.~\ref{figure:two_diagrams}.

In the conventional calculation of $\epsilon_K$ the entire contribution to the off-diagonal, $K^0-\overline{K}^0$ mixing term $M_{\overline{0}0}$ is expressed as the $K^0-\overline{K}^0$ matrix element of a local, $\Delta S=2$ operator $H_{\mathrm{eff,conv}}^{\Delta S = 2}$ given by
\begin{eqnarray}
H_{\mathrm{eff,conv}}^{\Delta S = 2} &=& \frac{G_F^2}{16 \pi^2} M_W^2 \left[
\lambda_c^2 \eta_1 S_0(x_c) + \lambda_t^2 \eta_2 S_0(x_t) 
+ 2\lambda_c \lambda_t \eta_3 S_0(x_c, x_t) \right] O_{LL} + {\rm h.c.}
\label{Eq:Heff_conv}
\end{eqnarray}
where
\begin{equation}
O_{LL} = (\overline{s} d)_{V-A} (\overline{s}d)_{V-A}
\label{eq:LL}
\end{equation}
and the subscript $V-A$ indicates the usual difference of vector and axial vector currents.
The functions $S_0(x)$ and $S_0(x,y)$ are Inami-Lim functions~\cite{Inami:1980fz} and their arguments are the mass ratios $x_q=m_q^2/m_W^2$ for $q = u$, $c$ and $t$.  The coefficients $\eta_i$, $i=1,2,3$ would each be unity in a lowest order calculation and incorporate corrections of first or higher order in $\alpha_s$.  In this conventional approach the charm quark is treated as heavy compared to the QCD scale and integrated out while the up quark is treated as a massless fermion with perturbative QCD couplings.  While we choose not to use this description, the contribution obtained in this way is often referred to as the short-distance part and long-distance corrections are computed by adding dimension-eight operators to better represent the charm physics in the three-flavor theory and a more refined treatment of the light quarks is given using chiral perturbation theory~\cite{Buras:2010pza}.

\subsection{Alternative application of CKM unitarity}
\label{sec:lattice-modify}

In the approach developed here we rely on these perturbative calculations of the $\eta_i$ but make two changes.  The most important, discussed in greater depth in Sec.~\ref{sec:lattice-method} and \ref{section:SD_correction}, is that we use QCD perturbation theory only above the charm quark scale and always work in the four-flavor theory with an active charm quark. The dominant contribution to $\epsilon_K$ will continue to be described by a local operator proportional to $O_{LL}$ given by a formula similar in structure to Eq.~\eqref{Eq:Heff_conv}.

The four-flavor effective theory is then renormalized in such a way that this $O_{LL}$ operator provides the entire standard model contribution, determined by electroweak and QCD perturbation theory, to a specific $\Delta S=2$, four-quark Green's function in which the external quark lines carry large non-exceptional momenta specified at a scale $\mu$ above the charm quark mass $m_c$.  In contrast, the conventional $\eta_i$ coefficients in Eq.~\eqref{Eq:Heff_conv} are chosen so that this operator will reproduce a perturbative calculation of the standard model contribution to a similar four-quark Green's function in the three-favor theory with vanishing external quark momenta. 

There are two important advantages to this modified approach: i) The modified $O_{LL}$ operator can be determined from perturbation theory in a infrared safe regime and the use of perturbation theory will become increasingly accurate as the scale $\mu$ is increased.  ii) All low-energy, standard-model matrix elements can be computed from this modified $\Delta S=2$, $O_{LL}$ operator combined with the usual $\Delta S=\pm1, \Delta C=\pm1$ four-quark weak effective theory evaluated at second order.  We can then use lattice QCD, including an active charm quark, to evaluate the matrix elements in this second-order effective theory, avoiding the use of QCD perturbation theory at or below the charm quark scale.  As is explained in Secs.~\ref{sec:lattice-method} and \ref{section:SD_correction}, these two partial representations of standard model physics can be combined to give the correct standard model prediction if we properly choose the low energy constants in the second-order effective theory.

The second change that we make to the conventional approach is theoretically less significant, but is important for numerical computation.  In the conventional calculation, the CKM unitarity relation given in Eq.~\ref{eq:CKM-unitarity} is used to remove the up quark contribution proportion $\lambda_u$, reducing the UV divergences in the terms proportional to $\lambda_c$ and $\lambda_t$~\cite{Inami:1980fz} as suggested in Eq.~\eqref{eq:up-subtraction}.  However, any quark flavor might have been chosen.   Instead of subtracting a term multiplied by the vanishing sum $\sum_i \lambda_i$ containing an up-quark line and eliminating the term proportional to $\lambda_u$, we instead subtract a term containing a charm-quark line to eliminate the original term proportional to $\lambda_c$, replacing the substitution shown in Eq.~\eqref{eq:up-subtraction} by that corresponding to a charm-quark subtraction:
\begin{equation}
	\sum_{i = u,c,t} \frac{\lambda_i\, \slashed{p} } {p^2 +  m_i^2} = \lambda_u\left\{ \frac{
	\slashed{p} }{p^2 + m_u^2} - \frac{ \slashed{p} }{p^2 + m_c^2} \right\}
	+ \lambda_t \left\{ \frac{\slashed{p}} {p^2 + m_t^2} - \frac{\slashed{p}} {p^2 - m_c^2} \right\} .
\label{eq:charm-subtraction}
\end{equation}

The effects of this alternative charm-quark subtraction may be easiest to discuss if we focus on the resulting change to the $\Delta S=2$ effective Hamiltonian given in Eq.~\eqref{Eq:Heff_conv}.  With this alternative subtraction $H_W^{\Delta S=2}$ will have the form:
\begin{eqnarray}
\label{Eq:Heff}
H_W^{\Delta S = 2} &=& \frac{G_F^2}{16 \pi^2} M_W^2 \left[
\lambda_u^2 \eta'_1 S_0(0,0,x_c) + \lambda_t^2 \eta'_2 S_0(x_t,x_t, x_c)  \right. \\\nonumber
&& \left. \hskip 2cm + 2\lambda_u \lambda_t \eta'_3 S_0(x_t,0,x_c) \vphantom{\lambda_u^2 } \right] O_{LL} + {\rm h.c.}
\end{eqnarray}
Since in Eq.~\eqref{Eq:Heff}, we have made the choice of subtracting the charm quark, we use a different notation to represent the general form of the Inami-Lim functions and the corresponding QCD corrections. 

Here we use a single Inami-Lim function showing three arguments $S_0(x_1, x_2, x_3)$ in a way that explicitly displays the internal quark structure.  Now  $S_0(x_1, x_2, x_3)$ gives the result from a box diagram where one up-type quark line involves the $(q_1 - q_3)$ difference of propagators while the other up-quark line has been replaced by the difference $(q_2 - q_3)$.  Thus the last argument, $x_3$ depends on the mass of the subtracted up-type quark. For simplicity we have made the usual choice $x_u = 0$. However, for clarity we do not drop this argument as is done by Inami and Lim. In this notation, the Inami-Lim functions with the standard $u$ quark subtraction are given by $S_0(x_c) = S_0(x_c,x_c,0)$ and $S_0(x_c,x_t) = S_0(x_c,x_t,0)$.  For the perturbative coefficients $\eta'_i$ which represent the QCD corrections, we have added a prime to distinguish those needed in our formulation from the conventional factors. 

We make this unconventional choice of charm quark subtraction because we now have only a single term (the $\lambda_u \lambda_t$ term) to calculate from lattice QCD, reducing the computational cost compared to the conventional subtraction. Following the usual choice of CKM phases the term proportional to $\lambda_u \lambda_u$ is purely real and does not contribute to $\Imag{M_{\overline{0}0}}$.  The remaining two terms give contributions of nearly the same size. This is because $S_0(x_t,0,x_c) = O(10^{-3})$, $S_0(x_t,x_t,x_c) = O(1)$ and $\lambda_t/\lambda_u \approx 0.0016$.  The term proportional to $\lambda_t \lambda_t$ can be calculated in QCD perturbation theory with very high precision since it is short-distance dominated.  One might think that this term also has a contribution coming from two internal charm quarks carrying small momenta which could not be accurately evaluated using QCD perturbation theory.  However, such a contribution is suppressed by the ratio $\lambda_t/\lambda_u$ compared to similar effects in the $\lambda_u \lambda_t$ term.  Only the $\lambda_u \lambda_t$ term has a long-distance contribution which is not suppressed by CKM factors and therefore this is the only term that requires a lattice QCD calculation.  Thus, this charm quark subtraction significantly reduces the cost of the lattice calculation.  

Since the coefficients $\eta_i$ and $\eta_i'$ and the Inami-Lim function $S(x,y,z)$ do not depend on the CKM matrix element products, it is straight-forward to relate the two sets of coefficients $\eta_i$ and $\eta_i'$ to arbitrary order in $\alpha_s$ by comparing Eqs.~\eqref{Eq:Heff_conv} and \eqref{Eq:Heff} viewed as second-order polynomials in two of the three quantities, $\lambda_q$, $q=u$, $c$ and $t$.  Using $\lambda_t$ and $\lambda_u$ we can express $\{\eta_i'\}_{i=1,2,3}$ in terms of $\{\eta_i\}_{i=1,2,3}$:
\begin{eqnarray}
\eta_1'                             &=& \eta_1 \label{Eq:eta1prime} \\
\eta'_2 S_0(x_t,x_t,x_c)  &=& \eta_2 S_0(x_t) + \eta_1 S_0(x_c) - 2\eta_3 S_0(x_c, x_t) \label{Eq:eta2prime} \\
\eta'_3 S_0(x_t,0,x_c)     &=& \eta_1 S_0(x_c) - \eta_3 S_0(x_c, x_t).  \label{Eq:eta3prime} 
\end{eqnarray}
As is required by their definitions, each of the six coefficients $\eta_i$ and $\eta_i'$ are one at zeroth order in $\alpha_s$.  In the context of Eqs~\eqref{Eq:eta2prime} and \eqref{Eq:eta3prime} this requires that 
\begin{eqnarray}
S_0(x_t,x_t,x_c)  &=& S_0(x_t) + S_0(x_c) - 2S_0(x_c, x_t), \label{Eq:Sttc} \\
S_0(x_t,0, x_c)    &=& S_0(x_c) - S_0(x_c, x_t). \label{Eq:Stuc}
\end{eqnarray}

\subsection{Lattice calculation of the long-distance contribution to $\epsilon_K$}
\label{sec:lattice-method}

We will now discuss in greater detail how lattice QCD can be used to calculate the $\lambda_u \lambda_t$ contribution to $\Imag{M_{\overline{0}0}}$ coming from energy scales on the order of the charm quark mass and below.  While we found it convenient to discuss the unconventional aspects of our calculation in the context of the $\Delta S=2$ effective Hamiltonian given in Eq.~\eqref{Eq:Heff}, our calculation requires a complete effective theory of the low-energy weak interactions including an active charm quark which we will now determine. 

\subsubsection{Effective theory of $\Delta S=2$ weak interactions}
\label{sec:eff-theory}

For completeness, we begin at the energy of the $W$ boson and top quark where the weak interactions are described by diagrams of the sort shown in Fig.~\ref{figure:two_diagrams}.  
In the first step the sums over $u$, $c$ and $t$ propagators shown in those figures are rearranged following the charm-quark subtraction scheme described in Eq.~\eqref{eq:charm-subtraction}.  Thus, each of the two up-type quark lines connecting the weak vertices in Fig.~\ref{figure:two_diagrams} will be replaced by a difference of $q$ minus $c$ quark propagators and each such difference will be associated with the CKM product $\lambda_q$ where $q=u$ or $t$.  It is those terms proportional to the product of $\lambda_t\lambda_u$ which we must represent in our low energy effective theory. 

These diagrams are simplified by integrating out the W boson and the top quark, resulting in an effective bilocal product of two $\Delta S = 1$ operators which correspond to the two exchanged $W$ bosons.  In addition the local $\Delta S=2$ operator $O_{LL}$ of Eq.~\eqref{eq:LL} will appear, correcting the contribution of the bilocal product when the locations of the two operators coincide.  This description will be valid at energies sufficiently far below the $W$ scale that the $W$ propagators cannot be distinguished from four-dimensional position-space delta functions.  The Wilson coefficients of the operators in the bilocal product and the local operator $O_{LL}$ can be reliably computed in perturbation theory provided the energy scale $\mu$ at which these operators are defined is not too far below the $W$ mass so that any logarithms of the form $\ln(M_W/\mu)$  that appear in the perturbative expansion will not be large. 

Next, renormalization group evolution can be used to replace the combination of bilocal and local operators renormalized at a scale below, but on the order of, $M_W$ by a similar combination of bilocal and local operators renormalized at a much lower energy on the order of the charm quark mass. Specifically, we will want to evolve to an energy scale accessible to lattice QCD, but sufficiently large that this perturbative treatment is accurate and the charm quark is still active.   Here the renormalization group can control the large logarithms that appear, summing all terms of order $\alpha_s^n \left(\alpha_s \ln(M_W/m_c)\right)^l$ for all $l\ge l_{\mathrm{min}}(n)$.   The coefficients of these terms are now known for $n=-1$ (LO),  0 (NLO) and partially for $n=1$ (NNLO)~\cite{Brod:2010mj,  Brod:2011ty}.  Note $ l_{\mathrm{min}}(-1) = 1$ so there is no term which behaves as $1/\alpha_s$.   The result of this perturbative analysis is an expression for $K^0$-$\overline{K}\,^0$ mixing that is written as the $K^0$-$\overline{K}\,^0$ matrix element of the sum a local and bilocal operator:
\begin{eqnarray}
M_{\overline{0}0} &=& \langle \overline{K}^0|\left\{\int d^4 x H_W^{\Delta S=1}(x,\mu) H_W^{\Delta S=1}(0,\mu)\right\}_\mu + H_W^{\Delta S=2}(0,\mu)|K^0\rangle.
\label{eq:2nd-order-eff}
\end{eqnarray}

The two $H_W^{\Delta S=1}(x,\mu)$ operators are the standard four-quark operators that appear in the low energy effective theory of the weak interactions at first order in the Fermi constant, $G_F$.  They must be renormalized and the argument $\mu$ specifies the scale at which this renormalization of $H_W^{\Delta S=1}$ is performed.  The second term in Eq.~\eqref{eq:2nd-order-eff} is a four-quark operator which changes strangeness by two units.  As discussed above, it serves two purposes.  First it represents the low-energy effects of high-energy phenomena, such as the contribution of a top quark loop.  Second it introduces a counter term or low-energy constant that corrects the unphysical singularity that is encountered in the space-time integral in Eq.~\eqref{eq:2nd-order-eff} as $x \to 0$.  The large curly brackets surrounding the integrated product of the two $\Delta S=1$ operators with the subscript $\mu$ indicates that this integral has been regulated at the scale $\mu$.  It is because of this singular part of the bilocal operator that we must specify the scale $\mu$ at which the product of the two $\Delta S=1$ operators is renormalized.  Note, for convenience we are using the same scale $\mu$ to renormalize the lattice operators which appear in $H_W^{\Delta S=1}(x,\mu)$ as is used to define the bilocal product.   In the usual application of these operators, the dependence on $\mu$ should drop out except for errors arising from the truncation of perturbation theory sums.

Here it may be useful to review in greater detail the further steps that would be taken in a conventional ``short-distance'' calculation of $\epsilon_K$.  In such a treatment the scale describing the kaon state is assumed to be small compared to the charm quark mass and the charm quark is also integrated out.  The result is approximated by the local operator $O_{LL}$ alone, multiplied by the coefficients in Eq.~\eqref{Eq:Heff} and the contribution of a remaining bilocal operator that involves only light quarks is assumed to be adequately approximated by the local operator $O_{LL}$.  A calculation of the $K^0-\overline{K}^0$ matrix element of this $O_{LL}$ operator is then performed using lattice QCD.  Since the charm quark mass ($\sim 1.2$ GeV) is close to the non-perturbative scale $\Lambda_{\rm QCD}$, this procedure can be subject to large errors from three sources:~i) neglect of higher orders of $\alpha_s$ when using QCD perturbation theory to integrate out the charm quark (truncation errors);  ii) omission of higher order terms in the expansion in $(\Lambda_{\rm QCD}/m_c)$ (errors from higher-dimension operators) and iii) neglect of the nonlocal effects associated with the exchange of light quarks between the two $\Delta S=1$ operators (long-distance light-quark effects).  For an estimate of the size of some of these effects see Refs.~\cite{Buras:1990fn, Buras:2010pza}.

These difficulties can be avoided and a result with errors that can be reliably estimated and systematically reduced can be obtained if we use lattice QCD to evaluate the matrix element of the combination of bilocal and the local operators given in Eq.~\eqref{eq:2nd-order-eff} and defined at a scale above the charm quark mass. This method requires an ``active'' charm quark in the calculation, which should be possible with controllable discretization errors using lattice spacings which are currently being studied.  The starting point for such a calculation must be the perturbative results for the coefficients of the operators which appear in Eq.~\eqref{eq:2nd-order-eff}.  These coefficients should be determined at a scale sufficiently far above the charm quark mass that perturbation theory is reliable.

We will now exploit the detailed perturbative treatment reviewed in Ref.~\cite{Buchalla:1995vs} to explicitly determine the operator $O_{LL}$ and the operators which appear in the $\Delta S=1$ effective weak Hamiltonian together with their Wilson coefficients that correspond to the $\lambda_u \lambda_t$ term which we wish to evaluate.   At this stage these operators and their Wilson coefficients should be evaluated in the $\MS$ scheme at an energy scale $\mu_{\MS}$ well above the charm quark threshold.
Here and in the following we will specialize ``$\mu$'' to refer to three renormalization scales: $\mu$ with no subscript will indicate a generic scale not connected with any particular renormalization scheme while $\mu_{\MS}$ and $\mu_{\RI}$ are the scales used in the $\MS$ and RI/SMOM renormalization schemes.

We begin by reviewing how the various terms in the effective field theory arise from the two types of diagrams shown in Fig.~\ref{figure:two_diagrams}.  Referring to that figure and keeping in mind our scheme to use CKM unitarity to subtract a vanishing term containing a charm quark from each of the up-type quark lines, we recognize that the term proportional to $\lambda_u \lambda_t$ has a $t-c$ propagator and a $u-c$ propagator in the place of the two internal quark lines for both types of diagram shown in that figure.  Since the energy scales of the top and charm quark are so different, it is useful to separate this $(t-c)\times(u-c)$ structure into two parts: $t\times(u-c)$ and $c\times(c-u)$. 

We first consider the $t\times(u-c)$ part.  For this case the large top quark mass leads to different behaviors for the connected and disconnected topologies shown in Fig.~\ref{figure:two_diagrams}.  For the connected topology the large mass of the top quark forces the two $W$ propagators to be separated by a distance of order $1/m_t$, reducing the entire graph to the effective four-quark operator $O_{LL}$ with a Wilson coefficient that can be reliably determined in a perturbative calculation.   

However, the situation for the disconnected topology is quite different since here the top quark and one $W$ propagator appear as a short-distance correction to a gluon vertex.  The result is a QCD penguin contribution described by a $\Delta S=1$ four quark operator composed of the two quarks joining the gluon vertex and the two quark which couple to the other end of the gluon line.  The second $W$ propagator leads to a second $\Delta S=1$ four-quark operator which could be of either the current-current or QCD penguin type.  (For an explanation of this standard nomenclature, see for example Ref.~\cite{Donoghue:1992dd}.)  Thus, the  $t\times(u-c)$ part of the $\lambda_u \lambda_t$ contribution is a combination of the operator $O_{LL}$ together with the product of two $\Delta S=1$ operators, one a QCD penguin operator and the other either a QCD penguin or a current-current operator.

For the $c\times(c-u)$ contribution the connected and disconnected diagrams in  Fig.~\ref{figure:two_diagrams} each generate terms that are given by the matrix elements of the local operator $O_{LL}$ as well as bilocal products of current-current and QCD penguin operators.

The detailed steps that relate the underlying two-$W$ exchange weak amplitudes illustrated in Fig.~\ref{figure:two_diagrams} and the second-order effective field theory that can be used in a lattice calculation are reviewed and summarized in Buchalla, {\it et al.}~\cite{Buchalla:1995vs}.  We begin with the four-flavor, first-order theory which when taken to second order will describe the $\lambda_u\lambda_t$ terms of interest:
  \begin{eqnarray}
	\label{Eq:HeffS1}
	H_W^{\Delta S=1} &=& \frac{G_F}{\sqrt{2}} \left(\sum_{q,q'=u,c} V^*_{q's} V_{qd}
	\sum_{i=1,2} C_i Q_i^{q'\bar q} - \lambda_t \sum_{i=3}^{6} C_i Q_i \right) \,,
	\label{eq:H-DS_1}
\end{eqnarray}
where the $C_i$ are Wilson coefficients and
\begin{eqnarray}
	Q_{1}^{q'\overline{q}} &=& (\overline{s}_a q'_b)_{V-A}(\overline{q}_b d_a)_{V-A} \quad\quad
	Q_{2}^{q'\overline{q}} = (\overline{s}_a q'_a)_{V-A}(\overline{q}_b d_b)_{V-A} 
\label{eq:current-current} 
\end{eqnarray}
\begin{eqnarray}
	Q_3 &=& (\overline{s}_a d_a)_{V-A} \sum_{q=u,d,s,c} (\overline{q}_b q_b)_{V-A}  \quad\quad
	Q_4 = (\overline{s}_a d_b)_{V-A} \sum_{q=u,d,s,c} (\overline{q}_b q_a)_{V-A}
\label{eq:QCD-penguin} \\
	Q_5 &=& (\overline{s}_a d_a)_{V-A} \sum_{q=u,d,s,c} (\overline{q}_b q_b)_{V+A}  \quad\quad
	Q_6 = (\overline{s}_a d_b)_{V-A} \sum_{q=u,d,s,c} (\overline{q}_b q_a)_{V+A} \nonumber.
\end{eqnarray}
where sums over the color indices $a$ and $b$ are understood.

The eight, four-quark operators $Q^{q'\bar q}_1$ and $Q^{q'\bar q}_2$ are current-current operators while $Q_3$, $Q_4$, $Q_5$ and $Q_6$ are QCD penguin operators. The electro-weak penguin operators have been dropped since they are suppressed by a factor of $\alpha_{\rm EM}$.  For the current-current operators $Q_i^{q'\bar{q}}$, (i=1,2), the label $q'\bar{q}$ can be any combination of up and charm quarks. The QCD penguin operators involve a symmetrical sum over the four relevant flavors in our calculation. Here the subscript $V-A$ indicates a left-handed vertex and $V+A$ a right-handed one.

The structure of $H_W^{\Delta S=1}$ shown in Eq.~\eqref{eq:H-DS_1} is a consequence of the GIM mechanism and the Wilson coefficients $\{C_i\}$, $1 \le i \le 6$  can be related to the six, 4-flavor Wilson coefficients $\{z_i\}$, $1 \le i \le 2$ and $\{v_i\}$, $3 \le i \le 6$ determined in Ref.~\cite{Buchalla:1995vs}.  (Note that $v_1=z_1$ and $v_2=z_2$ in Ref.~\cite{Buchalla:1995vs}.)  In order to understand the structure of Eq.~\eqref{eq:H-DS_1} we distinguish the irreducible representations of $SU(4)_L\times SU(4)_R$ and follow the renormalization group evolution as one moves from a high-energy scale close to but below $M_W$ down to a scale that is close to but above the charm quark mass.  Because of the mass-independence of the RI/SMOM and $\MS$ renormalization schemes which we use, this renormalization-group scale evolution will be symmetrical under $SU(4)_L\times SU(4)_R$.

A general four-flavor, four-quark ``left-left'' operator has the form:
\begin{equation}
T^{ab}_{cd}\left[\overline{q}_a\gamma^\mu(1-\gamma^5)q^c\right]\left[\overline{q}_b\gamma_\mu(1-\gamma^5)q^d\right],
\label{eq:SU(4)}
\end{equation}
where a sum over the flavor indices $a$, $b$, $c$ and $d$ as well as the space-time index $\mu$ is understood.  Here the color indices are not shown but are to be contracted within each of the square brackets, defining ``color-diagonal'' operators.  A similar family of ``color-mixed'' four-quark operators can be defined if each color index in one square bracket is contracted with the appropriate index in the other.  Such a left-left operator will be a singlet under $SU(4)_R$ while its representation under $SU(4)_L$ is determined by the properties of the tensor $T^{ab}_{cd}$.  For this left-left operator, Fierz symmetry implies that exchanging the indexes $a$ and $b$ on the color-diagonal operator results in the corresponding color-mixed operator with the original order of $a$ and $b$.  

It is consistent with $SU(4)_L$ symmetry to distinguish between tensors $T^{ab}_{cd}$ which are traceless, {\it e.g.} obeying $\sum_{d=1}^4 T^{ad}_{cd}=0$, from the trace term with $T^{ad}_{cd}=t^a_c \delta^b_d$ where $\delta^b_d$ is the usual Kronecker delta.  This trace term transforms as the product of the $4$ and $\overline{4}$ representations of $SU(4)_L$ and will belong to the $(1,1)$ or $(15,1)$ representations of $SU(4)_L\times SU(4)_R$.  For such trace terms this behavior under $SU(4)_L\times SU(4)_R$ is not changed if the appropriate $(1-\gamma^5)$ factor in Eq.~\eqref{eq:SU(4)} is changed to $(1+\gamma^5)$.  In this way one can identify the four distinct groups of fifteen (15,1)  operators to which the four gluonic penguin operators $\{Q_i\}_{3 \le i \le 6}$, belong: color-diagonal and color-mixed, left-left and left-right.

Of equal interest are the traceless tensors which are can be classified according to their symmetry under the exchange of indices of the same type: $T^{ab}_{cd} = \pm T^{ba}_{cd}$.  The symmetrical case defines the (84,1) representation of 
$SU(4)_L\times SU(4)_R$ while the anti-symmetrical case corresponds to the (20,1) representation.  Because of the Fierz symmetry mentioned above, the traceless parts of the combinations  $Q^{q'\overline{q}}_1\pm Q^{q'\overline{q}}_2$ belong to the (84,1) and (20,1) representations respectively.   Note, if operators containing the charm quark are omitted, the traceless part of the (84,1) operator $Q^{u\overline{u}}_1+ Q^{u\overline{u}}_2$ transforms in the (27,1) representation of $SU(3)_L\times SU(3)_R$.  However, there is no traceless part of $Q^{u\overline{u}}_1- Q^{u\overline{u}}_2$ so there is no $SU(3)_L\times SU(3)_R$ analogue to the (20,1) $SU(4)_L\times SU(4)_R$ representation.

Thus the eight current-current operators can be divided into two sets of four operators, $Q^{q',\overline{q}}_1 \pm Q^{q',\overline{q}}_2$ for $q',q\in\{u,c\}$ with components which transform in the symmetrical $(84,1)$ or anti-symmetrical $(20,1)$ representations of the flavor symmetry group $SU(4)_L\times SU(4)_R$.  In contrast, the four QCD penguin operators belong to four distinct irreducible $(15,1)$ representations of $SU(4)_L\times SU(4)_R$.  Thus, if we consider the set of twelve operators defined in Eqs.~\eqref{eq:current-current} and \eqref{eq:QCD-penguin} the eight current-current operators will enter with two distinct Wilson coefficients which evolve independently, each with its own anomalous dimension.  Because each of these eight operators contains a piece belonging to the $(84,1)$ or $(20,1)$ representations, none of these eight operators will be generated when the scheme with which or the scale at which the four QCD penguin operators are renormalized is changed. 

In contrast, the four QCD penguin operators will both mix among themselves and are generated when the renormalization scale of a current-current operator is changed.  However, it is only the $(15,1)$ components of the current-current operators that require the introduction of the QCD penguin operators when their renormalizaiton scale is changed.  Since the GIM mechanism implies that the $(15,1)$ components of the current-current operators are proportional to $\lambda_t$, the structure of $H_W^{\Delta S=1}$ shown in Eq.~\eqref{eq:H-DS_1} can be easily understood.  The two ingredients which require the presence of the QCD penguin operators are both proportional to $\lambda_t$.  The first arises from a virtual top quark whose contribution is necessarily proportional to $\lambda_t$.  The second comes from the four current-current operators which do not change charm and, using Eq.~\eqref{eq:CKM-unitarity}  can be written as 
\begin{equation}
C_i\left(\lambda_u Q^{u\bar u}_i + \lambda_c Q^{c\bar c}_i\right) = C_i(\lambda_u+\lambda_t)\left(Q^{u\bar u}_i - Q^{c\bar c}_i\right) -C_i\lambda_t Q^{u\bar u}_i,
\label{eq:GIM}
\end{equation}
for $i = 1, 2$.  Since the differences $Q^{u\bar u}_i - Q^{c\bar c}_i$ belong to the $(84,1)+(20,1)$ representation, when renormalized they cannot generate QCD penguin operators and the evolution of the Wilson coefficients will introduce the QCD penguin operators only from the final term on the right-hand side of Eq.~\eqref{eq:GIM} which is proportional to $\lambda_t$.

The above discussion reviews the origin of the structure of Eqs.~(6.5) and (6.21) in Buchalla, {\it et al.}~\cite{Buchalla:1995vs} and allows us to relate the 6 Wilson coefficients $C_i$, $1 \le i \le 6$ to the $z_i$ and $v_j$ introduced in that paper:
\begin{eqnarray}
C_i &=& z_i, \quad i\in\{1,2\} \label{eq:Wilson-coefs}\\
C_j &=& v_j, \quad j\in\{3,4,5,6\}.
\nonumber
\end{eqnarray}

\subsubsection{Identifying the $\lambda_u \lambda_t$ terms}
\label{sec:logarithmic_GIM}

In Eq.~\eqref{eq:2nd-order-eff}, after the replacement of $\lambda_c$ by -$(\lambda_u+\lambda_t)$, only terms proportional to a factor $\lambda_u \lambda_t$ contribute to the imaginary component of the matrix element $M_{\overline{0}0}$.  In this section we identify the terms proportional to $\lambda_u \lambda_t$ that contribute to the first term on the right-hand side of Eq.~\eqref{eq:2nd-order-eff}, containing the bilocal product of two local $H_W^{\Delta S=1}$ operators.  The terms proportional to $\lambda_u \lambda_t$ that contribute to the second term containing the local $H_W^{\Delta S=2}$ operator are either straight-forward to identify from the conventional short-distance perturbative calculation or are determined as counter terms required by the first, bilocal term which we will now examine.   These  $\lambda_u \lambda_t$ terms can appear in a number of ways depending on which pair of operators is being considered. When both operators are current-current, as shown in Eq.~\eqref{Eq:Qij_cc} below, a contribution occurs when both operators have the flavor content $q=q'=c$ or when each operator contains both an up and a charm quark.  A second contribution arises when one operator is current-current and the other is a QCD penguin, as shown in Eq.~\eqref{Eq:Qij_cp}. For these second contributions, the current-current operator must have the flavor structure $q=q'=u$ or $q=q'=c$, structures which occur with opposite signs as in the GIM mechanism.

Thus, collecting these terms proportional to $\lambda_u \lambda_t$, we obtain the explicit second-order effective operator  which can be used to evaluate the matrix element on the right-hand side of Eq.~\eqref{eq:2nd-order-eff}:
 \begin{eqnarray}
	\label{Eq:H_eff_dS2}
	H_{eff,ut}^{\Delta S = 2} &=& \frac{G_F^2}{2} \lambda_u \lambda_t  \sum_{i=1,2}
	\left\{\sum_{j=1}^{6} C_i C_j \sum_{x,y} [[\widetilde{Q}_i \widetilde{Q}_j(x,y)]] + C_{7i}\sum_x O_{LL}(x) \right\}\\\label{Eq:Qij_cc}
	[[\widetilde{Q}_i \widetilde{Q}_j (x,y) ]] & = & \frac{1}{2} T\{Q_i^{c\bar c}(x)(Q_j^{c\bar c}(y)-Q_j^{u\bar u}(y))
	                                                              + (Q_i^{c\bar c}(x)-Q_i^{u\bar u}(x))Q_j^{c\bar c}(y) \nonumber \\
                             && \hskip 1 cm - Q_i^{u\bar c}(x)Q_j^{c\bar u}(y) - Q_i^{c\bar u}(x) Q_j^{u\bar c}(y)\} ,\;\; (i,j=1,2)\\\label{Eq:Qij_cp}
[[\widetilde{Q}_i \widetilde{Q}_j(x,y)]] &=& \frac{1}{2} T \{ \left[Q_i^{c\bar c}(x) - Q_i^{u\bar u}(x) \right] Q_j(y) \\\nonumber
 && \hskip 1cm  + Q_j(x)\left[Q_i^{c\bar c}(y) -Q_i^{u\bar u}(y)\right]\} ,\;\; (i=1,2;j=3,...,6)
\end{eqnarray}
where the $T$ in Eqs.~\eqref{Eq:Qij_cc} and \eqref{Eq:Qij_cp} denotes time ordering.  

The notation introduced in the left-hand sides of Eqs.~\eqref{Eq:Qij_cc} and \eqref{Eq:Qij_cp} is intended to serve the following purposes.  First the tilde which appears over the local operator $Q_k$ for $1 \le k \le 6$ indicates that the corresponding factor appearing on the right hand sides of Eqs.~\eqref{Eq:Qij_cc} and \eqref{Eq:Qij_cp} may be linear combinations of products of the four operators  $Q_k^{cc}$, $Q_k^{uu}$, $Q_k^{cu}$ and $Q_k^{uc}$ when $k=1$ or 2.  The double square brackets indicate that these are specially defined bilocal operators, specified in Eqs.~\ref{Eq:Qij_cc} and \ref{Eq:Qij_cp} as linear combinations of multiple products of single pairs of local operators.  Finally, this notation will allow us to later indicate the renormalization scheme that has been imposed by adding two superscripts.  Thus, $[[\widetilde{Q}^U_i \widetilde{Q}^U_j (x,y) ]]^V$ denotes a bilocal operator constructed from individual four-quark operators renormalized in the scheme $U$ while the divergence which arises when the two local operators approach each other is renormalized in scheme $V$.  Here $U$ and $V$ could take the values $\Lat$, $\RI$ and $\MS$.

Referring to the discussion of Wilson coefficients and renormalization of local operators in Section~\ref{sec:eff-theory}, it is important to recognize the limitations on the meaning of the indices $i$ and $j$ in Eqs.~\eqref{Eq:Qij_cc} and \eqref{Eq:Qij_cp} which no longer identify single factors in a product of two local operators.  Because of Eq.~\eqref{eq:Wilson-coefs}, physical, scheme-independent operators can be constructed from the products $C^U_i C^U_j [[\widetilde{Q}^U_i \widetilde{Q}^U_j (x,y) ]]$ if $i$ is summed over 1 and 2 and $j$ summed over 1 through 6, provided the singularity when $x\to y$ is temporarily ignored.  However, because the indices 1 and 2 represent a sum of operators that transform differently under a change of renormalization scheme, we cannot multiply $[[\widetilde{Q}^U_i \widetilde{Q}^U_j (x,y) ]]$ from the left by a $6\times 6$ renormalization matrix $Z^{U\to V}_{ki}$ or $Z^{U\to V}_{kj}$ to change the left or right ``operator'' $\widetilde{Q}^U_i$ or $\widetilde{Q}^U_j$ from the scheme $U$ to the scheme $V$.

As discussed above, the bilocal operator product $[[\widetilde{Q}_i \widetilde{Q}_j (x,y)]]$ is singular as $x$ approaches $y$ and leads to a divergent position-space integral in Eq.~\eqref{Eq:H_eff_dS2} as the continuum limit is taken.  Because of the GIM mechanism and the short-distance chiral symmetry of the domain wall fermion formalism the singularity of the integral is only logarithmic and can be removed by adding an appropriate coefficient to the local $\Delta S=2$ operator $O_{LL}$.  Thus, the Wilson coefficients $C_{7i}$ multiplying $O_{LL}$ in Eq.~\eqref{Eq:H_eff_dS2} will be different from those in Ref.~\cite{Buchalla:1995vs} since these coefficients must now include counter terms to remove these lattice-regulated singularities. 

\subsubsection{Treatment of bilocal operators in lattice QCD}

To describe a second-order process using a Euclidean path integral, we introduce the product of two first-order effective Hamiltonians $H_W^{\Delta S=1}$, integrate their product over a time interval
$[t_a, t_b]$ and define the ``double-integrated'' correlator~\cite{Christ:2012se}:
\begin{equation}
	\label{Eq:intCorr}
	\mathcal{A} = \frac{1}{2} \sum\limits_{t_2 = t_a}^{t_b} \sum\limits_{t_1 
	= t_a}^{t_b} \langle  T \left\{ \overline{K}^0(t_f) 
	H_W(t_2) H_W(t_1) \overline{K}^0 (t_i) \right\} \rangle \,.
\end{equation}
After inserting a sum over intermediate states and performing the summation of $t_1$ and $t_2$, treated here for simplicity as integrations, we find 
\begin{eqnarray}
	\mathcal A = N_{K}^{2} e^{-M_{K} ( t_{f} -t_{i} )} \left\{ \sum_{n }
		\frac{\langle \overline{K}^{0} | H_{W} | n \rangle \langle n | H_{W} | K^{0}
		\rangle}{M_{K} -E_{n}} \left( -T+ \frac{e^{( M_{K} -E_{n} ) T}-1}{M_{K}
	-E_{n}} \right) \right\}\,,
\end{eqnarray}
where $T = t_b - t_a + 1$ is the length of the integration region.
The term proportional to $T$ is the contribution of the bilocal term in Eq.~\eqref{eq:2nd-order-eff}  to $M_{\overline{0}0}$
\begin{eqnarray}
	M_{\overline{0}0}^{\BL} =\sum \limits_{n } \frac{\langle \overline{K}^{0} | H_W | 
	n \rangle \langle n | H_W | K^{0}	\rangle}{M_{K} -E_{n}} \,.
	\label{Eq.ind_contr}
\end{eqnarray}

To determine $M_{\overline{0}0}^{\BL}$ from the integrated correlator, the same methods introduced in Refs.~\cite{Christ:2012se} and \cite{Bai:2014cva} for the calculation of $\Delta M_K$ can be used. The intermediate states $|n\rangle$ whose energy $E_n$ is less than the kaon mass are identified. These states result in exponential increasing terms proportional to $e^{(M_K-E_n)T}$ in the integrated correlator and these exponentially increasing contributions must be explicitly removed. For intermediate states that have an energy higher than the kaon mass, the choice of integration region $T$ must be large enough so that their contribution is exponentially suppressed. 

In the exploratory numerical study presented in Sec.~\ref{section:Results}, $M_K<2M_\pi$, so that the only intermediate states that we need to consider are the single-pion state and the vacuum state.  In a future calculation with a physical pion mass, two- and three-pion states will need to be dealt with as well. The three-pion state will be kinematically suppressed and should not contribute a significant exponential contamination. The matrix elements for the two-pion state will need to be calculated and subtracted as is done for the physical-mass $\Delta M_K$ calculation~\cite{Wang:2022lfq}.

Although the ``double-integration'' method described above (see Eq.~\eqref{Eq:intCorr}) is used in this paper, there is a more refined approach, developed after the current calculation was complete, referred to as the ``single-integration'' method~\cite{Bai:2018hqu, Wang:2022lfq} which has been observed to reduce the statistical error by approximately a factor of two.  In this single-integration approach one operator is held at a fixed time $t_1$ while the second operator at the time $t_2$ is integrated over the range $|t_2-t_1|<t_\mathrm{max}$.  Examining the behavior of the integrand as a function of $|t_2-t_1|$ one can identify a value of $t_\mathrm{max}$ which will capture the region within which the integrand is non-negligible.  By limiting the integration to this region the statistical noise may be reduced because we have omitting larger values of $|t_2-t_1|$ which contribute only noise to the result.  A final average over an appropriate range for $t_1$ then gives a more accurate result than that obtained from the double integration approach used here.

A final topic that must be addressed in a lattice calculation of either $\epsilon_K$ or $\Delta M_K$ is the effect of finite volume.  The infinite-volume expression for $M_{\overline{0} 0}$ on the right-hand side of Eq.~\eqref{Eq:M00} contains a continuous integral over the intermediate-state energy with a principal part prescription used to resolve a pole singularity.  In contrast, the finite-volume expression on the right-hand side of \eqref{Eq.ind_contr} involves a sum over discrete finite-volume energy eigenstates with energy denominators which depend on the difference of a finite-volume eigenvalue and the mass of the kaon.  For the case of two-particle intermediate states the potentially large difference between such a discrete finite-volume sum and the infinite-volume principal part integral is known and can be written in terms of on-shell matrix elements so that the necessary finite-volume correction~\cite{Christ:2015pwa} can be made.

\section{Short-distance divergence}
\label{section:SD_correction}

As discussed above, the important role of the top quark in indirect CP violation implies that the long-distance contributions to $\epsilon_K$ are less protected by the GIM mechanism than is the case for the CP conserving mass difference $\Delta M_K$.  Specifically, in Section~\ref{sec:logarithmic_GIM} we point out that the terms of interest, proportional to $\lambda_u\lambda_t$, will contain logarithmic divergences when computed to second order in the four-quark $\Delta S=1$ operators that must be used in a lattice QCD calculation.  In this section we present a method to control such divergent terms using a combination of non-perturbative techniques to remove these divergent terms from the lattice amplitudes and QCD plus electroweak perturbation theory, applied at short distances, to determine the low energy constants with which these divergent terms should be replaced.

\begin{figure}[ht]
	\centering
	\begin{tabular}{c|c}\hline
		\includegraphics[width=0.45\textwidth,height=0.18\textwidth]{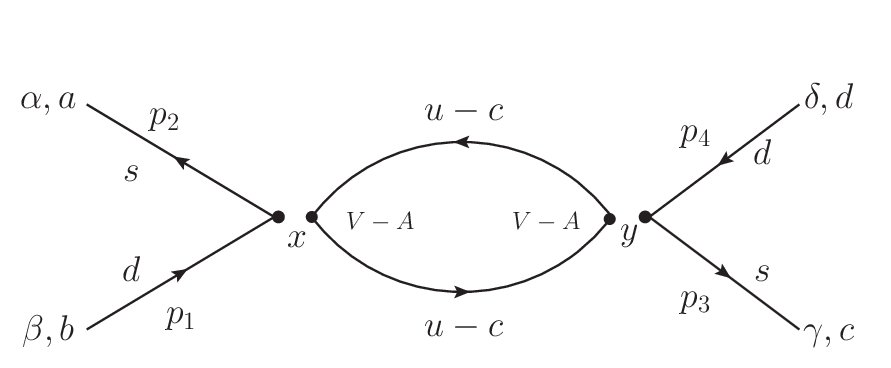} & 
		\includegraphics[width=0.5\textwidth,height=0.18\textwidth]{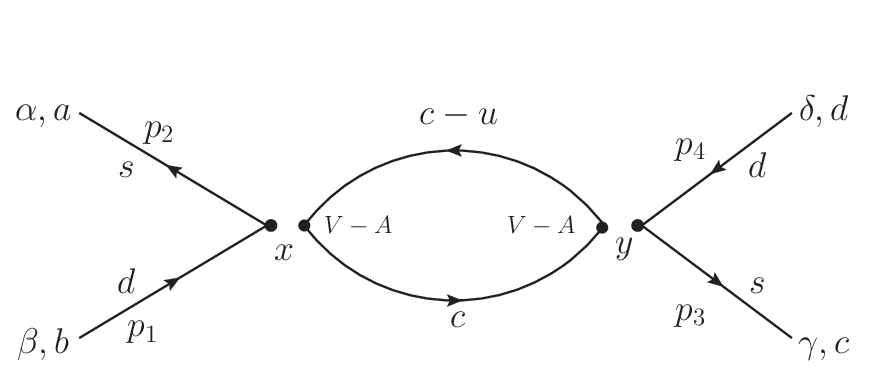}\\\hline
	\end{tabular}
	\caption{An example of a diagram which appears in the $\Delta M_K$ calculation (left) and a 
	similar diagram entering the calculation of $\epsilon_K$ (right).}
	\label{figure:Div_diag}
\end{figure}

As an example, in Fig.~\ref{figure:Div_diag} we compare two typical diagrams which appear in the calculation of $\Delta M_K$ (left) and $\epsilon_K$ (right), identified as diagrams of type 1 in the discussion below.  To study the ultraviolet behavior, we can ignore the momentum in the four external quark lines and consider the case of free field propagators.
The Feynman amplitude corresponding to the $\epsilon_K$ example diagram is given in Eq.~\eqref{Eq:epsilon_K_div}
while the amplitude for $\Delta M_K$ is given in Eq.~\eqref{Eq:DeltaMK_div}.
\begin{eqnarray}
	\label{Eq:epsilon_K_div}
	&&\int d^4p \gamma^{\mu} (1 - \gamma^5) (\frac{\slashed{p} - m_c}{p^2 + m_c^2}
	- \frac{\slashed{p} - m_u}{p^2 + m_u^2}) \gamma^{\nu} (1 - \gamma^5) 
	(\frac{\slashed{p} - m_c}{p^2 + m_c^2}) \\ \nonumber
	&& \hskip 1.0 in =  \int d^4p\gamma^{\mu} (1 - \gamma^5) \frac{\slashed{p} (m_u^2 - m_c^2)}
	{(p^2 + m_u^2)(p^2 + m_c^2)} \gamma^{\nu} (1 - \gamma^5)
	(\frac{\slashed{p}}{p^2 + m_c^2}) \\
	\label{Eq:DeltaMK_div}
	&&\int d^4p \gamma^{\mu} (1 - \gamma^5) (\frac{\slashed{p} - m_c}{p^2 + m_c^2}
	- \frac{\slashed{p} - m_u}{p^2 + m_u^2}) \gamma^{\nu} (1 - \gamma^5) 
	(\frac{\slashed{p} - m_c}{p^2 + m_c^2}
	- \frac{\slashed{p} - m_u}{p^2 + m_u^2})\\\nonumber
	&& \hskip 1.0 in =  \int d^4p\gamma^{\mu} (1 - \gamma^5) \frac{\slashed{p} (m_u^2 - m_c^2)}
	{(p^2 + m_u^2)(p^2 + m_c^2)} \gamma^{\nu} (1 - \gamma^5)
	(\frac{\slashed{p} (m_u^2 - m_c^2)}
	{(p^2 + m_u^2)(p^2 + m_c^2)} ),
\end{eqnarray}
where we have neglected the external momenta and kept only the leading terms for large loop momentum $p$.  By counting the powers of momenta in Eq.~\eqref{Eq:epsilon_K_div}, we can recognize a logarithmic ultraviolet divergence.  However, the expression in the lower equation for $\Delta M_K$ is ultraviolet finite because we have subtracted the charm and up quark propagators in both quark lines. 

\subsection{Renormalization overview}

In the lattice QCD calculation the ultraviolet divergence discussed in the previous paragraphs is regulated at the scale of the inverse lattice spacing ($1/a$) and we must identify and renormalize this unphysical, divergent piece to obtain physical results.  We use a generalization to bilocal operators~\cite{Christ:2015phf, Bai:2017fkh} of the Rome-Southampton, regularization-independent (RI/SMOM) method~\cite{Martinelli:1994ty, Aoki:2007xm, Sturm:2009kb} to perform this short-distance correction.  At energy scales below the lattice cutoff, the cutoff-dependent part of the bilocal operator can be represented by the single local operator $O_{LL}$ multiplied by a coefficient that depends logarithmically on the lattice spacing.  Thus, we will add to each bilocal operator a cutoff-dependent counter term proportional to the $O_{LL}$ operator so that each combined operator obeys an RI/SMOM normalization condition.  With the addition of these counter terms our lattice-determined bilocal operators become well-defined, with both the operator mixing among the $\Delta S=1$ operators entering each factor and the treatment of the singularity when the two factors collide well-defined. 

We will now determine that combination of these well-defined local and bilocal lattice operators, expressed in the RI/SMOM scheme, which corresponds to the physical second-order, $\Delta S=2$ effective weak Hamiltonian proportional to the $\lambda_u\lambda_t$ product determined by the standard model.  This operator whose determination is reviewed in Ref.~\cite{Buchalla:1995vs} is conventionally expressed in $\MS$ conventions.  More specifically, it is expressed as a sum over bilocal operators whose factors are defined in the $\MS$ scheme and the singularity when the positions of these two factor coincide is also defined using $\MS$ regularization.  Of course, in addition to these bilocal operators there is the usual local $O_{LL}$ operator representing the short-distance standard-model contribution to $K^0-\overline{K}^0$ mixing.  The result is a complete, $\Delta S=2$ effective Hamiltonian density defined perturbatively in the $\MS$ scheme, using the notation introduced in Eqs.~\eqref{Eq:Qij_cc} and \eqref{Eq:Qij_cp}:
\begin{eqnarray}
\mathcal{H}_{W,ut}^{\Delta S=2} = \frac{G_F^2}{2}\lambda_u\lambda_t\sum_{i=1,2}
   \left\{\sum_{j=1,6} \int d^4 x C_i^{\MS} C_j^{\MS} 
                          [[\widetilde{Q}^{\MS}_i(x)\widetilde{Q}^{\MS}_j(0)]]^{\MS}
                                          + C_{7i}^{\MS} O_{LL}^{\MS}(0)\right\}.
\label{eq:Heff-SM}
\end{eqnarray}
Here the $\MS$ superscript on the double square bracket surrounding the product of operators $\widetilde{Q}^{\MS}_i(x)\widetilde{Q}^{\MS}_j(0)$ indicates that the singularity encountered in the integral at $x=0$ is resolved using $\MS$ conventions.  The twelve operator products surrounded by square brackets are defined in Eqs.~\eqref{Eq:Qij_cc} and \eqref{Eq:Qij_cp}.

The effective Hamiltonian density given in Eq.~\eqref{eq:Heff-SM} should be viewed as a complete description of the physics of the standard model when studied at energies below the bottom quark mass where this four-flavor version is appropriate.  Of course, the reference to perturbation theory and specifically $\MS$ regularization prevents its direct use in a lattice QCD calculation.  However, with a change of normalization prescription from $\MS$ to the non-perturbatively-defined RI/SMOM scheme we can express the effective Hamiltonian defined in Eq.~\eqref{eq:Heff-SM} in terms of quantities that can be directly evaluated in lattice QCD.   

The first step is to replace the $\MS$ renormalization of the singularity as $x \to 0$ with that of the following generalized RI/SMOM scheme.   Given the GIM cancellation present in the quantities under discussion the singular terms present in the operator products that appear in Eq.~\eqref{eq:Heff-SM} correspond to singular constants multiplying the operator $O_{LL}$.   The $\MS$ scheme provides a particular choice for those constants.  In the RI/SMOM scheme, generalized to this case of bilocal operators, we instead require a choice of the constants multiplying $O_{LL}$ which makes the sum of the bilocal operators and the $O_{LL}$ counter terms vanish when evaluated in a Landau-gauge-fixed Green's function with four external quark lines with specific off-shell kinematics specified by the RI/SMOM renormalization scale $\mu_{\RI}$.   (The RI/SMOM scheme applied to bilocal operators is described in greater detail below.)  This can be done for each operator pair in Eq.~\eqref{eq:Heff-SM}, directly relating the $\MS$ and RI/SMOM schemes:
\begin{eqnarray}
\int d^4 x [[\widetilde{Q}^{\MS}_i(x)\widetilde{Q}^{\MS}_j(0)]]^{\MS} 
                 &=& \int d^4 x [[\widetilde{Q}^{\MS}_i(x)\widetilde{Q}^{\MS}_j(0)]]^{\RI} + Y^{\MS}_{ij}(\mu_{\MS},\mu_{\RI})O_{LL.}^{\MS}
\label{eq:bilocal-MSbar-RI}
\end{eqnarray}
Here the coefficient $Y^{\MS}_{ij}(\mu_{\MS},\mu_{\RI})$ is determined by applying the RI/SMOM condition to Eq.~\eqref{eq:bilocal-MSbar-RI} since the RI-normalized operator $[[\widetilde{Q}^{\MS}_i(x)\widetilde{Q}^{\MS}_j(0)]]^{\RI}$ will vanish at those RI/SMOM kinematics.  This determines $Y^{\MS}_{ij}(\mu_{\MS},\mu_\RI)$ as the appropriate spin-projected $\MS$ Green's function containing the bilocal operator and evaluated at the RI/SMOM-defining kinematics for the momenta carried by the four external quark lines.  As indicated, $Y^{\MS}_{ij}$ will depend on both the $\MS$ scale $\mu_{\MS}$ and the RI/SMOM scale $\mu_\RI$.  This step can also be found in the method we used in the rare kaon calculations~\cite{Christ:2016eae}.

The next step is to substitute Eq.~\eqref{eq:bilocal-MSbar-RI} into Eq.~\eqref{eq:Heff-SM} and then to replace the sums of $\MS$-renormalized operators with their Wilson coefficients in the RI/SMOM-normalized operator products with equivalent lattice operators multiplied by their lattice Wilson coefficients in the identical RI/SMOM-normalized operator products to obtain:
\begin{eqnarray}
\mathcal{H}_{W,ut}^{\Delta S=2} &=& \frac{G_F^2}{2}\lambda_u\lambda_t\sum_{i=1}^2
   \left\{\sum_{j=1}^6 \sum_x C_i^{\Lat} C_j^{\Lat} 
                          [[\widetilde{Q}^{\Lat}_i(x)\widetilde{Q}^{\Lat}_j(0)]]^{\RI} \label{eq:Heff-SM-RI-Lat} \right. \\
               &&\hskip 1.0 in+\left. \left(C_{7i}^{\MS}+\sum_{j=1}^6 C_i^{\MS} C_j^{\MS}Y^{\MS}_{ij}(\mu_{\MS},\mu_{\RI})\right) Z_{LL}^{\Lat\to\MS}O_{LL}^{\Lat}(0)\right\}.
\nonumber
\end{eqnarray}
Here we are using the usual conversion from $\MS$ to RI/SMOM conventions followed by conversion from RI/SMOM to lattice conventions for the individual four-quark operators as is described in greater detail in Appendix~\ref{sec:appendix}.  Since all of the operators appearing in Eq.~\eqref{eq:Heff-SM-RI-Lat} can be defined on a space-time lattice, we have replaced the integrals over the position $x$ with the sum over lattice sites $x$.

Next we express the RI/SMOM-renormalized product of lattice operators which appears in Eq.~\eqref{eq:Heff-SM-RI-Lat} by the lattice-regulated product using the relation:
\begin{eqnarray}
\sum_x [[\widetilde{Q}^{\Lat}_i(x)\widetilde{Q}^{\Lat}_j(0)]]^\RI 
                 &=& \sum_x [[\widetilde{Q}^{\Lat}_i(x)\widetilde{Q}^{\Lat}_j(0)]]^{\Lat} - X^{\Lat}_{ij}(1/a,\mu_{\RI})O_{LL.}^{\Lat}.
\label{eq:bilocal-Lat-RI}
\end{eqnarray}
Similar to Eq.~\eqref{eq:bilocal-MSbar-RI}, the coefficients $X^{\Lat}_{ij}$ are chosen so that the appropriate Green's function  containing the bilocal operator $[[\widetilde{Q}^{\Lat}_i(x)\widetilde{Q}^{\Lat}_j(0)]]^\RI$ vanishes when evaluated at external momenta obeying the RI/SMOM kinematics at the scale $\mu_\RI$.  Now the coefficients $X^{\Lat}_{ij}$ depends on both the lattice scale $1/a$ and the RI/SMOM scale $\mu_\RI$.  Equation~\eqref{eq:bilocal-Lat-RI} can be substituted into Eq.~\eqref{eq:Heff-SM-RI-Lat} to express $\mathcal{H}_{W,ut}^{\Delta S=2}$ entirely in terms of operators whose matrix elements can be computed using lattice QCD:
\begin{eqnarray}
\mathcal{H}_{W,ut}^{\Delta S=2} &=& \frac{G_F^2}{2}\lambda_u\lambda_t\sum_{i=1}^2
   \left\{\sum_{j=1}^6 C_i^{\Lat} C_j^{\Lat} 
            \Biggl(\sum_x [[\widetilde{Q}^{\Lat}_i(x)\widetilde{Q}^{\Lat}_j(0)]]^{\Lat} - X^{\Lat}_{ij}(\mu_\RI)O_{LL}^{\Lat}(0)\Biggr)
                          \right. \nonumber \\
               &&\hskip 0.8 in+\left. \Biggl(C_{7i}^{\MS}+\sum_{j=1}^6 C_i^{\MS} C_j^{\MS}Y^{\MS}_{ij}(\mu_{\MS},\mu_{\RI})\Biggr) Z_{LL}^{\Lat\to\MS}O_{LL}^{\Lat}(0)\right\}.
\label{eq:Heff-SM-Lat}
\end{eqnarray}

Essential to this approach of exploiting lattice QCD is the fact that the scale $\mu_\RI$ at which this perturbative matching between the operators in Eq.~\eqref{eq:Heff-SM} and the operators used on the lattice can be chosen to be large, typically above the charm quark mass giving control over the perturbation theory errors.  Thus, we can exploit the detailed standard model information encoded in the perturbative result given in Eq.~\eqref{eq:Heff-SM} while working at an energy scale that can be made sufficiently large that perturbation theory is accurate.  We will now describe the details of this procedure.

The RI/SMOM renormalization of the bilocal operators appearing in Eq.~\eqref{eq:Heff-SM} proceeds in two steps:  i) the  perturbative calculation of the coefficients $Y^{\MS}_{ij}$ to convert from the $\MS$ to RI/SMOM renormalization of the singularity that occurs when the operators in a bilocal pair coincide and ii) the non-perturbative determination of the coefficients $X^{\Lat}_{ij}$.  We first discuss the perturbative determination of coefficients $Y^{\MS}_{ij}$.

\subsection{Perturbative determination of $Y^{\MS}_{ij}$}
\label{sec:pert-renorm}

The low-energy constants $Y^{\MS}_{ij}$ are defined in Eq.~\eqref{eq:bilocal-MSbar-RI} and that equation can be used to calculate them in perturbation theory.  We must simply insert the integrated $\MS$ bilocal operator $[[\widetilde{Q}^{\MS}_i(x)\widetilde{Q}^{\MS}_j(0)]]^{\MS}$ into the appropriate five-point Green's function and evaluate the result with the external momenta used to define the RI/SMOM normalization condition.   

\begin{figure}[ht]
	\includegraphics[width=0.9\textwidth]{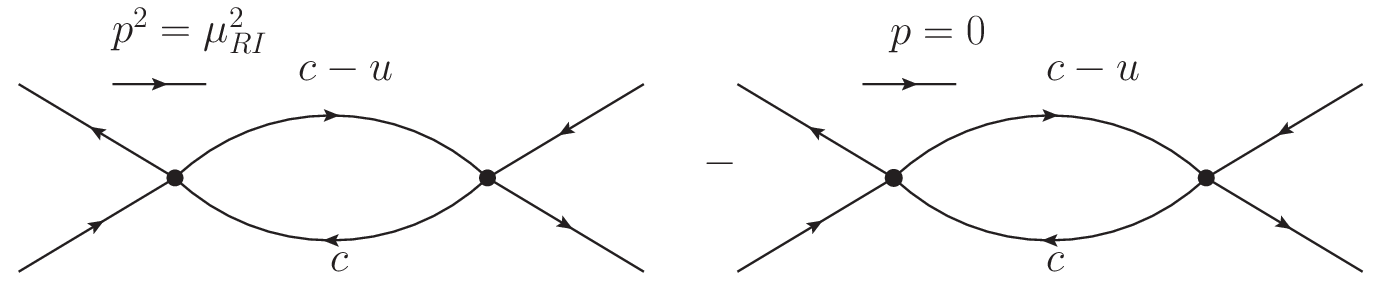}
	\caption{Illustration of the calculation of $\Delta Y$.}
	\label{Fig:deltaY}
\end{figure}

Thus, in this calculation of $Y^{\MS}_{ij}(\mu_{\MS}, \mu_\RI)$, the external momenta are set to the energy scale $p^2 = \mu^2_\RI$ and we perform the integration over the internal quark lines in the left-hand panel of Fig.~\ref{Fig:deltaY} . This bilocal operator Green's function is then equated to the Green's function containing the local operator $O_{LL}$ multiplied by the coefficient $Y^{\MS}_{ij}$.  Fortunately, in the conventional perturbative calculation of $\epsilon_K$~\cite{Buchalla:1995vs} something very similar is evaluated at NNLO~\cite{Brod:2010mj, Brod:2011ty}. In fact, exactly this calculation is performed except the external momenta are all set to the value zero.  We can therefore define a quantity $\Delta Y^{\MS}_{ij}$, which is the difference between $Y^{\MS}_{ij}(\mu_{\MS}, \mu_{\RI})$ evaluated for off-shell momenta at the scale $\mu_{\RI}$, minus $Y^{\MS}_{ij}(\mu_{\MS}, 0)$ evaluated at zero external momentum. The quantity $\Delta Y^{\MS}_{ij}(\mu_{\MS}, \mu_{\RI}) = Y^{\MS}_{ij}(\mu_{\MS}, \mu_{\RI}) - Y^{\MS}_{ij}(\mu_{\MS}, 0)$ is therefore a quantity that is both ultra-violet and infra-red finite at order $\alpha_s^0$ in four dimensions and is independent of the $\overline{MS}$ scale $\mu_{\MS}$, making it an straight-forward quantity to compute. The calculation of $\Delta Y^{\MS}_{ij}$ is illustrated in Fig.~\ref{Fig:deltaY}.  (The fact that this calculation of $Y^{\MS}_{ij}(\mu_{\MS}, \mu_{\RI})$ is carried out only to order $\alpha_s^0$, reduces the accuracy of the final numerical results presented in this paper from a NLO calculation containing all terms of order $\alpha_s^n \left(\alpha_s \ln(M_W/m_c)\right)^l$ for $n=-1$ (LO) and $n=0$ (NLO) to one that is incomplete at NLO.)

With this perturbation theory step Eq.~\eqref{eq:Heff-SM-Lat} can be written
\begin{eqnarray}
\mathcal{H}_{W,ut}^{\Delta S=2} &=& \frac{G_F^2}{2}\lambda_u\lambda_t\sum_{i=1}^2
   \left\{\sum_{j=1}^6 C_i^{\Lat} C_j^{\Lat} 
         \Biggl(\sum_x [[\widetilde{Q}^{\Lat}_i(x)\widetilde{Q}^{\Lat}_j(0)]]^{\Lat} - X^{\Lat}_{ij}(\mu_\RI)O_{LL}^\Lat(0)\Biggr)
                                                     \right. \nonumber \\
               &&\hskip 0.6 in+ \Biggl(\sum_{j=1}^6 C_i^{\MS} C_j^{\MS}
                      \left[Y^{\MS}_{ij}(\mu_{\MS},\mu_{\RI}) -Y^{\MS}_{ij}(\mu_{\MS},0)\right] \Biggr) Z_{LL}^{\Lat\to\MS}O_{LL}^{\Lat}(0)
                       \nonumber \\
               &&\hskip 1.0 in+\left. \Biggl(C_{7i}^{\MS}+\sum_{j=1}^6 C_i^{\MS} C_j^{\MS}Y^{\MS}_{ij}(\mu_{\MS},0)\Biggr) Z_{LL}^{\Lat\to\MS}O_{LL}^{\Lat}(0)\right\}.
\label{Eq:masterEq}
\end{eqnarray}
The first line of Eq.~\eqref{Eq:masterEq} involves the $\Delta S=1$ lattice operators and the coefficient $X^{\Lat}_{i,j}$
determined from non-perturbative renormalization (NPR), described in greater detail below. We refer to this term as the ``long-distance'' (LD) part and use $M_{\overline{0}0}^{ut,\LD}(\mu_{\RI})$ to denote its contribution to the kaon mixing matrix element.  The second line involves the coefficient $\Delta Y^{\MS}_{i,j}(\mu_{\RI})$ calculated from perturbation theory.  This term is described as the ``perturbative $\MS$ to RI/SMOM correction'' and we use $M_{\overline{0}0}^{ut,\MS\rightarrow\RI}(\mu_{\RI})$ to denote its contribution to the kaon mixing matrix element.  The last term is the conventional standard model result for $\epsilon_K$.  We will describe the combination of the second and third terms as the ``short-distance'' (SD) part of the standard model calculation of $\epsilon_K$ and use $M_{\overline{0}0}^{ut,\SD}(\mu_{\RI})$ to denote its contribution to the kaon mixing matrix element.  Thus, the scale $\mu_{\RI}$ separates the long- and short-distance parts. We anticipate that in the future $Y^{\MS}_{ij}$ will be computed directly in perturbation theory to NL or higher order, avoiding our use of the conventional standard model result for $\epsilon_K$ and the quantity $M_{\overline{0}0}^{ut,\MS\rightarrow\RI}(\mu_{\RI})$, whose current value is accurate only to order $\alpha_s^0$. making our full calculation incomplete at NLO.

\subsection{ Non-perturbative determination of $X^{\Lat}_{ij}$}

Finally we describe in greater detail the non-perturbative calculation of the twelve coefficients $X^{\Lat}_{ij}(\mu_\RI)$.  These coefficients are determined by solving the twelve independent equations:
\begin{equation}
	\left(\Gamma_{\alpha\beta\gamma\delta,ij}^{\BL,\mathrm{amp}}(p_1,p_2,p_3,p_4) - X^{\Lat}_{ij}(\mu_\RI) 
		\Gamma_{\alpha\beta\gamma\delta}^{\Loc,\mathrm{amp}} (p_1,p_2,p_3,p_4) \right)
		      P_{\alpha\beta\gamma\delta} = 0.
	 \label{Eq:calc_Xij}
\end{equation}
Here $\Gamma^{\BL,\mathrm{amp}}_{\alpha\beta\gamma\delta,ij}(p_1,p_2,p_3,p_4)$ is the five-operator Green's function:
\begin{eqnarray}
	\Gamma_{\alpha\beta\gamma\delta,ij}^{\BL,\mathrm{amp}}(p_1,p_2,p_3,p_4)
	        &=& \label{Eq:Gamma_BL} \\
	&& \hskip -0.75 in \langle 0|T\left\{ \bigl(s_{\alpha}(p_1)\overline{d}_{\beta}(p_2)\bigr) 
	                \left[ \sum_{x_1 x_2} [[\widetilde{Q}^{\Lat}_i(x_1)\widetilde{Q}^{\Lat}_j(x_2)]]^{\Lat} \right] 
	                                     \bigl(s_{\gamma}(p_3) \overline{d}_{\delta}(p_4)\bigr)
	                                                 \right\}|0 \rangle_{\mathrm{amp}}, 
	                                                 \nonumber
\end{eqnarray}
with its four external legs amputated.  The color indices of each spinor pair enclosed in curved brackets, $(\ldots)$ are contracted; $\alpha\beta\gamma\delta$ are spinor indices.  The choice of the external momenta is discussed in Section~\ref{sec:RI_SMOM-renorm}.

The second Green's function in Eq.~\eqref{Eq:calc_Xij} is similar but with the local operator $O_{LL}^\Lat(x)$
replacing the  bilocal operator $[[\widetilde{Q}^{\Lat}_i(x_1)\widetilde{Q}^{\Lat}_j(x_2)]]^{\Lat}$:
\begin{eqnarray}
	\Gamma^{\Loc,\mathrm{amp}}_{\alpha\beta\gamma\delta}(p_1,p_2,p_3,p_4)
	        = \langle 0|T\left\{ \bigl(s_{\alpha}(p_1)\overline{d}_{\beta}(p_2)\bigr) 
	                \sum_x O^\Lat_{LL}(x)
	                                     \bigl(s_{\gamma}(p_3) \overline{d}_{\delta}(p_4)\bigr)
	                                                 \right\}|0 \rangle_{\mathrm{amp}}, \label{Eq:Gamma_SD}
\end{eqnarray}
using a notation similar to that in Eq.~\eqref{Eq:Gamma_BL}.  The spinor projector $P_{\alpha\beta\gamma\delta}$ appearing in Eq.~\eqref{Eq:calc_Xij} is defined by:
\begin{equation}
P_{\alpha\beta\gamma\delta} = \sum_\mu \left[(1-\gamma_5)\gamma_\mu\right]_{\alpha\beta}
                                                                     \left[(1-\gamma_5)\gamma_\mu\right]_{\gamma\delta}.
\label{Eq:gamma_HW}
\end{equation}

With the choice of non-exceptional momenta entering the amplitude $\Gamma^{\BL,\mathrm{amp}}_{\alpha\beta\gamma\delta,ij}(p_1,p_2)$ this quantity is infrared safe and the corresponding perturbative calculation of $Y(\mu_{\MS},\mu_\RI)_{ij}$ performed in Section~\ref{sec:pert-renorm} is given by an expansion in $\alpha_s(\mu_\RI)$ which should make the perturbation theory increasingly accurate as the scale $\mu_\RI$ is increased.  In the exploratory calculation reported here we choose $\mu_\RI=2.11$ GeV for the renormalization of the bilocal operators and $\mu_\RI=2.15$ GeV for the renormalization of the individual four-quark operators, each of which may be sufficiently large to allow the use of NLO QCD perturbation theory.  Phrased differently, when the energy scale $\mu_\RI$ of the external momentum that enters the renormalization condition is sufficiently high, the integrated correlator will be dominated by short-distance contributions. We will test this statement in Section~\ref{section:Results}.

\section{Lattice Implementation}
\label{section:Results}

This calculation is carried out using 200 gauge configurations from a $2+1$ flavor ensemble generated by the RBC and UKQCD collaborations~\cite{Allton:2008pn} using the domain wall fermion (DWF) and Iwasaki gauge actions. These configurations have a $24^3 \times 64$ lattice volume with an inverse lattice spacing $1/a=1.78$ GeV.  Throughout we use the Shamir variant of the DWF formulation with an extent in the fifth dimension of $L_s=16$.  The pion and kaon masses are 339 MeV and 592 MeV.  We use a valence charm quark with mass renormalized in the $\MS$ scheme at the scale of 2 GeV with the value $m_c(\mbox{2 GeV}) = 968$ MeV.   These unphysical values for the quark masses are chosen to reduce the computational cost of this first, exploratory calculation.  Having demonstrated that this calculation is practical, we anticipate future calculations with physical masses and several smaller values of the lattice spacing to allow a continuum extrapolation and quantifiable systematic errors.

\subsection{Correlation Function Construction}\label{sec:latt_impl}

To calculate the long-distance contribution to $M_{\overline{0}0}$, we evaluate four-point functions with the bilocal operators of Eq.~\eqref{Eq:H_eff_dS2} appearing between two $K^0$ interpolating operators as in Eq.~\eqref{Eq:intCorr}.  There are five types of four-point diagrams to calculate on the lattice. Each type of diagram, except type 5, can have either a pair of current-current operators or a combination of a current-current operator and a QCD penguin operator. The flavors of the internal quark lines will depend on the type of diagram and the specific weak operators which appear.

\begin{figure}[ht!]
	\centering
	\begin{tabular}{c|c}\hline
		\includegraphics[width=0.4\textwidth]{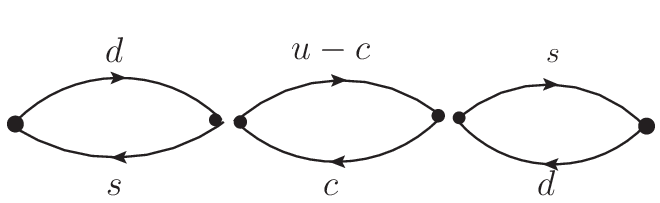} & 
		\includegraphics[width=0.4\textwidth]{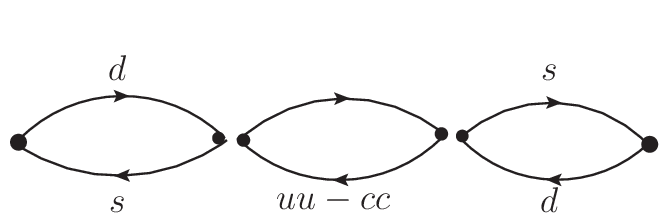} \\
		type 1, $C-C$ &type 1, $C - P$ \\ \hline 
		\includegraphics[width=0.3\textwidth]{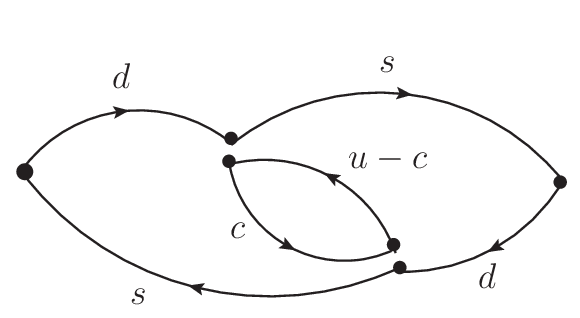} & 
		\includegraphics[width=0.3\textwidth]{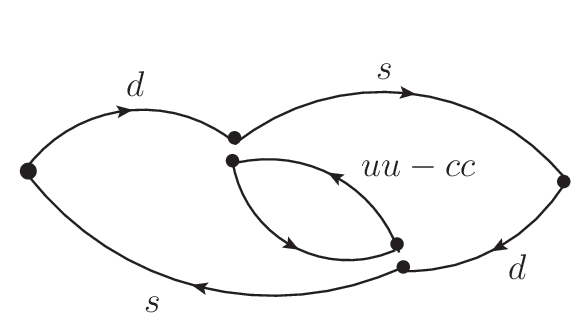} \\
		type 2, $C-C$ & type 2, $C-P$ \\\hline
	\end{tabular}
	\caption{Type 1 and type 2 four-point diagrams. In the captions of the subfigures $C$ indicates a current-current operator 
	and $P$ labels a QCD penguin operator.}
	\label{figure:type12}
\end{figure}

The type 1 and 2 diagrams are shown in Fig.~\ref{figure:type12}.  If both operators are current-current operators then the structure of Eq.~\eqref{Eq:H_eff_dS2} requires that there is a single charm quark propagator in one internal quark line and a charm minus up propagator difference in the other. If one of the operators is a current-current operator and the other is a QCD penguin operator, we will have the difference of two diagrams where in one diagram the two internal quark lines are both charm quarks while in the other diagram both internal quark lines are up quarks.  These are the only types of diagram without disconnected quark loops, so they will be more statistically precise than the rest.

\begin{figure}[ht!]
	\centering
	\begin{tabular}{c|c}
	\hline
		\includegraphics[width=0.3\textwidth]{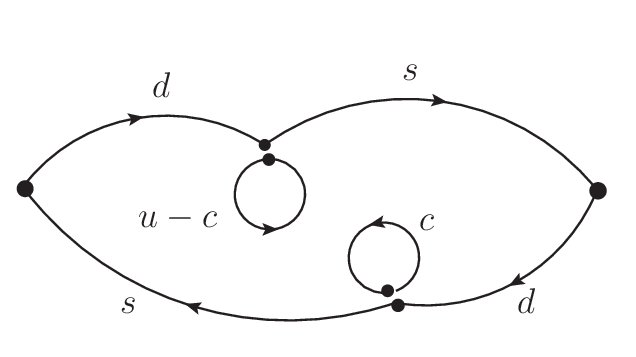} & 
		\includegraphics[width=0.3\textwidth]{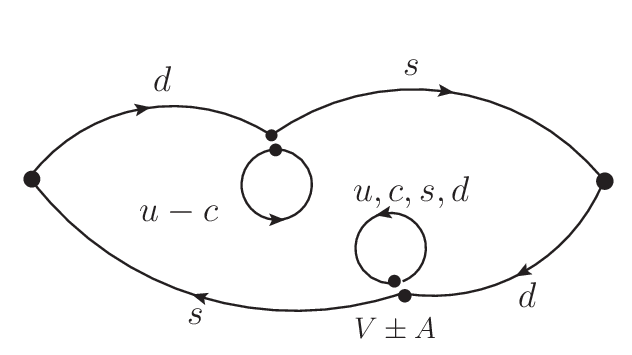} \\
		type 3, $C-C$ &type 3, $C - P$ \\\hline
		\includegraphics[width=0.4\textwidth]{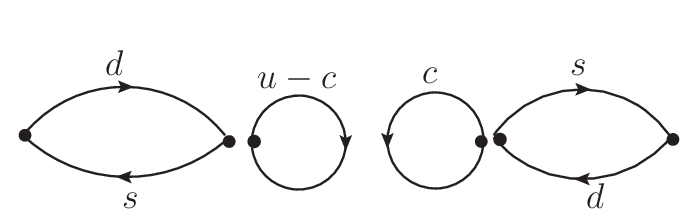} & 
		\includegraphics[width=0.4\textwidth]{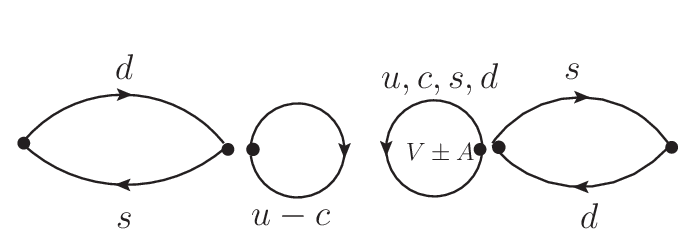} \\
		type 4, $C-C$ & type 4, $C-P$ \\\hline
	\end{tabular}
	\caption{Type 3 and type 4 four-point diagrams. In the captions of the subfigures $C$ indicates a current-current operator and $P$ labels a QCD penguin operator.}
	\label{figure:type34}
\end{figure}

The type 3 and 4 diagrams are shown in Fig.~\ref{figure:type34}. If both operators are current-current operators, we have a single charm quark self-loop connected to one vertex and the difference of a charm quark and an up quark propagator connected to the other.   If one of the vertices is a QCD penguin operator, we have the difference of charm and up quark self-loops connected to one vertex and a sum over all four flavors connected to the other self-loop.

\begin{figure}[h!]
	\centering
	\begin{tabular}{c|c}
	\hline
		\includegraphics[width=0.4\textwidth]{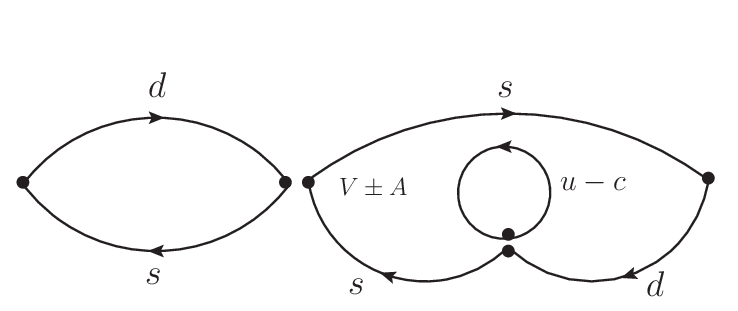} & 
		\includegraphics[width=0.4\textwidth]{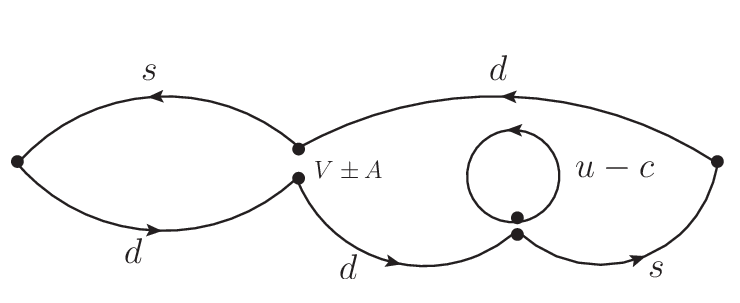} \\\hline
	\end{tabular}
	\caption{Type 5 four-point diagrams. These have a current-current operator at one vertex while the
	other vertex must come from a QCD penguin operator of the form $(\overline{s}d)_{V-A} (\overline{d} d)_{V\pm A}$  
	or $(\overline{s}d)_{V-A} (\overline{s} s)_{V\pm A}$.}
	\label{figure:type5}
\end{figure}

The type 5 diagrams are shown in Fig.~\ref{figure:type5}.  The type 5 diagrams are absent in the $\Delta M_K$ calculation because one of the vertices of a type 5 diagram must be a QCD penguin operator. We have two varieties of
type 5 diagram.  One variety contains an $(\overline{s}d)_{V-A} (\overline{d} d)_{V\pm A}$ QCD penguin operator while the other contains the combination $(\overline{s}d)_{V-A} (\overline{s} s)_{V\pm A}$.  In each case any one of the four QCD penguin operators $\{Q_i\}_{3 \le i \le 6}$ can appear.
 
For all five types of diagrams, a wall source propagator is used for each kaon. The two kaon wall sources are separated by a fixed distance of 28 lattice units and each of the two weak vertices is required to have a minimum time separation of six lattice units from each wall to reduce excited-state contamination. Therefore, the times at which the two vertices are inserted are integrated over a range of 16 lattice units.

In the calculation of type 1 and type 2 diagrams, we used a point source propagator located at one of the weak vertices while at the other weak vertex we combined the sinks of four propagators and summed the location of this second vertex over the relevant space-time volume.  The point sources are chosen to have the space-time coordinates $(4t,4t,4t,t)$, and periodic boundary conditions are used when an $(x,y,z)$ coordinate crosses a lattice boundary. Thus, we place the point source on the time slice $t$ at the spatial point (4$t$ mod $L$, 4$t$ mod $L$, 4$t$ mod $L$).  In the calculation of the type 3 and type 4 diagrams, all-to-all propagators are used for the self-loop. 
 
To construct the self-loop, $N_{\mathrm{ev}} = 450$ eigenvectors generated using the Lanczos algorithm are used and the propagator is calculated as
\begin{eqnarray}
	\label{Eq:self_loop}
	D^{-1}(x;x) &=& \sum_{i=1}^{N_{\mathrm{ev}}} \frac{h_i(x)h^\dagger_i(x)}{\lambda_i} + 
	\sum_{j=1}^{N_{\mathrm{hit}}} (D_{\mathrm{defl}}^{-1} \eta_j)_x \eta_j^\dagger (x) \\
	D_{\mathrm{defl}}^{-1} &=& D^{-1} - \sum_{i=1}^{N_{\mathrm{ev}}} \frac{h_i(x)h^\dagger_i(x)}{\lambda_i} \,,
\end{eqnarray}
where $h_i$ is the $i^{th}$ eigenvector and $\lambda_i$ the corresponding eigenvalue.  The $\eta_j$ are random vectors which which are functions of both space and time and satisfy $\langle \eta_i(x) \eta^\dagger_j(y)\rangle = \delta_{ij} \delta_{xy}$.  This procedure will have the advantage that the low-mode part of these self-loop propagators is more accurate than the simpler random source propagators that were used in Ref.~\cite{Bai:2014cva}. For each gauge configuration we have averaged over $N_{\mathrm{hit}}$ such random volume source vectors with $N_{\mathrm{hit}} = 80$.  For the self-loops in the type 5 diagrams, we used the same point source propagators that were used for the type 1 and 2 diagrams.  The other vertex is treated as a sink for the four propagators and summed over the space-time integration region.

\subsection{Details of Bilocal Operator Renormalization} \label{sec:RI_SMOM-renorm}

\begin{figure}[ht]
	\centering
	\begin{tabular}{ccc}
		\includegraphics[width=0.3\textwidth]{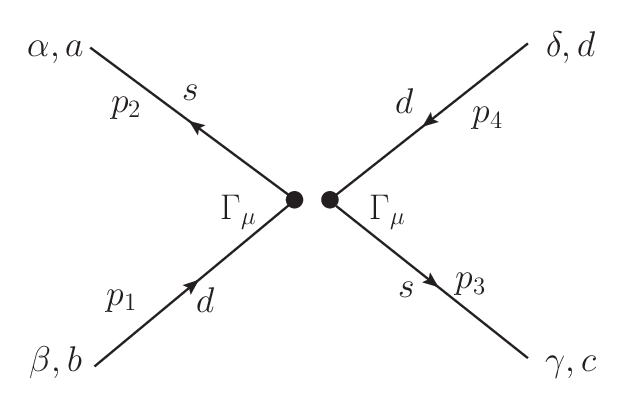}&
		\includegraphics[width=0.4\textwidth,height=0.18\textwidth]{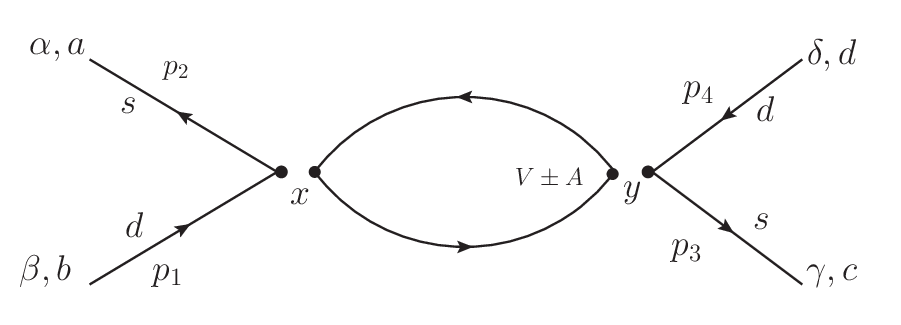} &
		\includegraphics[width=0.3\textwidth,height=0.35\textwidth]{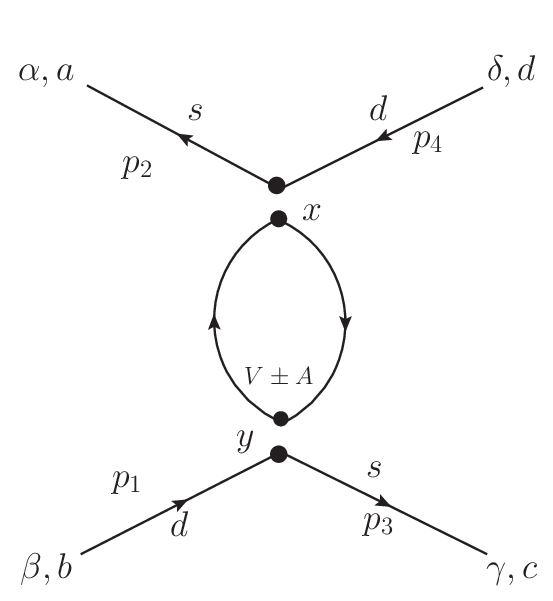}
	\end{tabular}
	\caption{Some example diagrams that determine the RI/SMOM counter terms $X^{\Lat}_{ij}$.  The diagram on the left determines the RI-SMOM normalization of the operator $O_{LL}$ while the diagrams in the center and on the right have two $\Delta S = 1$ operators. The symbol $\Gamma_{\mu}$ stands for $\gamma_\mu(1-\gamma_5)$ and $V \pm A$ stands for $\gamma_\mu(1\pm \gamma_5)$. The right hand vertex in the central diagram could come from the operators $Q_3$ and $Q_5$.  The Greek and Roman indices on the external quark lines represent spin and color respectively.}
	\label{figure:SD_diag}
\end{figure}

Figure~\ref{figure:SD_diag} shows three examples of the diagrams used to determine the RI/SMOM counter term $X^{\Lat}_{ij}$.  The diagram on the left represents the calculation of $\Gamma^{\Loc,\mathrm{amp}}_{\alpha\beta\gamma\delta}$, defined in Eq.~\eqref{Eq:Gamma_SD}, determining the off-shell normalization of the local operator $O_{LL}$.  The diagrams in the center and on the right in the figure represent the calculation of $\Gamma^{\BL,\mathrm{amp}}_{\alpha\beta\gamma\delta,ij}$ defined in Eq.~\eqref{Eq:Gamma_BL}, from a Green's function containing two $\Delta S = 1$ operators.  The diagrams shown correspond to the operator combination $Q_1 Q_1$ if both the vertices are $V-A$ and the internal quark lines are $c\times(c-u)$. The diagrams will correspond to $Q_1 Q_3$ if both vertices are $V-A$, but the internal quark lines are $c\times c- u\times u$. The diagram will correspond to $Q_1 Q_5$ if it has one $V-A$ vertex and one $V+A$ vertex and the internal quark lines $c\times c- u\times u$. We can also obtain the diagrams with the operators $Q_2$, $Q_4$ and $Q_6$ by making one or both of the vertices color mixed.

Here we choose momentum-conserving kinematics:
\begin{equation}
	p_1 + p_4 = p_2 + p_3 , \label{Eq:mom_cons}
\end{equation}
and all the momenta have the same energy scale $\mu_{\RI}$. Our choice of the 
momentum has the form:
\begin{eqnarray}
	\label{Eq:mom_choice}
	p_1 &=& \frac{2\pi}{L}(M,M,0,0) \\ \nonumber
	p_2 &=& \frac{2\pi}{L}(M,0,M,0) \\ \nonumber
	p_3 &=& \frac{2\pi}{L}(0,M,0,M) \\ \nonumber
	p_4 &=& \frac{2\pi}{L}(0,0,M,M) .
\end{eqnarray}
This choice of non-exceptional momenta (in which no partial sum of incoming momentum vanishes) leads to better controlled IR behavior. Exceptional momenta would allow the internal propagators to carry small momenta resulting in increased sensitivity to infrared effects. When $M$ is not an integer, twisted boundary conditions are used to obtain a quark propagator carrying a momentum not allowed by periodic boundary conditions. To study the 
scale dependence, calculations will be performed with scales $\mu_{\rm RI} = 2\pi\sqrt{2} M/L$ between 1.41 and 2.56~GeV.

The short-distance correction is performed only for type 1 \& 2 diagrams.  We have done a similar study for the other three types of diagrams with four external quark lines but made no short-distance correction because the amplitude $\Gamma^{\BL,\mathrm{amp}}_{\alpha\beta\gamma\delta,ij}$ obtained for the type 3, 4 and 5 diagrams is much smaller than that for the type 1 \& 2 diagrams and is consistent with 0 within statistical errors.

\subsection{Standard Model Inputs}

Before presenting the numerical results from our lattice calculation, we give a brief introduction to how we define the
operators on the lattice and their corresponding Wilson coefficients. We can find the $\overline{MS}$ values of the Wilson coefficients by using Eq.~(12.43) - Eq.~(12.61) in Ref.~\cite{Buchalla:1995vs}.   The strong coupling $\alpha_s$ is evaluated using Eq.~(3.19) in Ref.~\cite{Buchalla:1995vs}. To obtain $\Lambda_{QCD}$, we use $\alpha_s(M_Z) = 0.1184$ to find $\Lambda_{QCD}^{5}$ in the five-flavor theory and then by requiring $\alpha_s(M_b)$ to be the same in the five- and four-flavor theories we can find $\Lambda_{QCD}^{4}$ in the four-flavor theory.  The standard model input parameters are summarized in Table~\ref{Table:input} while the values we use for $\Lambda_{QCD}^{4}$ and $\alpha_s$ are summarized in Table~\ref{Table:alphas}.  We renormalize the $\MS$ operators at $\mu_{\MS}=2.15$ GeV and the six Wilson coefficients are listed in Eq.~\eqref{Eq:C_MS}:
\begin{eqnarray}
	C^{\overline{MS}}(\mbox{2.15 GeV}) = (-0.2967,1.1385, 0.0217 ,-0.0518, 0.0102, -0.0671).
	\label{Eq:C_MS}
\end{eqnarray}

\begin{table}[ht]
	\centering
	\begin{tabular}{c|c|c|c|c}\hline\hline
		$m_t$ & $M_W$ & $M_Z$ & $\alpha_s(M_Z)$ & $m_b$ \\\hline
		172.2 GeV & 80.379 GeV & 91.1876 GeV & 0.1184 & 4.19 GeV \\\hline\hline
	\end{tabular}
	\caption{Standard model input parameters~\cite{PhysRevD.98.030001} used for the evaluation of the six $\Delta S=1$ Wilson coefficients.}
	\label{Table:input}
\end{table}

\begin{table}[ht]
	\centering
	\begin{tabular}{c|c|c|c}\hline\hline
		$\alpha_s(m_b)$ & $\alpha_s(\mu_{\MS})$ & $\Lambda_{QCD}^5$ & $\Lambda_{QCD}^4$ \\\hline
		0.2265 & 0.2974 & 231 MeV & 330 MeV \\\hline\hline
	\end{tabular}
	\caption{The values for $\alpha_s$ at different energy scales and $\Lambda_{QCD}$ for different numbers
		of active quark flavors.}
	\label{Table:alphas}
\end{table}

The generic products of current-current operators $[[\widetilde{Q}_i\widetilde{Q}_j]]$ for $i$ and $j$ equal one or two appearing in Eqs.~\eqref{Eq:H_eff_dS2} have multiple flavor structures. However, the operators with different flavor structures but the same values for $i$ and $j$ have the same Wilson coefficients and hence common values for the product $C_iC_j$.  We use the NPR procedure described in Appendix~\ref{sec:appendix} to obtain the six Wilson coefficients $C_i^\Lat$ that obey
\begin{equation}
	\sum_{i=1}^{6} C_i^{\overline{MS}} Q_i^{\overline{MS}} = \sum_{i=1}^{6} C_i^\Lat Q_i^\Lat.
	\label{eq:DeltaS=1-Lat}
\end{equation}
This allows us express the $\Delta S=1$ effective weak Hamiltonian $\mathcal{H}_W^{\Delta S=1}$ directly in terms of lattice operators.  As described in Appendix~\ref{sec:appendix} this is done by introducing an intermediate non-perturbative RI/SMOM scheme at the energy scale $\mu_\RI=2.15$ GeV and applying the QCD perturbation theory results of Ref.~\cite{Lehner:2011fz} to  express $\mathcal{H}_W^{\Delta S=1}$ in terms of operators renormalized in the RI-SMOM scheme:
\begin{equation}
	\sum_{i=1}^{6} C_i^{\overline{MS}} Q_i^{\overline{MS}} = \sum_{i=1}^{6} C_i^\RI Q_i^\RI.
\end{equation}
Specifically we use the ($\gamma_\mu$, $\slashed{q}$) RI-SMOM scheme as described in Section V.A. of Ref.~\cite{RBC:2020kdj}.  Finally a non-perturbative lattice QCD calculation is used to express the RI-SMOM-normalized $\mathcal{H}_W^{\Delta S=1}$ in term of lattice operators and 
Wilson coefficients which will therefore obey Eq.~\eqref{eq:DeltaS=1-Lat}.  The resulting six lattice Wilson coefficients $C^\Lat_i$ are given in Eq.~\eqref{Eq:lat_Wilson}, where the numbers in the parenthesis are the statistical errors:
\begin{eqnarray}
	\label{Eq:lat_Wilson}
	C^{\Lat} &=&
%	-0.2373(1) & 0.6885(1) & 0.0113(6) & -0.0213(7) & 0.0085(6) & -0.0256(6) \\
%	\left(-0.2219(1), 0.6448(2), 0.0134(8),-0.0266(11),0.0103(9),-0.0302(9)\right)
	\left(-0.2290(1), 0.6654(2), 0.0138(8),-0.0275(11),0.0106(9),-0.0312(9)\right).
\end{eqnarray}

\subsection{Evaluation of $X^{\Lat}_{ij}$}

To remove the unphysical lattice-regulated, short-distance divergence present in our evaluation of the product of two $\Delta S=1$ weak operators, we must calculate the short-distance artifact represented by quantity $X^{\Lat}_{ij}(\mu_\RI)$ defined in Eq.~\eqref{Eq:calc_Xij}.  By evaluating $X^{\Lat}_{ij}(\mu_\RI)$ using large non-exceptional external momenta, we force all of the internal momenta in the five-point function that defines $X^{\Lat}_{ij}(\mu_\RI)$ to be large.  This in turn requires that the separation between the positions of the two operators $x_1$ and $x_2$ must be small, on the order of $1/\mu_\RI$.  This can be easily demonstrated in our calculation of $X^{\Lat}_{ij}(\mu_\RI)$ if we introduce a upper limit $R$ into the summation over $x_1$ and $x_2$ in Eq.~\eqref{Eq:Gamma_BL} and sum only the points $x_1$ and  $x_2$ that satisfy $(x_1 - x_2)^2 \le R^2$. The amputated Green's function will now depend on the space-time cutoff $R$:
\begin{equation}
	\Gamma^{BL}_{\alpha \beta \gamma \delta,ij}(p_1,p_2,p_3,p_4,R) = \langle s_{\alpha}(p_1)\overline{d}_{\beta}(p_2)\left[ \sum_
	{x_1 x_2 \atop (x_1 - x_2)^2 \le R^2}  Q_i(x_1) Q_j(x_2)\right] s_{\gamma}(p_3) \overline{d}_{\delta}(p_4) \rangle . 
	\label{Eq:Gamma_R}
\end{equation}
Note, the sum over $x_1$ in the definition of the quantities $\Gamma^{BL}_{\alpha \beta \gamma \delta,ij}$ and $\Gamma^{SD}_{\alpha \beta \gamma \delta}$ introduces a simple factor of the space-time volume because the total incoming momenta is zero: $p_1+p_4-(p_2+p_3)=0$, see \eqref{Eq:mom_cons}.  This sum over $x_1$ is included  never-the-less, to better represent the actual calculation in which the external lines correspond to volume-source quark propagators with the specified momenta and the space-time sums over both $x_1$ and $x_2$ in the case of $\Gamma^{BL}_{\alpha \beta \gamma \delta,ij}$ are performed in order to exploit the added precision that comes from volume-averaging.

We then use Eq.~\eqref{Eq:calc_Xij} to find the $X^{\Lat}_{ij}$ for different values of the upper limit $R$ and different operator combinations. The results are shown in Table~\ref{table:Eij_R} for an external momentum scale $\mu_{\RI} = 1.41$ GeV.  We have dropped the statistical errors because they are very small in this calculation.  We can see that for $R\ge 4$, the results are close to those without the cutoff, indicating a very small contribution from larger distances. This conclusion will become stronger at larger momenta.

\begin{figure}[ht]
		\centering
		\includegraphics[width=0.6\textwidth]{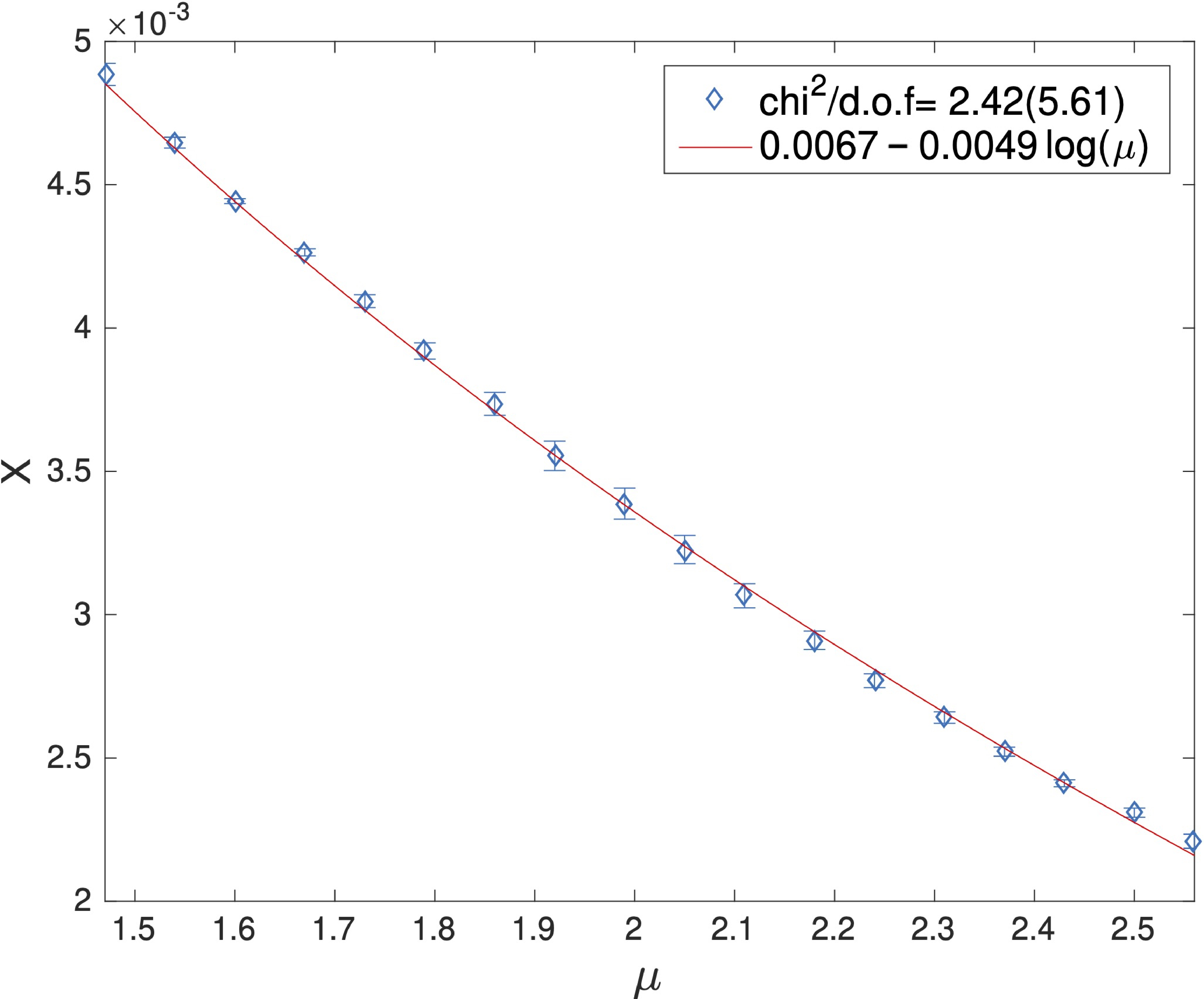} 
		\caption{A plot of the quantity $X(\mu_{\RI})$ as a function of momentum scale $\mu_\RI$ obtained from 3 
		gauge configurations.   Also shown is the result of an uncorrelated logarithmic fit.  The abscissa is plotted 
		in units of GeV.}
		\label{figure:log_fit_E}
	\end{figure}

\begin{table}[h]
	\centering
	\begin{tabular}{c|c|c|c|c|c|c}\hline\hline
		$R$ & 3 & 4 & 5 & 6 & 7 & none \\\hline
		$X^\Lat_{1,1}$ & -0.0491 &   -0.0530  &  -0.0534  &  -0.0533  &  -0.0533  &  -0.0533  \\\hline
		$X^\Lat_{1,2}$ & -0.0240 &   -0.0254  &  -0.0255  &  -0.0254  &  -0.0254  &  -0.0254  \\\hline
		$X^\Lat_{2,2}$ & -0.0140 &   -0.0148  &  -0.0148  &  -0.0148  &  -0.0148  &  -0.0148  \\\hline
		$X^\Lat_{1,3}$ & -0.1098 &   -0.1222  &  -0.1237  &  -0.1233  &  -0.1229  &  -0.1226  \\\hline
		$X^\Lat_{1,4}$ & -0.0258 &   -0.0275  &  -0.0275  &  -0.0274  &  -0.0273  &  -0.0272  \\\hline
		$X^\Lat_{1,5}$ & 0.1340&   0.1370 &  0.1371 &  0.1372 &  0.1374 &  0.1375 \\\hline
		$X^\Lat_{1,6}$ & 0.0547&   0.0561 &  0.0561 &  0.0562 &  0.0564 &  0.0567 \\\hline
		$X^\Lat_{2,3}$ & -0.0258 &   -0.0275  &  -0.0275  &  -0.0273  &  -0.0273  &  -0.0273  \\\hline
		$X^\Lat_{2,4}$ & -0.0302 &   -0.0324  &  -0.0325  &  -0.0323  &  -0.0322  &  -0.0322  \\\hline
		$X^\Lat_{2,5}$ & 0.0357&   0.0364 &  0.0364 &  0.0364 &  0.0364 &  0.0363 \\\hline
		$X^\Lat_{2,6}$ & 0.0444&   0.0451 &  0.0451 &  0.0451 &  0.0451 &  0.0448 \\\hline\hline
	\end{tabular}
	\caption{The short-distance subtraction constant $X^{\Lat}_{ij}(\mu_\RI)$ for various values of the space-time cutoff $R$ and operator combinations $[[\widetilde{Q}_i\widetilde{Q}_j]]$.  The external momenta have the scale $\mu_\RI = 1.41$~GeV.  We see the expected independence of $R$ as it is increased above 4 lattice units.}
	\label{table:Eij_R}
\end{table}
By summing over these coefficients multiplied by the lattice Wilson coefficients we can determine $X(\mu_{\RI}) = \sum_{i,j} C^{\Lat}_i C^{\Lat}_j X^{\Lat}_{ij}(\mu_{\RI})$ for each choice of momentum scale $\mu_{\RI}$.  To obtain results with a non-integer momentum, we have used twisted boundary conditions.  We show the quantity $X(\mu_{\RI}) $ in Table~\ref{table:E_lat_mu}.  Because this quantity is logarithmically divergent, we expect it to behave as $\ln(\mu_\RI a)$ when $\mu_\RI$ and $1/a$ are both larger than the charm quark mass.  The dependence of $X$ on $\mu_\RI$ is shown in Fig.~\ref{figure:log_fit_E} together with an uncorrelated logarithmic fit.

\begin{table}[ht]
	\centering
	\begin{tabular}{c|c|c|c|c|c|c|c|c|c}\hline\hline
		$\mu_{\RI}$ (GeV) & 1.47& 1.54& 1.60& 1.67& 1.73& 1.79& 1.86& 1.92& 1.99 \\\hline
		$X$($\times 10^{-3}$) & -5.5788 &   -5.3028  &  -5.0661  &  -4.8582  &  -4.6607  &  -4.4588 &
		-4.2453  &  -4.0362   & -3.8439 \\\hline \hline
		$\mu_{\RI}$ (GeV)& 2.05& 2.11& 2.18& 2.24& 2.31&  2.37& 2.43& 2.50& 2.56 \\\hline
		$X$($\times 10^{-3}$) &-3.6596 &  -3.4741  &  -3.2959 &   -3.1340  &  -2.9859 &   -2.8489 &  
		-2.7225 &   -2.6045 &   -2.4904 \\\hline\hline
	\end{tabular}
	\caption{Values of $X(\mu_\RI) =  \sum_{i,j} C^{\Lat}_i C^{\Lat}_j X^{\Lat}_{ij}(\mu_{\RI})$ for different values of the
	        momentum scale $\mu_\RI$.  We do not show the statistical errors because they are less than 1\%.}
	\label{table:E_lat_mu}
\end{table}

\subsection{Evaluation of $Y^{\MS}_{ij}$}
\label{sec:evaulate-Y}

As explained in Section~\ref{sec:pert-renorm}, instead of evaluating $Y^{\MS}_{ij}(\mu_{\MS}, \mu_{\RI})$, we evaluate the more accessible quantity $\Delta Y^{\MS}_{ij}(\mu_{\MS}, \mu_{\RI}) = Y^{\MS}_{ij}(\mu_{\MS}, \mu_{\RI}) - Y^{\MS}_{ij}(\mu_{\MS}, 0)$ which is evaluated at one-loop to zeroth order in QCD perturbation theory and contributes at NLO because the subtraction has removed the large logarithm $\ln(M_W/m_c)$.  The Wilson coefficients which multiply $\Delta Y(\mu_\RI)$ in Eq.~\eqref{Eq:masterEq} introduce some of the terms needed for a complete NLO sum over terms of $\mathcal{O}\bigl(\alpha_s \ln(M_W/\mu_\RI)\bigr)^l$.  However, here we do not attempt to determine additional NLO terms that appear in  $Y^{\MS}_{ij}(\mu_{\MS}, \mu_{\RI})$ arising for example from the external momentum dependence of the higher order QCD corrections to the bilocal operators $[[\widetilde{Q}_i\widetilde{Q}_j]]$.  Consequently our final result does not include all NLO logarithms.  These, as well as potentially important NNLO terms~\cite{Brod:2010mj, Brod:2011ty}, are omitted from the present unphysical-mass calculation which is intended only to demonstrate the practicality of the proposed approach.

Because of the convergence resulting from the subtraction defining $\Delta Y^{\MS}_{ij}$, we can perform this one-loop calculation in perturbation theory as suggested in Fig.~\ref{Fig:deltaY} without the use of dimensional regularization or the introduction of the scale $\mu_{\MS}$. The calculation of $Y(\mu_{\MS}, 0)$ with zero momentum on the external legs can be found in Eqs.~(12.63)-(12.66) of Ref.~\cite{Buchalla:1995vs}.  We have also listed the result here:  
\begin{eqnarray}
	\label{Eq:Y0}
	Y^{\MS}_{ij}(\mu_{\MS}, 0) = \frac{m_c^2}{8\pi^2} r_{ij}(\mu_{\MS}) \\
	r_{ij} =  \begin{cases}
		(-4\ln(\mu_{\MS}/m_c) + 2)\tau_{ij},\;\; & j = 1,2 \\
		(-8\ln(\mu_{\MS}/m_c) + 4)\tau_{ij},\;\; & j = 3,4 \\
		(8\ln(\mu_{\MS}/m_c) - 4)\tau_{ij},\;\; & j = 5,6
	\end{cases} \\
	\tau_{1,1} = \tau_{1,3} = \tau_{1,5} = 3 \\
	\tau_{1,2} = \tau_{1,4} = \tau_{1,6} = 1\\\label{Eq:Y0_last}
	\tau_{2,j} = 1,\; \text{for any j}.
\end{eqnarray}
We have made the necessary modifications to these formulae required by our use of CKM unitarity to eliminate $\lambda_c$ instead of $\lambda_u$.  We note that even for a NLO calculation, we do not need to take the scale dependence of the charm quark mass into consideration and use a constant charm quark mass given by the input lattice quark mass (0.363) converted to $\overline{MS}$: $m_c = 0.363 \times 1.78 (\mathrm{GeV}) \times 1.498 = 968$ MeV, where 1.78 GeV is the inverse lattice spacing and 1.498 is the mass renormalization factor $Z_m^{\Lat\rightarrow \MS}(2 \mathrm{GeV})$ taken from Ref.~\cite{Aoki:2010dy}.

Our results for $\Delta Y^{\MS}_{ij}(\mu_{\RI})$ are given by:  
\begin{eqnarray}
	\label{Eq:DeltaY}
	\Delta Y^{\MS}_{ij}(\mu_{\RI}) = \frac{m_c^2}{8\pi^2} \Delta r_{ij}(\mu_{\RI}) \\
	\Delta r_{ij} =  \begin{cases}
		\left[ \frac{\mu_{\RI}^2 + m_c^2}{m_c^2}\times c(m_c,\mu_{\RI}) 
		- b(m_c,\mu_{\RI}) - 1\right] \tau_{ij},\;\; & j = 1,2 \\
			-\left[ \frac{\mu_{\RI}^2}{m_c^2} \times d(m_c,\mu_{\RI}) + 
			2\times b(m_c,\mu_{\RI}) \right]\tau_{ij} ,\;\; & j = 3,4 \\
				4b(m_c, \mu_{\RI})\tau_{ij},\;\; & j = 5,6 
	\end{cases} \\
	b(m_c, \mu_{\RI}) = \int_0^1 \text{dx} \ln\frac{m_c^2}{x(1-x)\mu_{\RI}^2 + m_c^2} \\
	c(m_c, \mu_{\RI}) = \int_{0}^{1} 	\text{dx}   \ln \frac{x ( 1-x ) \mu_{\RI}^{2} +m_{c}^{2}}{x ( 1-x )  
	\mu_{\RI}^{2} + ( 1-x ) m_{c}^{2}} \\\label{Eq:DeltaY_last}
	d(m_c, \mu_{\RI}) = \int_0^1 \text{dx} \ln\frac{x(1-x)\mu_{\RI}^2 }{x(1-x)\mu_{\RI}^2 + m_c^2}. 
\end{eqnarray}
We have done this calculation in two ways.  The first is to analytically perform the Feynman integral over the internal quark loop.  The second is to perform a numerical free-field calculation, using the same projector as in Eq.~\eqref{Eq:calc_Xij}.  We have checked that they give the same result when computed at the same the values for $\mu_{\RI}$ and $m_c$.  The results given in Eqs.~\eqref{Eq:DeltaY} - \eqref{Eq:DeltaY_last} depend only on the energy scale of the external momenta $\mu_{\RI}$ and are independent of the specific choice of the four external momenta in Eq.~\eqref{Eq:mom_choice}, provided they have the same energy scale and momentum conservation is satisfied, as in Eq.~\eqref{Eq:mom_cons}. 

We also provide the numerical values for $\Delta Y(\mu_{\RI}) = \sum_{ij} C_i^{\overline{MS}}C_j^{\overline{MS}} \Delta Y^{\MS}_{ij}(\mu_{\RI})$ in Table~\ref{table:E_cont_mu} for the same set of energy scales that we used to calculate $X(\mu_\RI)_{ij}$.  We note that these values for $\Delta Y(\mu_{\RI})$ can not be directly compared to the results for $X(\mu_{\RI})$ given in Table~\ref{table:E_lat_mu} because $X(\mu_{\RI})$ is multiplied by an operator with lattice normalization while $\Delta Y(\mu_{\RI})$ multiplies an $\overline{MS}$ operator.

	\begin{table}[h]
		\centering
		\begin{tabular}{c|c|c|c|c|c|c|c|c|c}\hline\hline
		$\mu_{\RI}$ (GeV) & 1.47& 1.54& 1.60& 1.67& 1.73& 1.79& 1.86& 1.92& 1.99 \\\hline
			$\Delta Y$($\times 10^{-2}$) & 2.3032 & 0.9698 & 1.1117 & 1.2425& 1.4059 & 1.5552 & 1.7132&  1.9086 &
			2.3032 \\\hline \hline
		$\mu_{\RI}$ (GeV)& 2.05& 2.11& 2.18& 2.24& 2.31&  2.37& 2.43& 2.50& 2.56 \\\hline
			$\Delta Y$($\times 10^{-3}$) & 2.4993 & 2.7043 & 2.9547 & 3.1790& 3.4520  &3.6956 &3.9481 
			&4.2541 &4.5260\\\hline\hline
	\end{tabular}
	\caption{Numerical value for $\Delta Y(\mu_{\RI})$, at the same scales used to evaluate $X(\mu_{\RI})$.}
		\label{table:E_cont_mu}
	\end{table}

\subsection{Lattice results for the long-distance contribution to $\epsilon_K$} 

We have measured all five types of four-point contractions using lattice QCD.   Similar to what was done in Ref~\cite{Christ:2012se} for $\Delta M_K$ when computing the four-point correlator, we compute separately the parity conserving and the parity violating parts.  This separation is useful for identifying which intermediate states are present and need to be subtracted if lighter than the kaon.  This is achieved by separating the spin structure of each of the two weak vertices into the part that conserves parity and the part that violates parity. We use $(V-A)\times(V-A) = (VV + AA) - (AV + VA)$ and $(V-A)\times(V+A) = (VV - AA) + (VA - AV)$,  where $V$ corresponds to a $\gamma_\mu$ vertex and $A$ to a $\gamma_\mu \gamma_5$ vertex.  The $AA$ and $VV$ structures conserve parity while the $AV$ and $VA$ violate parity.
	 
For the parity conserving part, the single-pion intermediate state is lighter than the kaon.  For the parity violating part, only the intermediate vacuum state is lighter than the kaon. We note that the single pion states contributes to type 1,3,4,5 diagrams while the vacuum state contributes only to type 4 diagrams. We can sum over all five types of diagrams and then perform a subtraction of the lighter-than-kaon intermediate states, or we can do an intermediate state subtraction for each type of diagram and then combine them. We note that the second approach is less well-defined because we have to determine the single-pion contribution to each type of diagram independently when only the sum of the diagrams is guaranteed to correspond to actual Hilbert space matrix elements where intermediate states can be identified.  Thus, in our final result, we use the first approach while when we show how each type of diagram contributes, we must attempt the second approach. 

\begin{figure}[H]
		\centering
		\begin{tabular}{cc} 
			\includegraphics[width=0.44\textwidth]{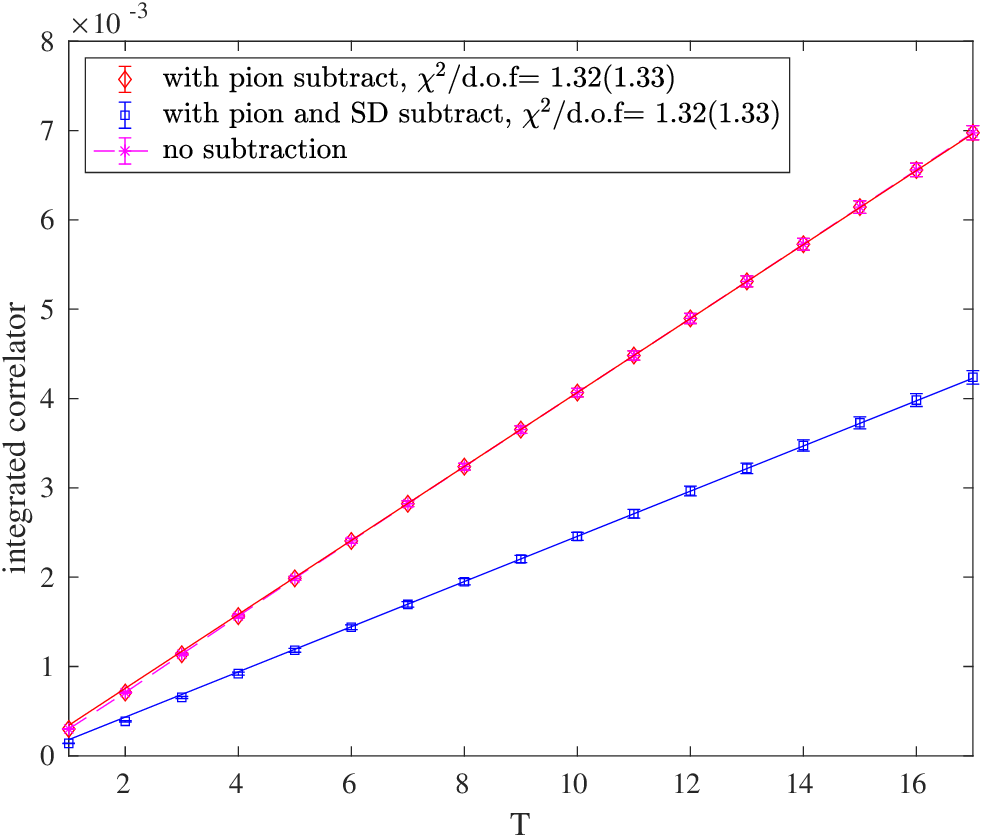}& 
			\includegraphics[width=0.44\textwidth]{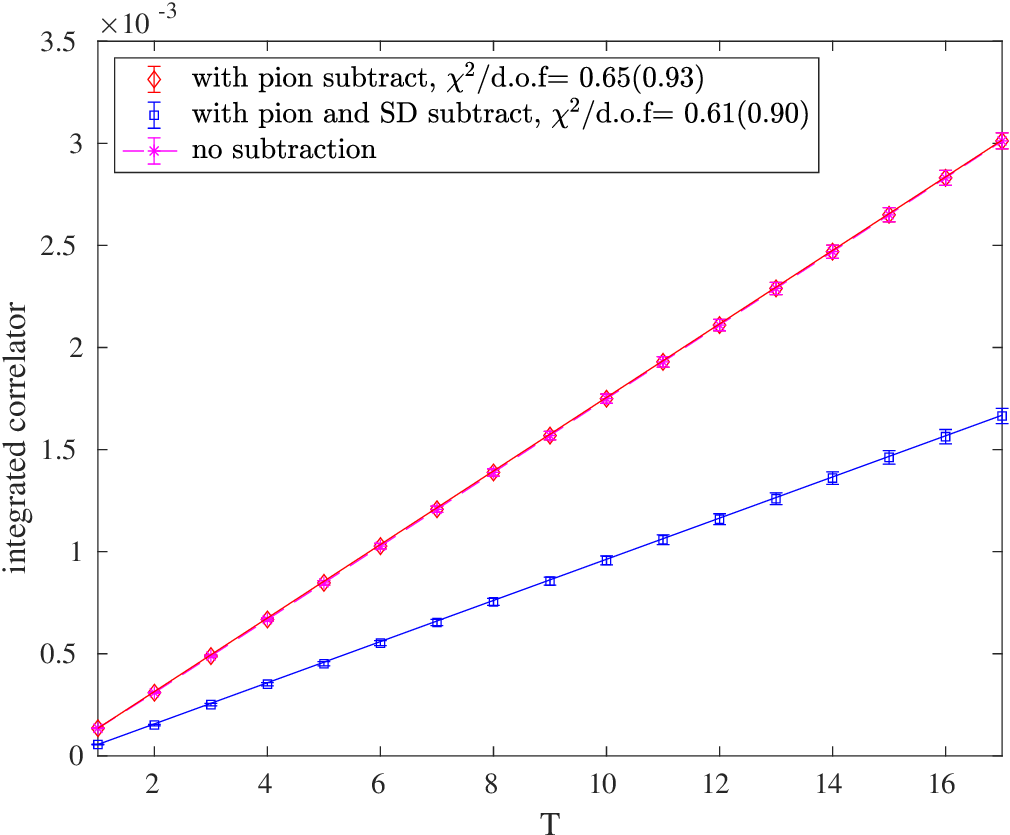}\\
			$Q_1Q_1$ & $Q_1Q_2$ \\
			\includegraphics[width=0.44\textwidth]{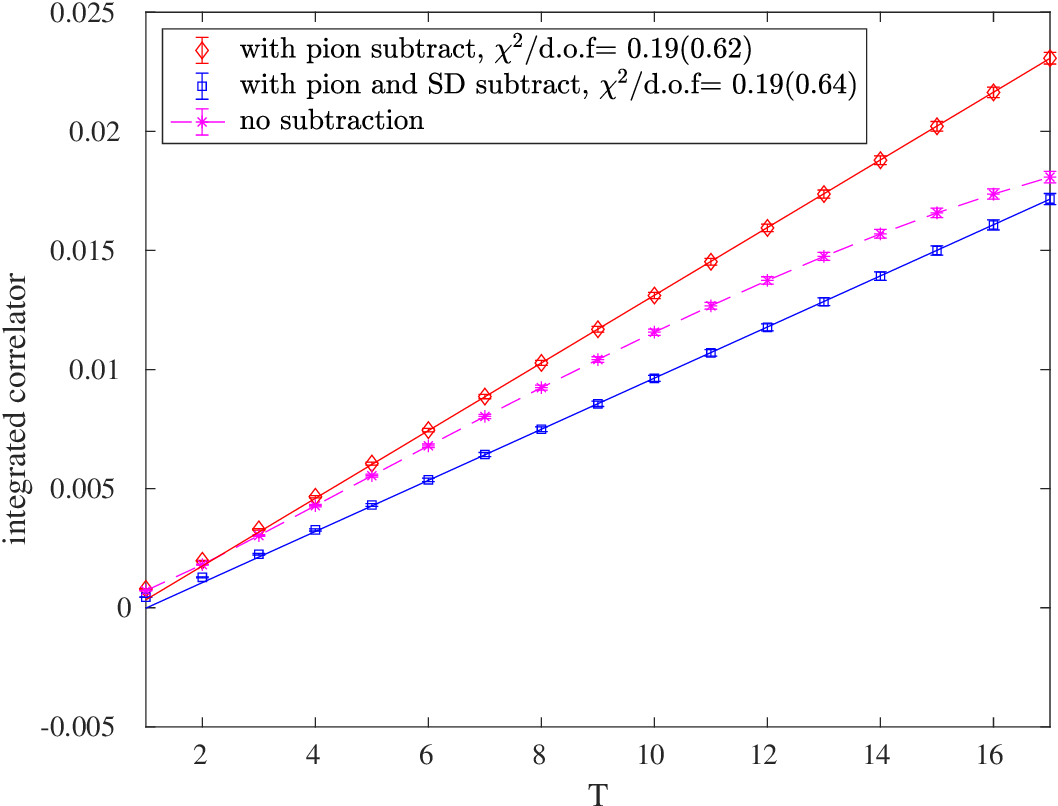}& 
			\includegraphics[width=0.44\textwidth]{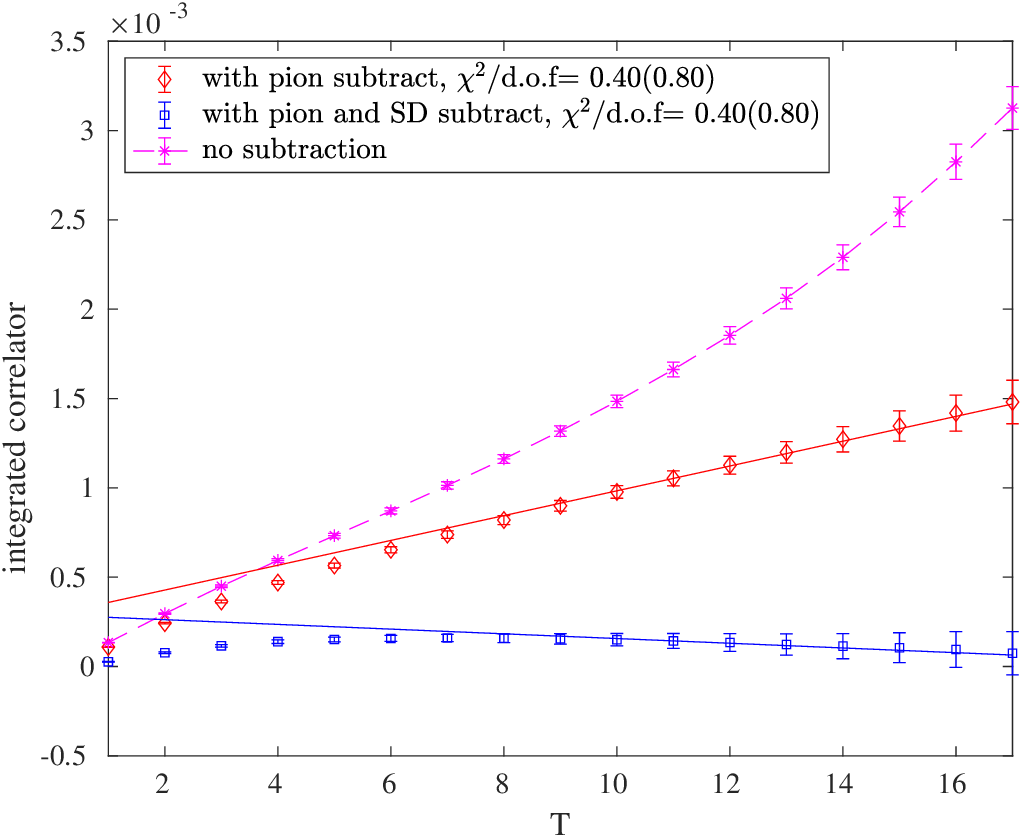}\\
			$Q_1Q_3$ & $Q_1Q_4$ \\
			\includegraphics[width=0.44\textwidth]{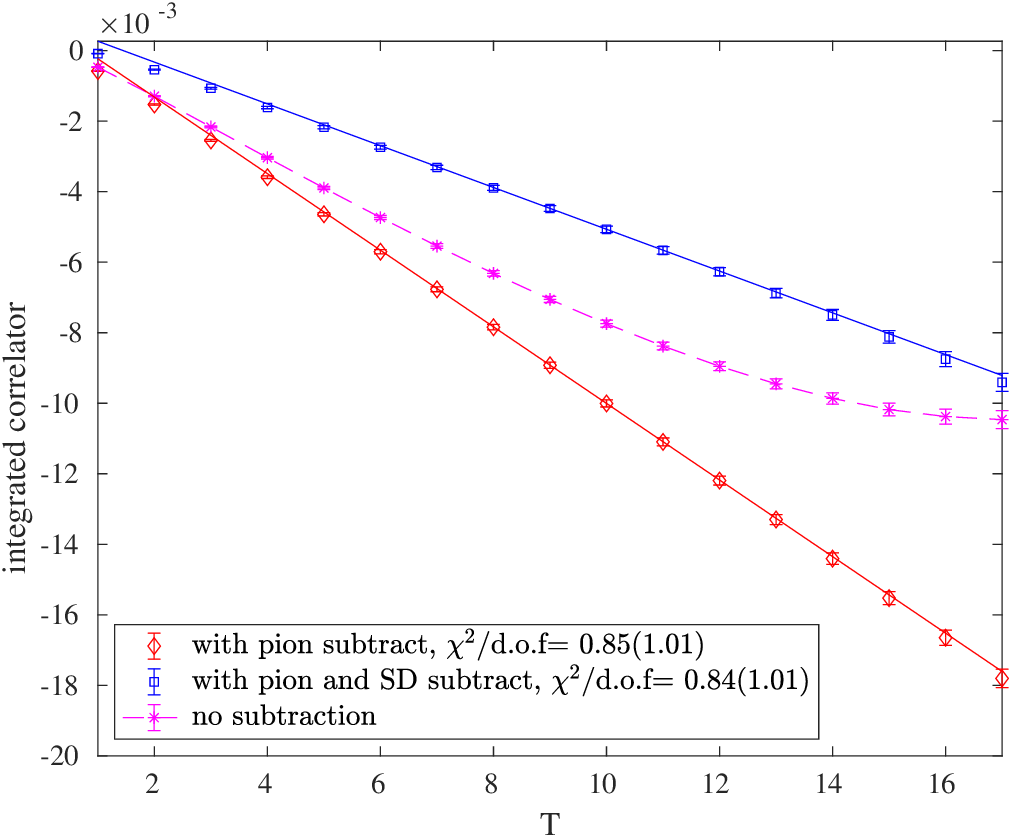}& 
			\includegraphics[width=0.44\textwidth]{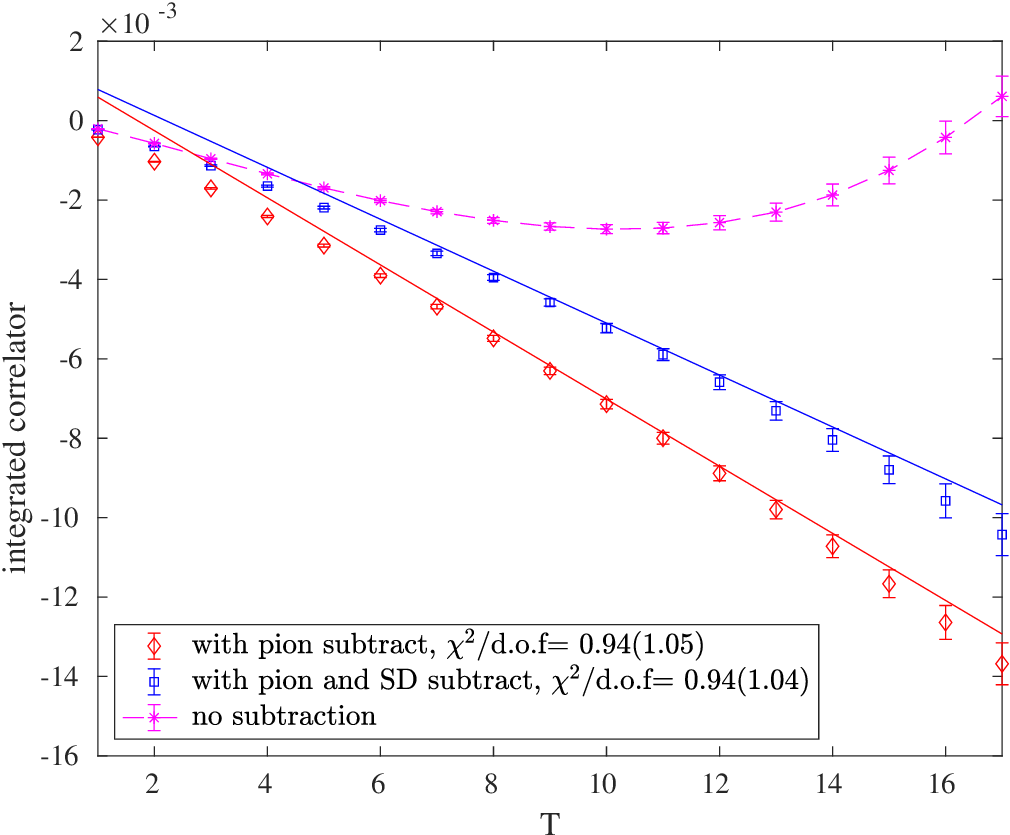}\\
			$Q_1Q_5$ & $Q_1Q_6$ \\
		\end{tabular}
		\caption{Integrated correlators for the products $Q_1Q_j$ with $j=1\ldots6$ including only 
		type 1 and type 2 diagrams. We show the results without subtraction, with subtraction of
		only the single-pion state, with subtraction of both the pion and the short-distance part. 
		We use a correlated fit with the fitting range $12 \le T \le 16$.  The Wilson coefficients are 
		not included.}
		\label{figure:fit_epsK12_1}
	\end{figure}

\begin{figure}[H]
		\centering
		\begin{tabular}{cc} 
			\includegraphics[width=0.44\textwidth]{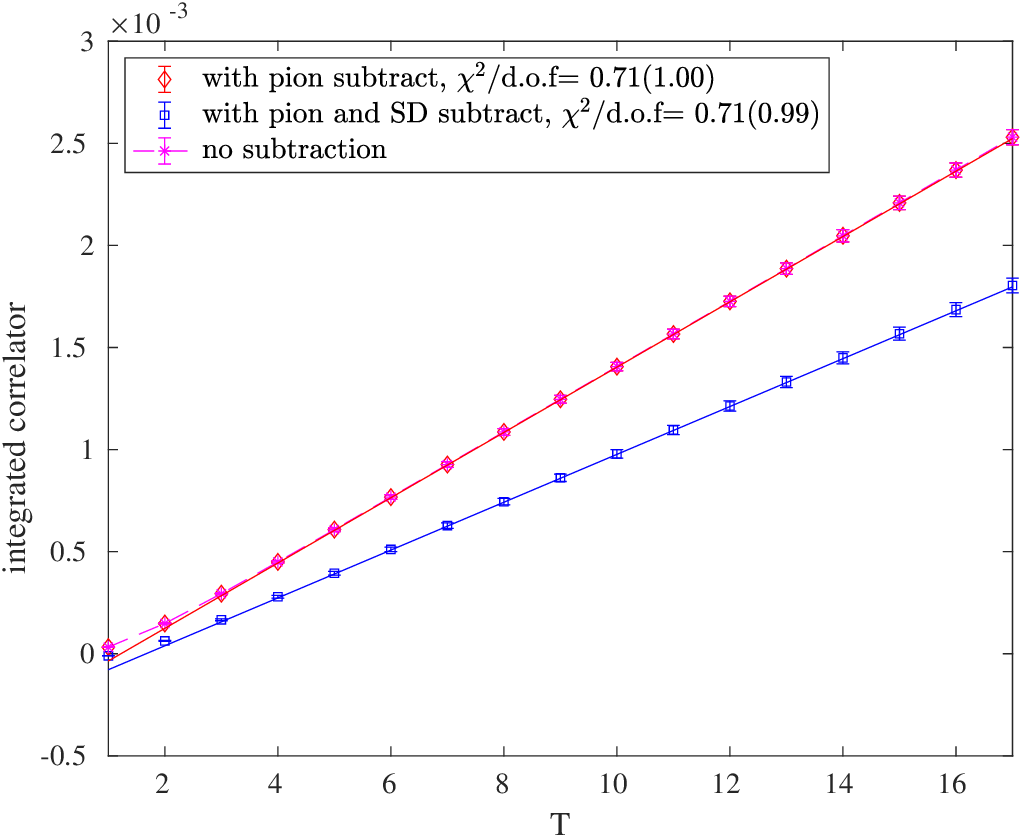}& 
			\includegraphics[width=0.44\textwidth]{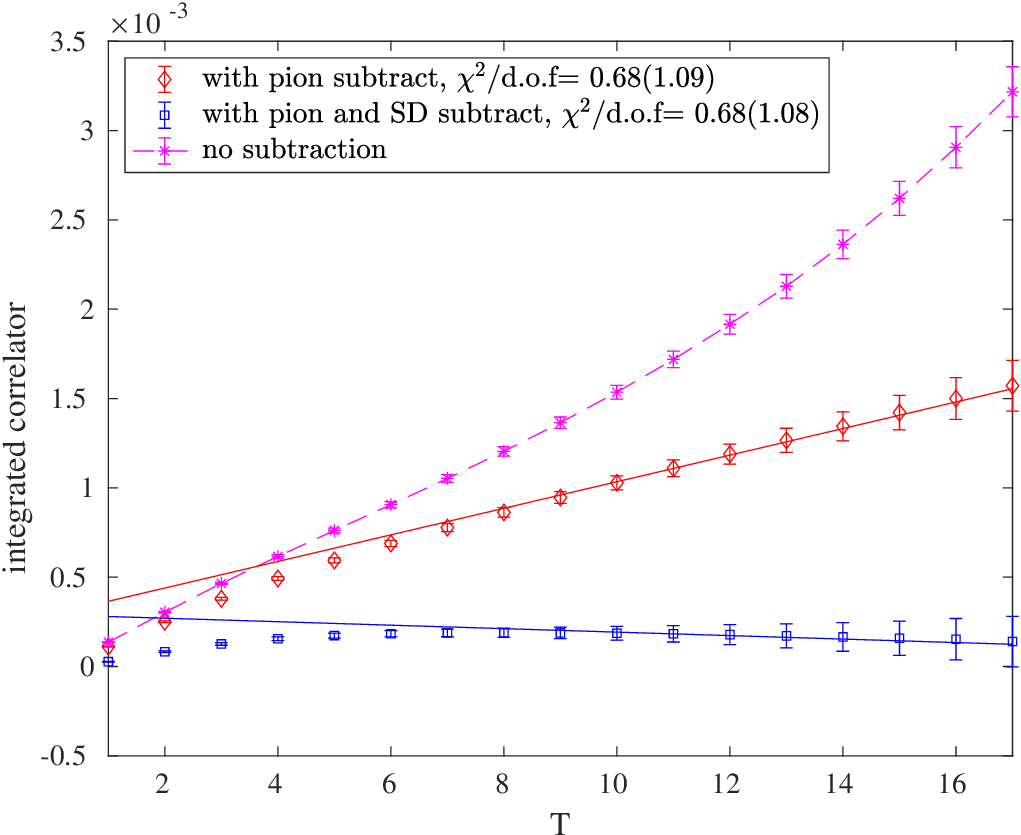}\\
			$Q_2Q_2$ & $Q_2Q_3$ \\
			\includegraphics[width=0.44\textwidth]{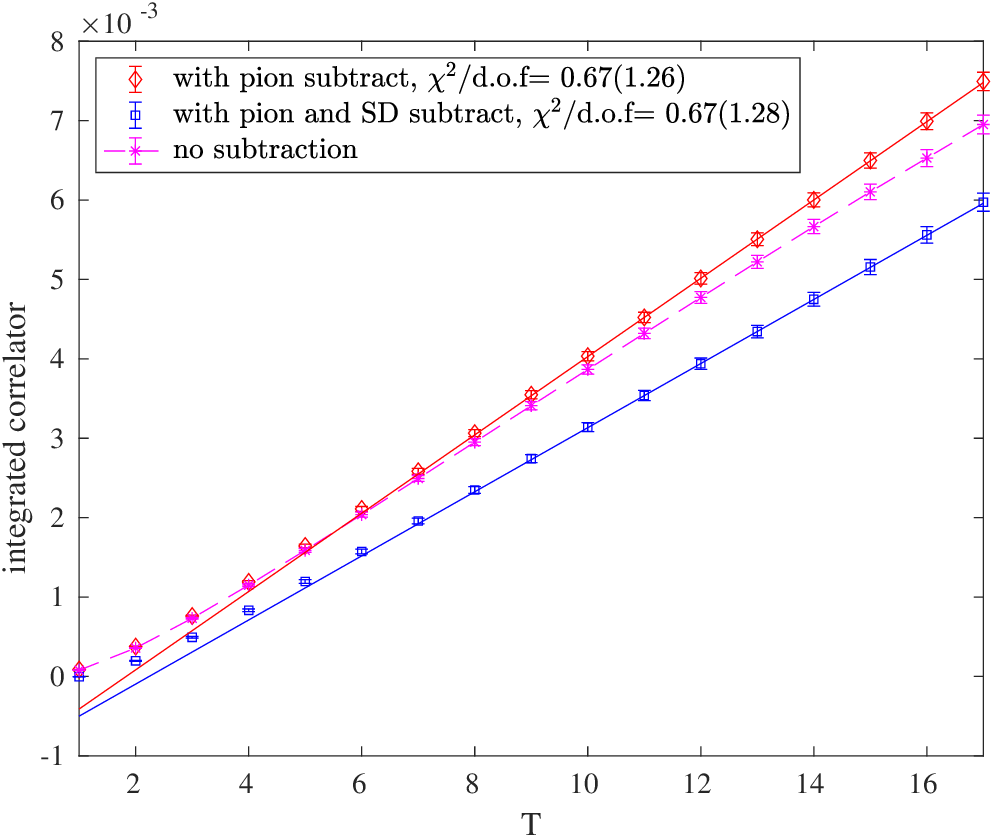}& 
			\includegraphics[width=0.44\textwidth]{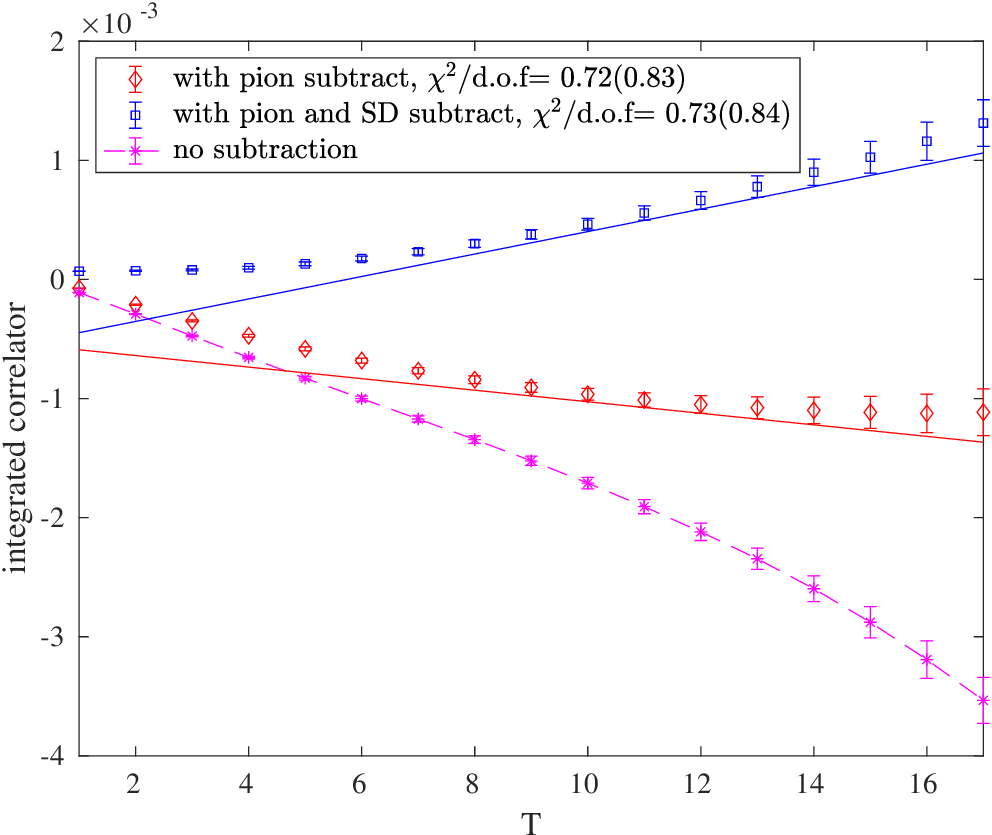}\\
			$Q_2Q_4$ & $Q_2Q_5$ \\
			\includegraphics[width=0.44\textwidth]{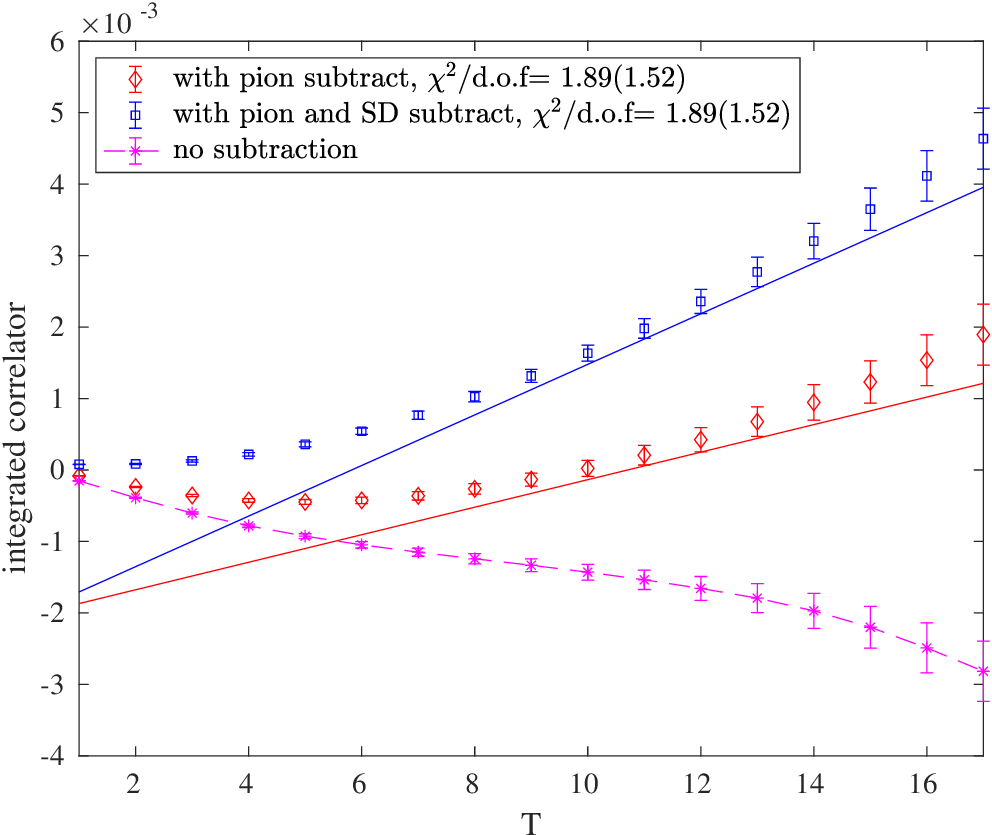}&\\
			$Q_2Q_6$ & \\
		\end{tabular}
		\caption{Integrated correlators for the products $Q_2Q_j$ with $j=2\ldots6$ including only 
		type 1 and type 2 diagrams. We show the results without subtraction, with subtraction of
		only the single-pion state, with subtraction of both the pion and the short-distance part.
		We use a correlated fit with the fitting range $12 \le T \le 16$.  The Wilson coefficients 
		are not included.}
		\label{figure:fit_epsK12_2}
	\end{figure}

After subtracting the intermediate states that are lighter than the kaon from our integrated correlator, we can do a linear fit to the dependence of the integrated correlator on the length $T$ of the integration region.  We show these linear fits to the type 1 and type 2 diagrams for all eleven bilocal operators in Figs.~\ref{figure:fit_epsK12_1} and \ref{figure:fit_epsK12_2}. We plot three versions of the integrated correlators: without any subtraction, after only the pion state has been subtracted and after we subtract both the pion state and the unphysical short-distance part determined by coefficient $X^{\Lat}_{ij}$.

Next we include the contributions of the remaining diagrams, those of type 3, 4 and 5.  These include quark-line disconnected topologies which increase the statistical noise and require the subtraction of the vacuum state.  For the parity violating parts of the integrated correlator, we have added the pseudo-scalar operator $\overline{s} \gamma_5 d$ to each $\Delta S=1$ operator with a coefficient chosen to cancel the vacuum intermediate state.  For the parity conserving part, we add the scalar operator $\overline{s} d$ to cancel the pion intermediate state. We determine the coefficients $c^s$ and $c^p$ of these two operators by requiring that the new operator $Q_i' = Q_i - c^s_i \overline{s} d - c^p_i \overline{s} \gamma_5 d$ have a zero matrix element between both the kaon and vacuum states, $\langle 0 | Q'| K^0 \rangle=0$, and between the kaon and single pion states, $\langle \pi | Q' | K^0 \rangle=0$. Thus, $c_i^s$ and $c_i^p$ are determined by the equations
\begin{eqnarray}
	\label{Eq:cs}
	\langle \pi | Q_i - c^s_i \overline{s} d | K^0 \rangle = 0 \\ \label{Eq:cp}
	\langle 0 | Q_i - c^p_i \overline{s} d | K^0 \rangle = 0. 
\end{eqnarray}

We can make these alterations to the $\Delta S=1$ effective weak Hamiltonian without changing its physical predictions because the scalar operator and pseudo-scalar operators are proportional to the divergence of a vector and axial current respectively, which implies that any process in which the initial and final four-momenta are equal will not be changed by adding these two operators.  With this construction we can remove the contribution of the vacuum intermediate state from amplitudes of type 4 and the contribution of the single-

\noindent
pion intermediate from amplitudes of type 1, 3, 4 and 5.  We point out that these scalar and pseudo-scalar two-quark operators enter only in diagrams of type 3, 4 and 5.
	
The subtraction of the pseudo-scalar operator $\overline{s} \gamma_5 d$ is particularly important because of the large coupling 
to the vacuum state, especially when we have a right-handed vertex. Without the subtraction of this operator, we must determine the kaon to vacuum matrix element $\langle 0 | Q_i | K^0 \rangle$ and directly subtract it using $\langle \overline{K}^0 |Q_i | 0 \rangle  \langle 0 | Q_j |K^0 \rangle/M_K$.  Because of the large size of the kaon to vacuum matrix element, the subtraction term is very large and after subtraction, the size of the integrated correlator is reduced by a factor on the order of 100. Fortunately, this $\overline{s} \gamma_5 d$ subtraction makes a comparable reduction in the statistical error.  On the other hand, the subtraction of the scalar operator $\overline{s} d$ is less important because the kaon to pion matrix element $\langle \pi | Q_i | K^0 \rangle$ is not that large. However, the scalar operator subtraction still reduces the error by roughly a factor of 5. 

\begin{figure}[H]
		\centering
		\begin{tabular}{cc} 
			\includegraphics[width=0.43\textwidth]{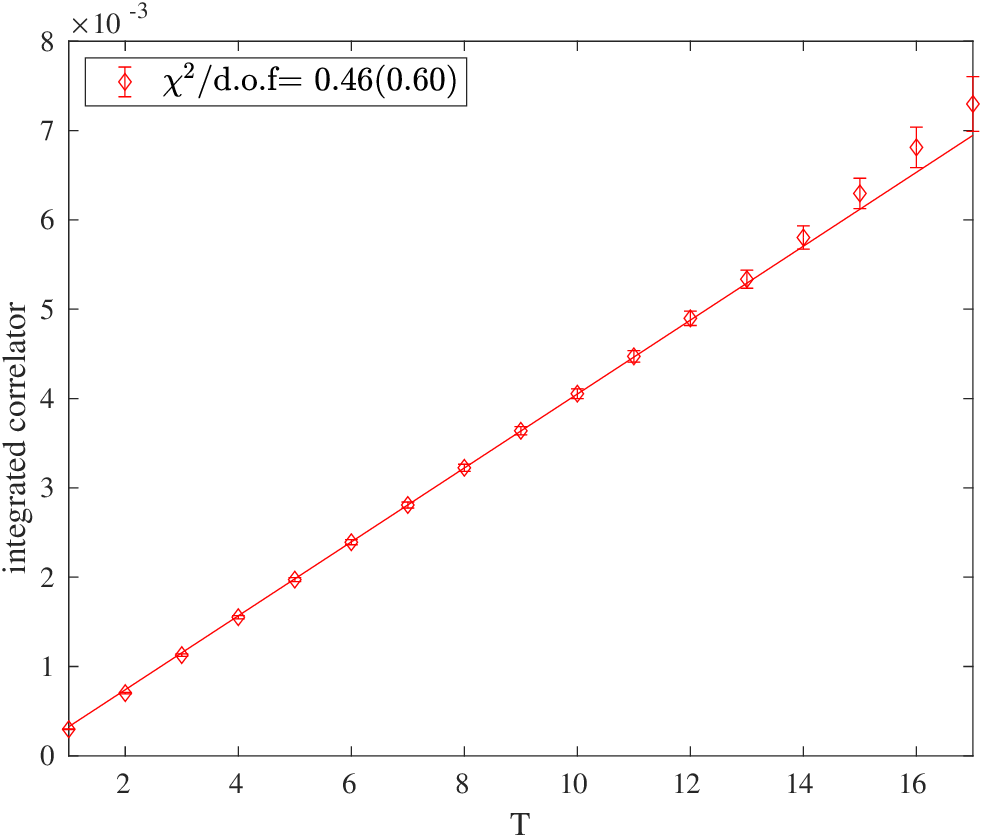}& 
			\includegraphics[width=0.43\textwidth]{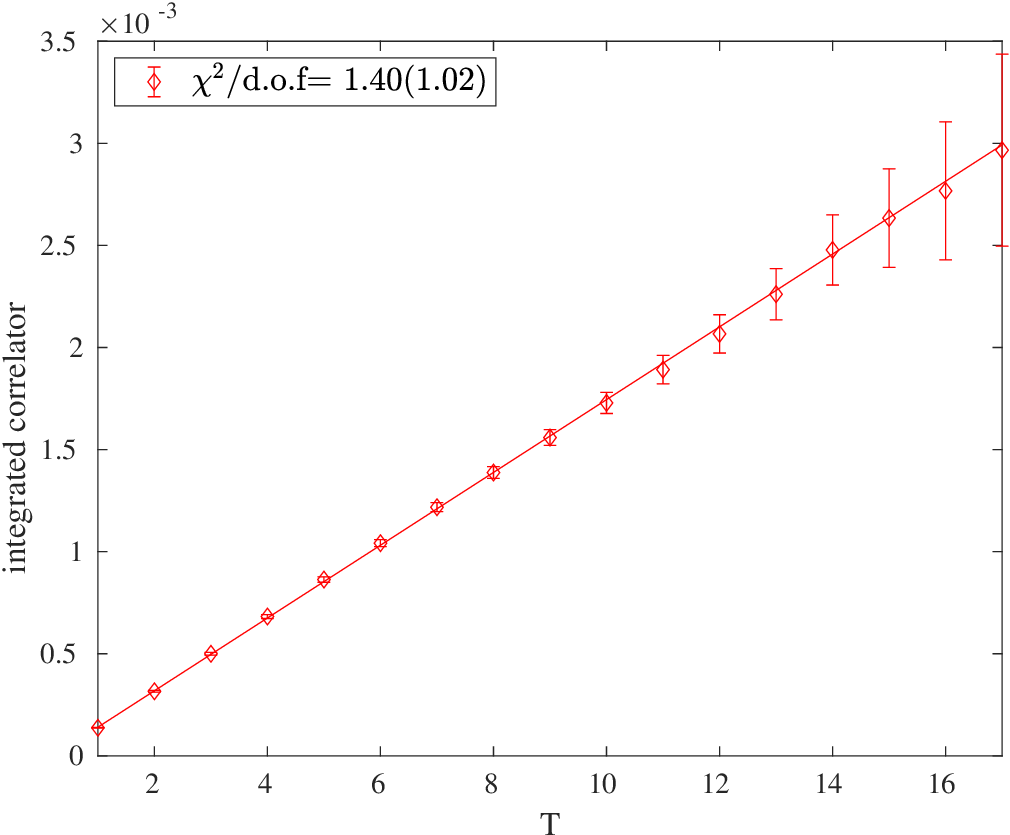}\\
			$Q_1Q_1$ & $Q_1Q_2$ \\
			\includegraphics[width=0.43\textwidth]{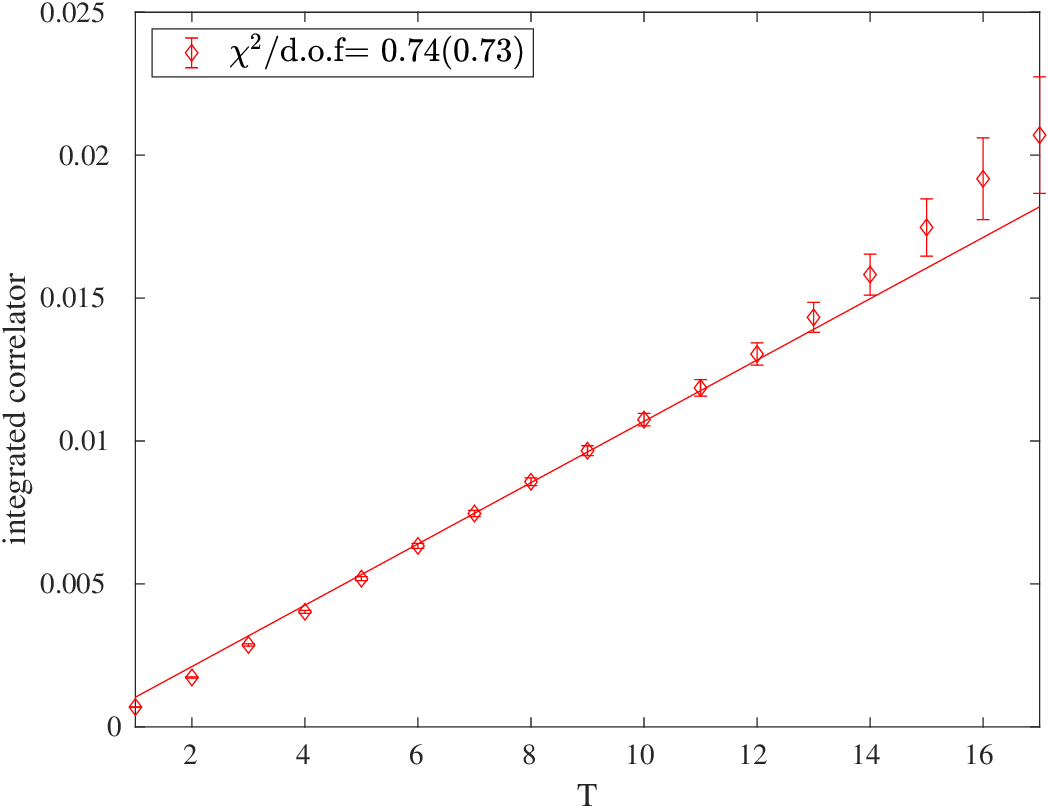}& 
			\includegraphics[width=0.43\textwidth]{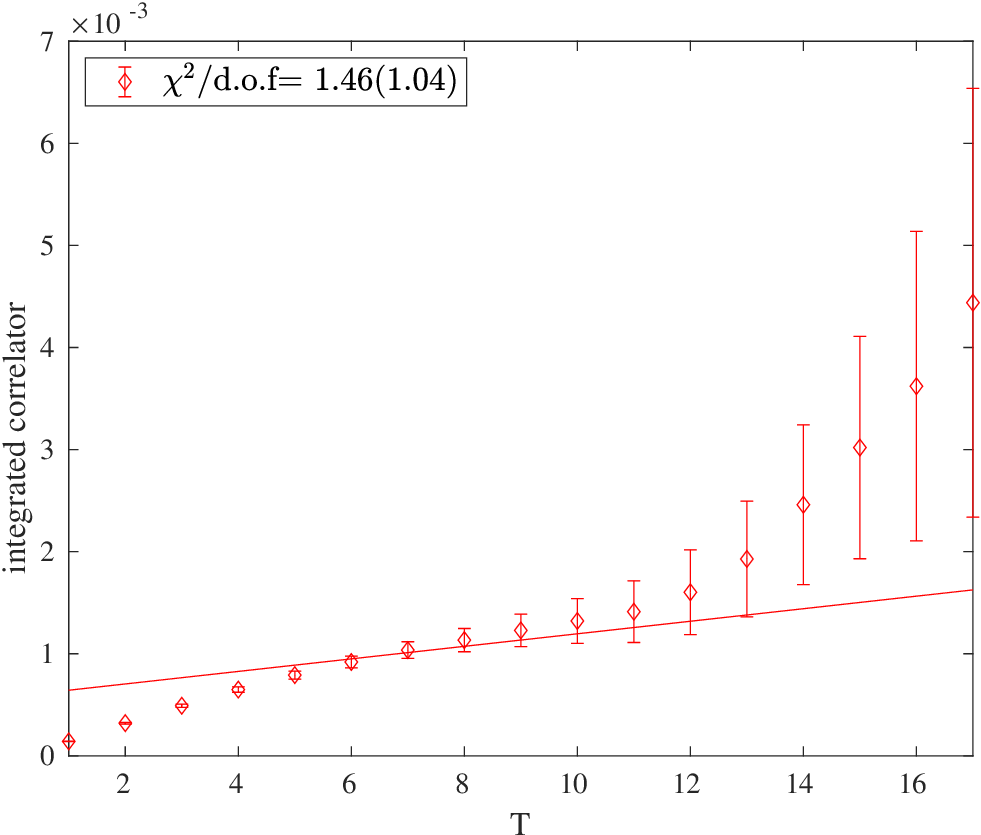}\\
			$Q_1Q_3$ & $Q_1Q_4$ \\
			\includegraphics[width=0.43\textwidth]{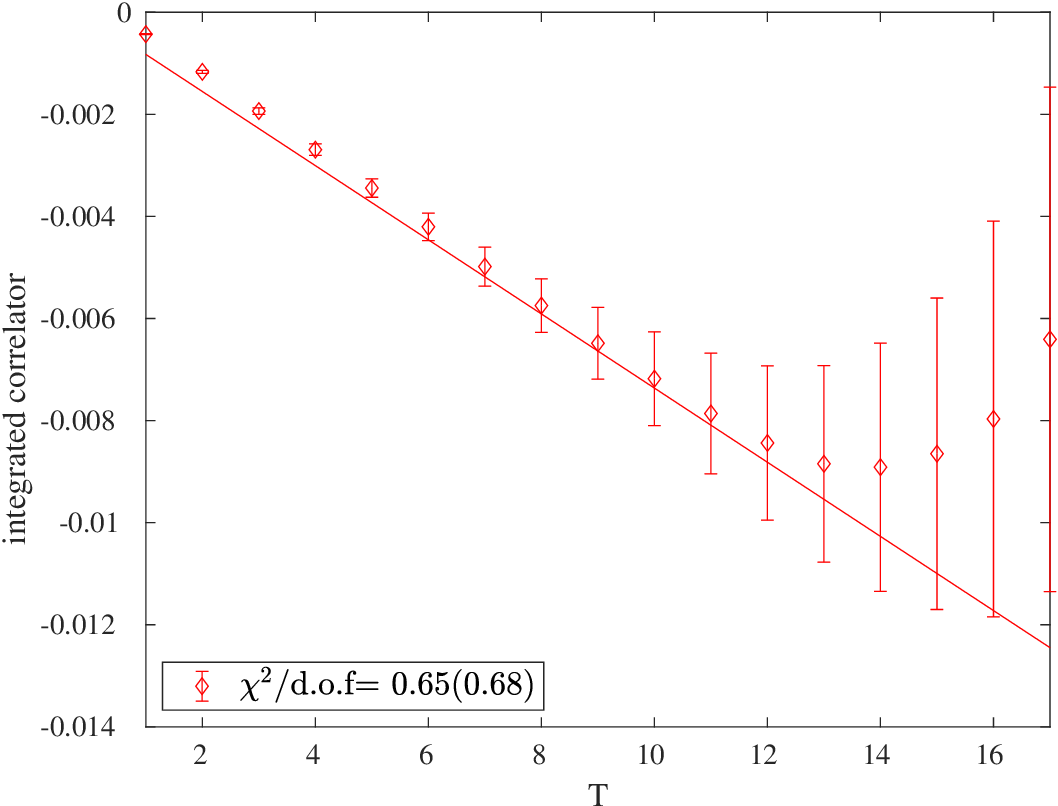}& 
			\includegraphics[width=0.43\textwidth]{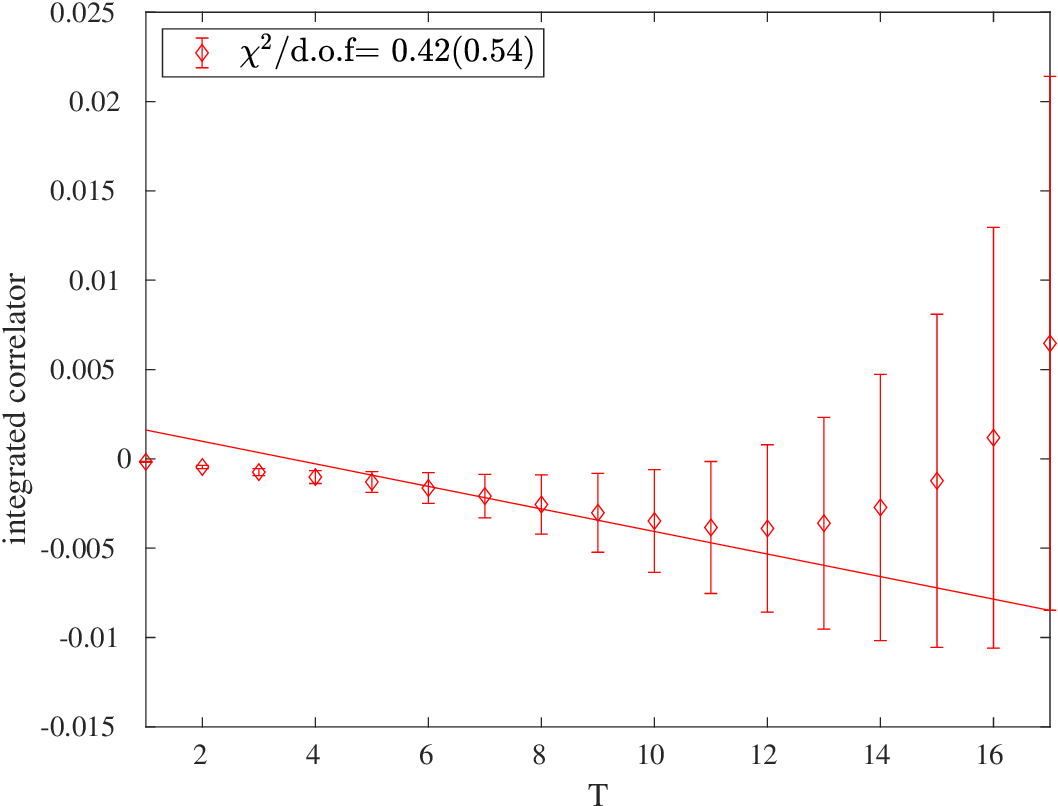}\\
			$Q_1Q_5$ & $Q_1Q_6$ \\
		\end{tabular}
		\caption{Integrated correlators including all five types of diagrams for the products $Q_1Q_j$ with 
		$j=1\ldots6$. We use a correlated fit with fitting range $10 \le T \le 16$. The Wilson coefficients are 
		not included.  Here we show only the result after subtraction of the short-distance piece and the 
		contributions of the vacuum and single-pion intermediate states.}
		\label{figure:all_diag_1}
	\end{figure}

\begin{figure}[H]
		\centering
		\begin{tabular}{cc} 
			\includegraphics[width=0.45\textwidth]{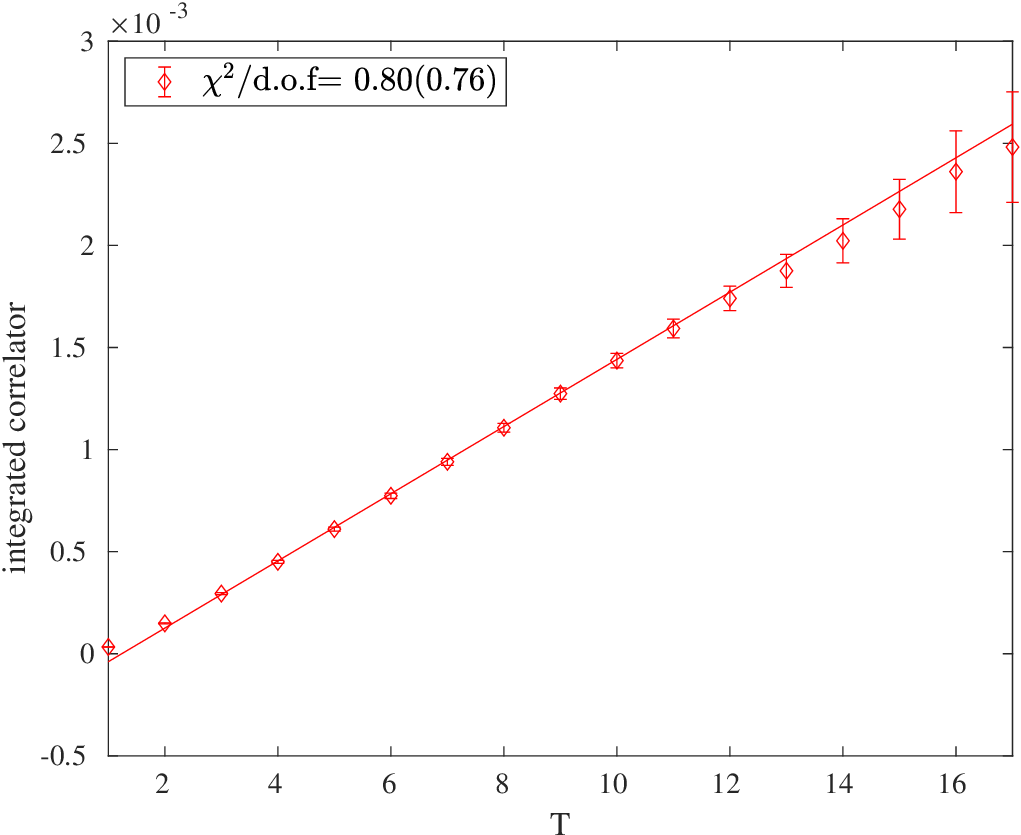}& 
			\includegraphics[width=0.45\textwidth]{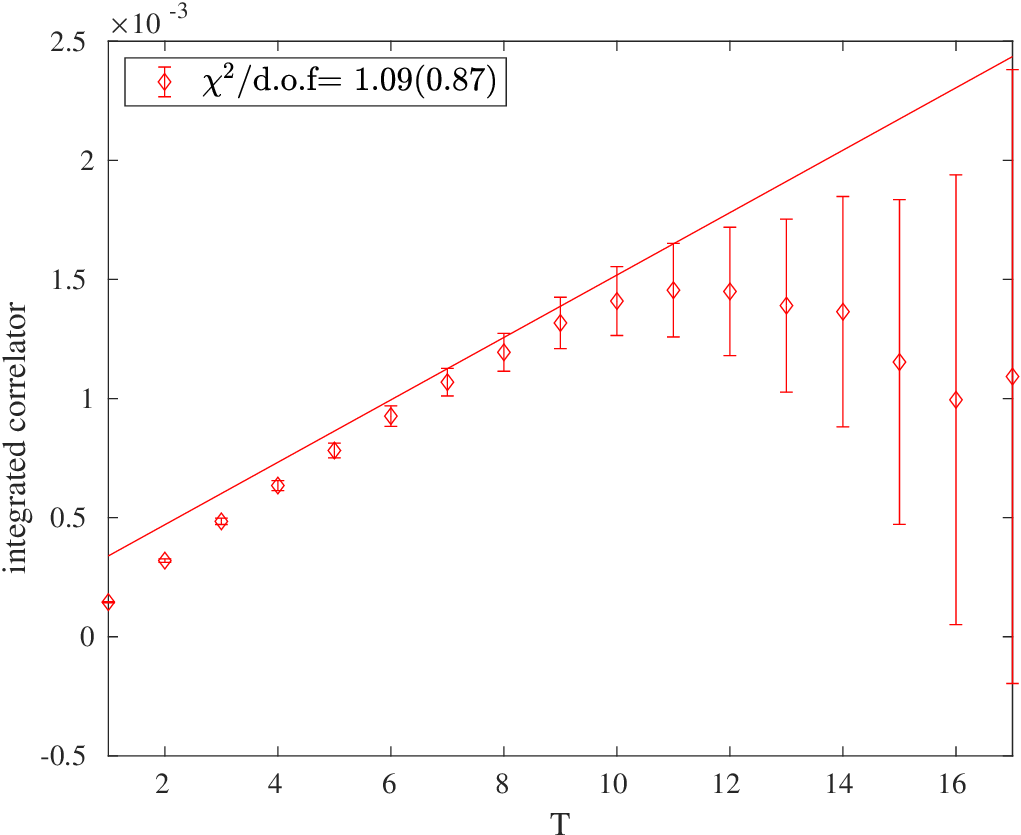}\\
			$Q_2Q_2$ & $Q_2Q_3$ \\
			\includegraphics[width=0.45\textwidth]{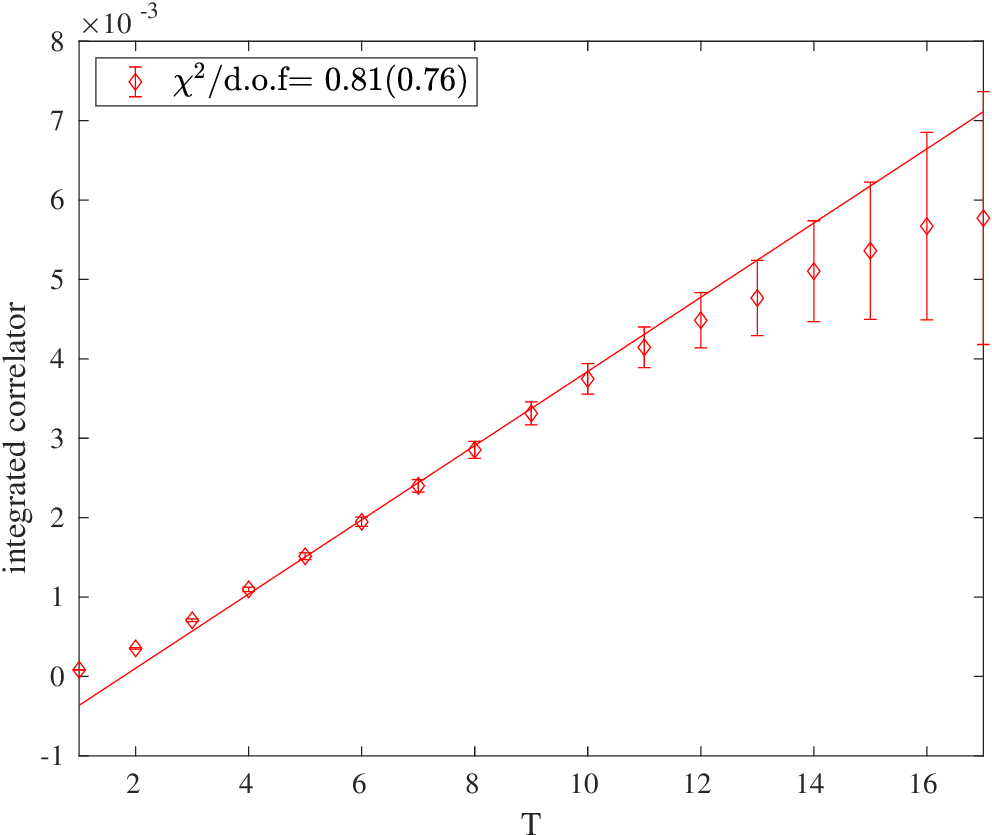}& 
			\includegraphics[width=0.45\textwidth]{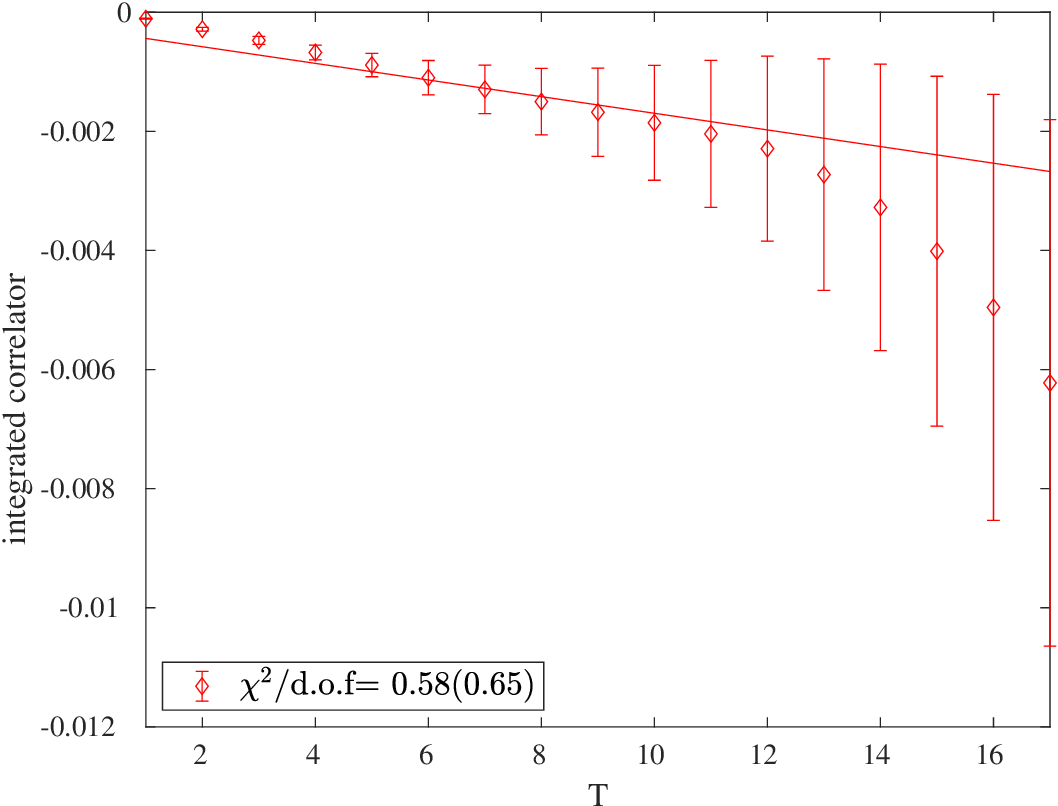}\\
			$Q_2Q_4$ & $Q_2Q_5$ \\
			\includegraphics[width=0.45\textwidth]{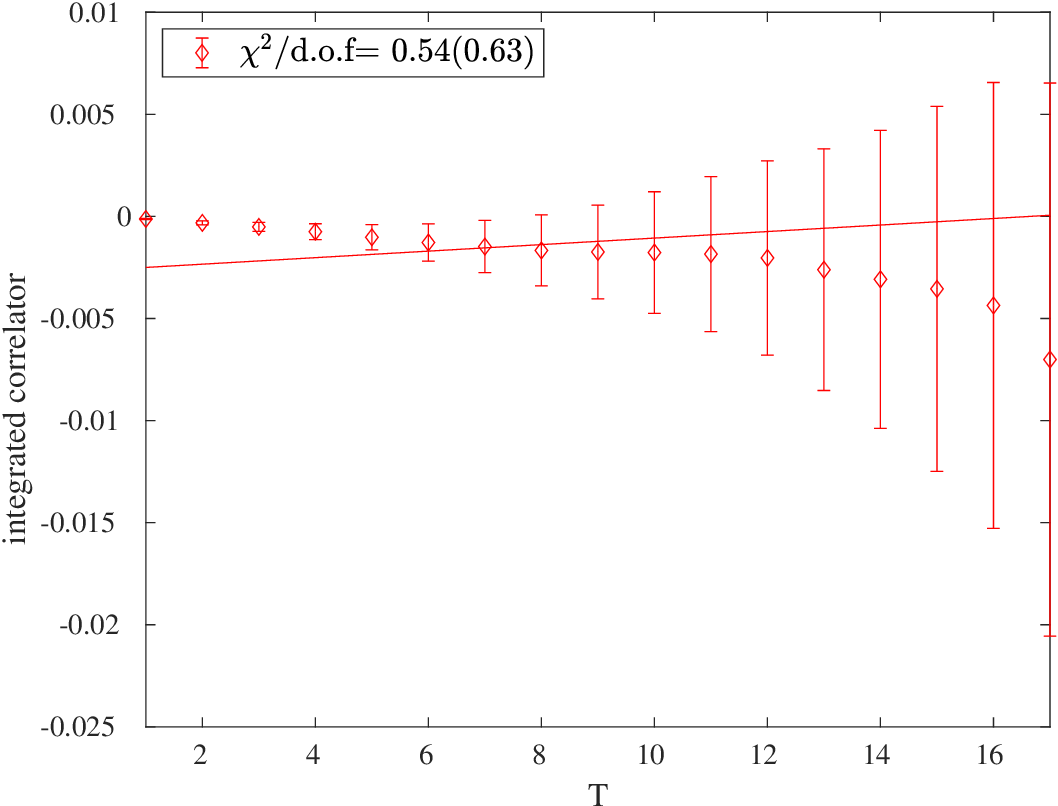}&\\
			$Q_2Q_6$ & \\
		\end{tabular}
		\caption{Integrated correlator including all five types of diagrams for the products $Q_2Q_j$ with 
		$j=2\ldots6$. We use a correlated fit with fitting range $10 \le T \le 16$. The Wilson coefficients are not 
		included.  Here we show only the result after subtraction of the short-distance piece and the 
		contributions of the vacuum and single-pion intermediate states.}
		\label{figure:all_diag_2}
	\end{figure}

In Figs.~\ref{figure:all_diag_1} and \ref{figure:all_diag_2} we plot the analog to Figs.~\ref{figure:fit_epsK12_1} and \ref{figure:fit_epsK12_2} but now show the result only after the subtraction of the short-distance piece $X^{\Lat}_{ij}$ and both the pion and vacuum intermediate states.  Shown also are linear fits to the matrix elements of the eleven, three-times-subtracted bilocal operators as functions of the length $T$ of integration interval.

In the results plotted above, we have removed the short-distance lattice-regulated contribution to the bilocal lattice operator using our intermediate RI/SMOM scheme with scale $\mu_\RI = 2.11$~GeV. The sum of all the contributions from different operator combinations $Q_i Q_j$ will be the total lattice result for  $\Imag{M^{ut}_{\overline{0}0}}$, which includes all the low energy contributions up to a high-energy cutoff determined by $\mu_{\RI}$. As explained earlier, we label this as $\Imag{M^{ut,\LD}_{\overline{0}0}}$.  This corresponds to the contribution to $\Imag{M^{ut}_{\overline{0}0}}$, from the $RI$ operator defined in Eq.~\eqref{eq:bilocal-Lat-RI}, or from the first line in the total $\Delta S =2$ weak Hamiltonian given in Eq.~\eqref{Eq:masterEq}.

In the fits shown in of Figs~\ref{figure:all_diag_1} and ~\ref{figure:all_diag_2} we have used a correlated fit with fitting range 10-16.  We show the $\chi^2$ per degree of freedom in the figure. In fitting the connected diagrams in Figs.~\ref{figure:fit_epsK12_1} and \ref{figure:fit_epsK12_2} we used a fitting range of 12-16. This reduced fitting range was needed because a linear fit did not represent the data well for smaller $T$, giving a poor $\chi^2$ (with $\chi^2/$d.o.f of order 5 or more).  While this choice of fitting range results in a larger statistical error when compared to 10-16, it gives more reliable results because of the better $\chi^2$.

We tabulate the contributions to $\Imag{M_{\overline{0}0}}$ from each operator combination in Table~\ref{table:epsK-all}.  The three sections of the table show three sets of results.  The top section shows the contributions from the type 1 and 2 diagrams before we remove the short-distance divergent 

\noindent
 part $X^{\Lat}_{ij}$.   The middle section shows the contributions from the type 1 and 2 diagrams after this lattice-regulated contribution has been removed.  Finally the bottom section shows the contribution of each bilocal operator product including all five types of diagram after the $X^{\Lat}_{ij}$ piece has been removed.  As is required, the appropriate Wilson coefficient factors have been included.  We note that the imaginary part comes only from the $\lambda_t$ factor because all the Wilson coefficients and $\lambda_u$ are real.  The values of $X^{\Lat}_{ij}$ are given in Table~\ref{table:Elat_192}.

By comparing middle and bottom sections of Table~\ref{table:epsK-all} we can see that the inclusion of the disconnected diagrams does not change the result for $\Imag{M^{ut}_{\overline{0}0}}$ significantly for most of the operators combinations $Q_i Q_j$. This is different from our experience in the $\Delta M_K$ calculation with similar unphysical quark masses where inclusion of the disconnected diagrams partially cancels the connected diagram result, decreasing the final result by roughly a factor of 2. 
Finally in Table~\ref{table:diff_diag} we list the contributions from different types of diagrams to the imaginary part of $M^{ut}_{\overline{0}0}$ in which the contributions from the eleven different bilocal operator products have been combined. 

	\begin{table}[ht]
		\centering
		\begin{tabular}{c|c|c|c|c|c}\hline\hline
			$X^{\Lat}_{1,1}$ & $X^{\Lat}_{1,2}$ & $X^{\Lat}_{1,3}$ & $X^{\Lat}_{1,4}$ &
			$X^{\Lat}_{1,5}$ & $X^{\Lat}_{1,6}$ \\\hline
			0.0374  &  0.0183  &  0.0818 &   0.0193&   -0.1092 &  -0.0432 \\\hline
		  & $X_{2,2}^{\Lat}$ & $X_{2,3}^{\Lat}$ & $X_{2,4}^{\Lat}$ &
			$X_{2,5}^{\Lat}$ & $X_{2,6}^{\Lat}$ \\\hline
	  	&0.0101  &  0.0196  &  0.0214  & -0.0310  & -0.0359 \\\hline\hline
		\end{tabular}
		\caption{The values for $X^{\Lat}_{ij}$ expressed in lattice units calculated using the external momentum 
		scale $\mu_\RI = 2.11$ GeV.  In the calculation of $X^{\Lat}_{ij}$, we only calculate $i < j$, since the value 
		for $X^{\Lat}_{ij}$ with $i > j$ is the same and is included in these coefficients.}
		\label{table:Elat_192}
	\end{table}

	\begin{table}[ht]
		\centering
		\begin{tabular}{c|c|c|c|c|c}\hline\hline
			$Q_1Q_1$ & $Q_1Q_2$ & $Q_1Q_3$ &$Q_1Q_4$& $Q_1Q_5$ & $Q_1Q_6$\\ \hline
		 	                   & $Q_2Q_2$ & $Q_2Q_3$ & $Q_2Q_4$ & $Q_2Q_5$ & $Q_2Q_6$\\ \hline\hline
			 -0.629(0.007)& 0.795(0.011)& 0.131(0.006)& -0.013(0.002)& -0.077(0.005)& 0.175(0.008)\\\hline
    			                     & -2.054(0.030)& -0.020(0.002)& 0.261(0.010)& 0.010(0.002)& 0.116(0.017)\\\hline\hline
			 -0.385(0.007)& 0.445(0.010)& 0.099(0.005)& 0.002(0.002)& -0.042(0.003)& 0.135(0.007)\\\hline
			                      & -1.505(0.029)& 0.003(0.002)& 0.214(0.008)& -0.019(0.002)& 0.213(0.017)\\\hline\hline
			 -0.384(0.016)& 0.438(0.069)& 0.067(0.006)& 0.004(0.012)& -0.016(0.013)& 0.091(0.113)\\\hline
			                      & -1.565(0.121)& -0.013(0.011)& 0.200(0.027)& -0.001(0.038)& 0.193(0.328)\\\hline\hline
		\end{tabular}
		\caption{Results for the imaginary part of $M^{ut}_{\overline{0}0}$ including the Wilson coefficients 
		and statistical errors.  All numbers are in units of $10^{-15}$ MeV.  The results are divided into three 
		sets of two rows separated by a double line. The top set shows the contribution of type 1 and 2 
		diagrams before the subtraction of the short-distance divergent part.  The middle set gives the 
		contribution of the type 1 and type 2 diagrams after the subtraction of the short-distance, divergent 
		part.  The bottom set contains the contribution of all five types of diagram after the subtraction of the 
		short-distance divergent	part.}
		\label{table:epsK-all}
	\end{table}

	\begin{table}[ht]
		\centering
		\begin{tabular}{c|c|c|c}\hline\hline
			$\Imag{M^{ut,lat,type\, 1+2}_{\overline{0}0}}$ & 
			$\Imag{M^{ut,\LD,type\, 1+2}_{\overline{0}0}}$ &
			$\Imag{M^{ut,\LD}_{\overline{0}0}}$ & $\Imag{M^{ut,\LD, \MS\to\RI}_{\overline{0}0}}$ \\\hline
			-1.328(0.038) & -0.865(0.037) & -0.986(0.389) & -0.552(0.389) \\\hline\hline
		\end{tabular}
		\caption{Combined contributions from the eleven bilocal operator combinations to 
		$\Imag{M^{ut}_{\overline{0}0}}$, in units of $10^{-15}$ MeV. The first column shows the results 
		before we remove the short-distance divergence from our lattice result and only includes the 
		type 1 \& 2 diagrams.  The second column again shows the contributions of only type 1 and 2
		diagrams but with the short-distance divergence removed. The third column gives the 
		contribution from all five types of diagram after the lattice-regulated short-distance contribution 
		has been removed. The last column is the long-distance correction to the conventional perturbative 
		short-distance result for $\Imag{M^{ut}_{\overline{0}0}}$, with the perturbative $\MS$ to RI/SMOM 
		correction included.}
		\label{table:diff_diag}
	\end{table}

To obtain the long-distance correction to the conventional short-distance result for $\epsilon_K$, we must finally add the perturbative $\MS$ to RI/SMOM matching factor, which corresponds to the second line of Eq~\ref{Eq:masterEq}.  We have evaluated the lowest order contribution to $\Delta Y^{\MS}_{ij}$, which is independent of the $\overline{MS}$ scale $\mu_{\MS}$, following the procedure specified in Section~\ref{sec:evaulate-Y}.  The amplitude $\Delta Y^{\MS}_{ij}$ is defined in the $\MS$ scheme and must be multiplied by the $K^0-\overline{K}^0$ matrix element of the $O_{LL}$ operator which is also normalized in the $\MS$ scheme.  This matrix element is most easily obtained from the $\MS$ kaon bag parameter $B_K(\mu_{\MS})$ using its definition:
\begin{eqnarray}
	\langle K^{\overline{0}}| O_{LL}(\mu_{\MS}) | K^0 \rangle =   \frac{4}{3} F_K^2 M_K B_K(\mu_{\MS}). \label{Eq:Bk}
\end{eqnarray}
We note that this equation is different from the conventional formula by a factor of $2M_K$ because we have used a different renormalization for the kaon energy eigenstate.  To perform this perturbative $\MS$ to RI/SMOM correction consistent with the other terms in our lattice calculation, we choose to use values for the parameters $B_K$, $F_K$ and $M_K$ appearing in Eq.~\eqref{Eq:Bk} that were determined from the same gauge ensemble that we have used for the lattice calculation, rather than more accurate values from more recent calculations.  We take the value for $B_K$ and $F_K$ from Ref.~\cite{Blum:2014tka}, which gives
\begin{equation}
	B_K^{RGI} = 0.750(15),\;\;
	F_K = 155.5(8)\, \mbox{MeV}.
\end{equation}

The RGI value for $B_K$ is the renormalization-group-invariant value, which is defined in Eq.~(18.4) of Ref.~\cite{Buchalla:1995vs}. We can use this formula to find $B_K$ at any energy scale, $\mu_{\MS}$.  We could use this ability to vary the $\mu_{\MS}$ scale to equal that used in the conventional short-distance result which we were correcting.  Of course, this scale dependence of $B_K$ will be of order $\alpha_s$, so including this $\mu_{\MS}$ dependence would be required only in an NNLO calculation.  In the right-most column of Table~\ref{table:diff_diag} and later tables we have used the value of $B_K$ at $\mu_{\MS}=2.11$ GeV to find the perturbative $\MS$ to RI/SMOM correction. Thus, the number in the final column of Table~\ref{table:diff_diag} is the sum of this $\MS$ to RI/SMOM correction which is proportional to $B_K$ and the long-distance result given in the third column that we obtained from the lattice calculation.  In Table~\ref{table:SM-input2} we list additional  standard model parameters that were used in these calculations.

\begin{table}[H]
	\centering
	\begin{tabular}{c|l}\hline\hline
    $G_F$       & $1.16637 \times 10^{-5}\;\mathrm{GeV}^{-2}$ \\
    $F_K$       & 0.1562 GeV \\
    $M_K$       & 0.4976 GeV \\
    $m_c(m_c)$  & 1.29 GeV \\
    $m_t(m_t)$  & 1.70 GeV \\
    $\Delta M_K$& $3.484 \times 10^{-15}$ GeV \\
    $\lambda_u$ & 0.2196 \\
    $\lambda_c$ &$-0.2193 - 1.1572 \times 10^{-4}i$ \\
    $\lambda_t$ &$-2.9565 \times 10^{-4} +1.1572 \times 10^{-4}i$ \\
    \hline\hline
    \end{tabular}
    \caption{Additional standard model parameters used in this calculation~\cite{PhysRevD.98.030001}.}
    \label{table:SM-input2}
\end{table}

In Table~\ref{table:M00_total}, we show the result of this calculation for five different intermediate RI/SMOM scales $\mu_{\RI}$ and in Table~\ref{table:M00_12only} we show the same quantities but include only the results from type 1 and type 2 diagrams.  The fourth and fifth columns in these tables do not contain the complete short-distance contributions to $\Imag{M^{ut}_{\overline{0}0}}$ or $\epsilon_K^{ut}$.  However, by including the perturbative $\MS$ to RI/SMOM correction $Y^{\MS}(\mu_\RI)$, these quantities do combine the RI/SMOM-renormalized (and therefore $\mu_{\RI}$-dependent) long-distance contribution with the $\mu_\RI$-dependent part of the missing short-distance contribution giving a quantity which should not depend on the long-to-short distance matching scale $\mu_\RI$.  

Examining Tables~\ref{table:M00_total} and \ref{table:M00_12only}, we can see this RI/SMOM matching appears successful because these combinations, $\Imag{M^{ut,\LD,\MS\to\RI}_{\overline{0}0}}$ or $\epsilon_K^{ut,\LD,\mathrm{corr}}$, have only a small dependence on $\mu_{\RI}$.

\begin{table}[ht]
	\centering
	\begin{tabular}{c|c|c|c|c}\hline\hline
		$\mu_{\RI}$ & $\Imag{M^{ut,\LD}_{\overline{0}0}}$ & 
		$\Imag{M^{ut,\MS\to\RI}_{\overline{0}0}}$
		& $\Imag{M^{ut,\LD,\MS\to\RI}_{\overline{0}0}}$ & 
		$\epsilon_K^{ut,\LD,\mathrm{corr}}$ \\\hline
    1.54 GeV & -0.746(0.389) & 0.282 & -0.464 (0.389) & 0.0911(0.076)\\\hline
    1.92 GeV & -0.912(0.389) & 0.384 & -0.527 (0.389) & 0.104(0.076)\\\hline
    2.11 GeV & -0.986(0.389) & 0.434 & -0.551 (0.389) & 0.108(0.076)\\\hline
    2.31 GeV & -1.050(0.390) & 0.486 & -0.565 (0.390) & 0.111(0.077)\\\hline
    2.56 GeV & -1.115(0.390) & 0.548 & -0.568 (0.390) & 0.111(0.077)\\\hline\hline
	\end{tabular}
	\caption{The long-distance contributions to the conventional short-distance part of 
	        $\Imag{M^{ut}_{\overline{0}0}}$ (in units of $10^{-15}$ MeV) and the corresponding 
	        contribution to $\epsilon_K$ as we vary $\mu_\RI$. The second column presents 
	        our results from the lattice calculation, after the removal of the short-distance 
	        divergence.  The third column is the perturbative $\MS$ to RI/SMOM correction that 
	        involves $\Delta Y^{\MS}$.  The fourth column is the final long-distance correction 
	        to the conventional short-distance contribution to $\Imag{M^{ut}_{\overline{0}0}}$, which 
	        is the sum of the previous two columns. The last column is the corresponding 
	        contribution to $\epsilon_K$, in units of $10^{-3}$.}
		\label{table:M00_total}
\end{table}

\begin{table}[ht]
	\centering
	\begin{tabular}{c|c|c|c|c}\hline\hline
		$\mu_{\RI}$ & $\Imag{M^{ut,\LD}_{\overline{0}0}}$ & 
		$\Imag{M^{ut,\MS\to\RI}_{\overline{0}0}}$
		& $\Imag{M^{ut,\LD,\MS\to\RI}_{\overline{0}0}}$ & 
		$\epsilon_K^{ut,\LD,\mathrm{corr}}$ \\\hline
  	1.54 GeV &  -0.620(0.036) & 0.282 & -0.337(0.036) & 0.066(0.007)\\\hline
  	1.92 GeV &  -0.786(0.036) & 0.384 & -0.401(0.036) & 0.079(0.007)\\\hline
  	2.11 GeV &  -0.860(0.037) & 0.434 & -0.425(0.037) & 0.084(0.007)\\\hline
  	2.31 GeV &  -0.924(0.037) & 0.486 & -0.439(0.037) & 0.086(0.007)\\\hline
  	2.56 GeV &  -0.989(0.037) & 0.548 & -0.442(0.037) & 0.087(0.007)\\\hline\hline
	\end{tabular}
	\caption{Results similar to those in Table~\ref{table:M00_total} but including only the contributions 
		from the diagrams of type 1 and 2.}
		\label{table:M00_12only}
\end{table}

\section{Conclusions and outlook}
\label{section:conclusion}

We have described in detail a method based on lattice QCD to calculate the long-distance contribution to the indirect CP violation parameter $\epsilon_K$.  In such a lattice calculation the weak interaction must be represented by its low-energy effective theory described by a dimension-six Hamiltonian density $\mathcal{H}_W$ written as the sum of twelve four-quark operators given in Eqs.~\eqref{eq:H-DS_1}-\eqref{eq:QCD-penguin}.  The parameter $\epsilon_K$ is determined by the $K^0-\overline{K}^0$ mixing matrix element $M_{\overline{0}0}$, a quantity that is second order in 
$\mathcal{H}_W$.  We separate long- and short-distances at the inverse energy scale above which QCD perturbation theory should be adequately accurate and below which the methods of lattice QCD can at present be applied.  Currently this energy scale may be 2-3 GeV.  As a result the lattice calculation described here is performed in the four-flavor theory, including an active charm quark.

We use the identity $\lambda_u + \lambda_c + \lambda_t = 0$ of Eq.~\eqref{eq:CKM-unitarity} to eliminate $\lambda_c$, expressing $M_{\overline{0}0}$ as a combination of terms proportional to $\lambda_t^2$, $\lambda_u^2$ and $\lambda_t\lambda_u$.  The last term contains the long-distance contribution to $\epsilon_K$.  For this CP violating quantity the GIM cancellation is incomplete and the singularity in the  second-order product of the two factors $\mathcal{H}_W(x)\mathcal{H}_W(y)$ as $x \to y$ results in a logarithmic singularity. The presence of this singularity requires that we combine our lattice calculation with a short-distance calculation which replaces this short-distance singularity with the actual short-distance contribution of the standard model.

Such a combination of a lattice calculation using the low-energy effective theory and a QCD and electroweak perturbative calculation which involves the $W$, $Z$, Higgs, top- and bottom-quark degrees of freedom is achieved by imposing an RI-SMOM condition on the second-order lattice calculation of an infrared-safe, off-shell, $\Delta S=2$ four quark Green's function at a scale $\mu_\RI$.  Imposing this condition requires the addition of a $\mu_{\RI}$-dependent counter-term proportional to the operator $O_{LL}$ of Eq.~\eqref{eq:LL}.  If the scale $\mu_\RI$ is chosen to be sufficiently large, then the usual QCD and electroweak perturbation theory calculation of this same off-shell, four-quark Green's function at a scale $\mu_\RI$ then can be used to determine the term proportional to $O_{LL}$ that must be added to the lattice result to obtain $M_{\overline{0}0}^{ut}$ to any specific order in QCD perturbation theory.

In the preceding sections we have determined the steps needed to carry out this lattice QCD calculation of $\epsilon_K$ including the needed four-quark operators and their Wilson coefficients.  With the exception of the incomplete GIM cancellation that appears in the calculation of $\epsilon_K$, the lattice calculation of the long-distance contribution to $\epsilon_K$ is similar to the calculation of $\Delta M_K$.  Of course, with the need to compute eleven instead of three bilinear operator combinations, the calculation of $\epsilon_K$ is significantly more difficult.  A calculation of the long-distance component of $\epsilon_K$ can naturally be combined with future calculations of $\Delta M_K$ and such combined calculations are being actively pursued by the RBC and UKQCD collaborations.  

In order to explore all of the issues involved in such a calculation of the long-distance contribution to $\epsilon_K$ we have carried out a complete lattice calculation using a $24^3\times 64$ lattice with an inverse lattice spacing $1/a=1.78$ GeV and unphysical light and strange quark masses which result in $M_\pi = 339$ MeV and $M_K=592$ MeV.   Given the relatively coarse lattice spacing we chose a lighter-than-physical charm quark mass of 968 MeV (renormalized in the $\MS$ scheme at 2 GeV).  Just as in the calculation of $\Delta M_K$ the connected contribution, coming from diagrams of type 1 and 2, can be calculated quite precisely with statistical errors from 200 configurations on the order of 5\%.  However, as in the $\Delta M_K$ case, the disconnected graphs are much more difficult with statistical errors of order 40\%.  We expect that a calculation with physical quark masses will be practical with statistical errors on the order of 10\% as is the case for $\Delta M_K$~\cite{Wang:2022lfq} where improved methods and a focus on obtaining increased statistics for the disconnected parts has given results with 10\% statistical errors.  As have been found in the calculation of $\Delta M_K$, we must expect relatively large discretization errors on the order of 40\% for a calculation performed on a lattice with $1/a=2.38$ GeV.  This suggests that results with less uncertainty than present perturbative or phenomenological estimates for both $\Delta M_K$ and the long-distance part of $\epsilon_K$ will require at least a second lattice spacing and a continuum extrapolation.

A complete result for $\epsilon_K$ requires that a lattice QCD calculation of the long-distance contribution to $\epsilon_K$ of the sort described here be accurately joined to a perturbative calculation of the much larger short-distance part (which also requires a lattice calculation of the single long-distance, hadronic amplitude $B_K$).  The short-distance QCD and electroweak perturbation theory calculation, reviewed for example in Ref.~\cite{Buchalla:1995vs}, has been carried out to NLO~\cite{Herrlich:1996vf} and partially to NNLO~\cite{Brod:2010mj, Brod:2011ty}.  While we anticipate that in the future such a calculation will be performed to evaluate the needed off-shell four-quark Green's function at an energy scale $\mu_{\RI}$ to NLO, at present such a result is not available.  However, present results do provide this four-quark Green's function at NNLO evaluated at the scale $\mu_{\RI}=0$.  This allows us to obtain the required Green's function for a larger value of $\mu_{\RI}$ to order $(\alpha_s)^0$ by evaluating a simple convergent one-loop integral. 

This one-loop calculation allows the RI-normalized lattice calculation to be matched to the $\MS$ perturbative result, providing an ``$\MS$ to RI/SMOM correction'', denoted $\Imag{M^{ut,\LD,\MS\to\RI}_{\overline{0}0}}$, is added to the lattice result.  At the order we are working this depends on $\mu_{\RI}$ but not on $\mu_{\MS}$.  However, when this correction is added to our RI-normallized lattice calculation the result should be independent of the scale $\mu_{\RI}$.  This $\mu_{\RI}$-dependence can seen in Tables~\ref{table:M00_total} and \ref{table:M00_12only}.   If we examine the more accurate result in Table~\ref{table:M00_12only} from diagrams of type 1 and 2 only (appropriate since the omitted disconnected diagrams only enter at higher order in QCD perturbation theory) we see the $14\%$ dependence of lattice result for Im $M_{\overline{0}0}^{ut,\RI}$ as $\mu_{\RI}$ is varied from 2.11 GeV to 2.56 GeV decreases to $4\%$ when combined with this $\MS$ to RI/SMOM correction.

The unphysical quark masses and single lattice spacing used in our calculation make the present result an unreliable long-distance correction to $\epsilon_K$.  Nevertheless it is of interest to compare the size of this correction to the current short-distance result for $\epsilon_K$:
\begin{eqnarray}
\epsilon_K^{LD}(\mu_\RI=2.11\,\mathrm{GeV})               &=& 0.195(0.077) e^{i\phi_\epsilon} \times 10^{-3} \\
\epsilon_K^{SD}   &=& 1.446(0.154)e^{i\phi_\epsilon} \times 10^{-3} \quad  \mbox{Ref.~\cite{Lee:2023lxz}} \\
\epsilon_K^{\RI\to\MS}(\mu_{\RI}=2.11\,\mathrm{GeV}) &=& -0.086e^{i\phi_\epsilon}  \times 10^{-3}.
\end{eqnarray}
Here the first number is our result for the long-distance contribution to $\epsilon_K$ including connected and disconnected diagrams with the bilinear operator product renormalized in the RI-SMOM scheme with $\mu_\RI=2.11$ GeV.  (We have explicitly included the phase of $\epsilon_K$ introduced in Eqs.~\eqref{Eq:epsK1} and \eqref{Eq:epsK2} so that we can display the magnitude of $\epsilon_K$ and still combine the quantities shown algebraically, including their relative signs.)  The second number is a recent result for $\epsilon_K$ without long-distance correction~\cite{Lee:2023lxz}.  The third number is the $\mathcal{O}(1)$ correction that should be added to the second number giving a sum which represents the complete RI-SMOM-normalized short-distance contribution, also evaluated at $\mu_\RI=2.11$ GeV.  This sum could then be added to the first line to obtain a consistent prediction for $\epsilon_K$, had these quantities been computed with consistent quark masses and other weak interaction input parameters.  The 8\% relative size of the difference of the sum of the first and the third lines compared to the second is somewhat larger than the phenomenological estimate of 5\%~\cite{Buras:2010pza} because we are comparing to a short-distance prediction which itself is somewhat smaller than the measured result $|\epsilon_K| = 2.228(0.011)\times 10^{-3}$.  The discrepancy between the experimental result and the standard model short-distance prediction given in Ref.~\cite{Lee:2023lxz} is not understood but may be related to the significant discrepancy between the exclusive and inclusive experimental results for the CKM matrix element $V_{cb}$.

\section{Acknowledgement}

We thank our RBC and UKQCD collaboration colleagues for extensive discussions and support. ZB, NHC, JK and BW were supported by U.S. Department of Energy (DOE) grant \#DE-32SC0011941.  JK was also supported by U.S. DOE grant \#DE-SC0011941 and in part by the U.S. DOE Contract No.~DE-AC05-06OR23177, under which Jefferson Science Associates, LLC operates Jefferson Lab.  CTS was partially supported by an Emeritus Fellowship from the Leverhulme Trust and by STFC (UK) grant ST/T000775/1.  AS was supported by U.S. DOE awards \#DE-SC0012704 and \#DE-SC001704.  The calculation reported here was carried out on the BG/Q computing facilities provided by the RIKEN Brookhaven Research Center and the Brookhaven National Laboratory.

\appendix
\section{Four-quark operator mixing and non-perturbative renormalization}\label{sec:appendix}

The four-flavor RI-SMOM non-perturbative renormalization scheme used here to determine the Wilson coefficients for the four-quark operators defined on the lattice was first used in the calculation of $\Delta M_K$ in Ref.~\cite{Christ:2012se}.  In the case of $\Delta M_K$, only the coefficients of the two current-current operators are needed.  For the calculation of $\epsilon_K$, the nonperturbative renormalization (NPR) will be more challenging because we must also include the QCD penguin operators which mix among themselves and also appear when the renormalization scheme or scale for the two current-current operators is changed. The procedure is similar to what we have done in Refs.~\cite{Bai:2015nea} and \cite{RBC:2020kdj} when computing $K\to\pi\pi$ decay but with the difference that we are now working in the four-flavor theory and do not need to include the electroweak penguin operators.  We impose the RI/SMOM condition specified in Ref.~\cite{Lehner:2011fz} on Landau gauge fixed, amputated Green's functions with off-shell external momenta $p_1$ and $p_2$. These momenta obey $p_1^2 = p_2^2 = (p_1 - p_2)^2 = \mu_\RI^2$. We have chosen to use $\mu_\RI = 2.15$ GeV, the same as our choice in Ref.~\cite{Christ:2012se}. Four different RI-SMOM schemes are studied in Ref.~\cite{Lehner:2011fz}: the ($\gamma_\mu,\gamma_\mu$), ($\gamma_\mu,\slashed{q}$), ($\slashed{q},\slashed{q}$) and ($\slashed{q},\gamma_\mu$) where the first factor indicates the structure of the four-quark projector used in the RI-SMOM condition while the second determines the scheme used for the quark operator renormalization.  Here we use the ($\gamma_\mu, \gamma_\mu$) and ($\gamma_\mu, \slashed{q}$) schemes.

We begin by rewriting the effective four-flavor weak Hamiltonian in Eq.~\eqref{eq:H-DS_1} as:
\begin{eqnarray}
	\label{Eq:HW_mix2}
	H_W &=& \frac{G_F}{\sqrt{2}} \left\{ V^*_{us} V_{ud} \left[ (1-\tau) \sum_{i=1,2} z_i(\mu)
	(Q_i^{u\overline{u}} - Q_i^{c\overline{c}})	+ \tau \sum_{i = 1}^{6} v_i(\mu) Q_i \right]  \right. \\\nonumber
	&& \left. + V^*_{us} V_{cd} \sum_{i=1,2} z_i Q_i^{u\overline{c}}  + V^*_{cs} V_{ud} \sum_{i=1,2} z_i Q_i^{c\overline{u}} \right\},
\end{eqnarray}
where $\tau=-\lambda_t/\lambda_u$.  The operators $\{Q_i^{q\overline{q}'}\}_{i=1,2}$ represent four distinct current-current operators for each value of $i=1$ or 2 depending on the combination $qq'$ of up and charm quarks which appears.  The operators $Q_1$ and $Q_2$ without superscripts indicate $Q_1^{u\overline{u}}$ and  $Q_2^{u\overline{u}}$.    We note that a single Wilson coefficient $C_i$ can be used for all four operators with the subscript $i$ for the case $i=1$ or 2.

In order to apply the RI-SMOM intermediate renormalization procedure, we need to specify a minimal complete set of operators which transform into themselves when either the renormalization scheme or the scale is changed.  Examining the operators which appear in Eq.~\eqref{Eq:HW_mix2} we recognize that first term in that equation, $Q_i^{u\overline{u}} - Q_i^{c\overline{c}}$, does not mix with the QCD penguin operators $\{Q_j\}_{j=3,4,5,6}$ because of GIM cancellation and the terms on the second line of Eq.~\eqref{Eq:HW_mix2} also do not mix with the QCD penguin operators because of their flavor structure (either $Q_i^{u\overline{c}}$ or $Q_i^{c\overline{u}}$).  As we have observed above, the Wilson coefficients for the current-current operators do not depend on the flavor structure. So we have $z_1 = v_1$ and $z_2 = v_2$, as is explained in Section~\ref{sec:eff-theory} and in Ref.~\cite{Buchalla:1995vs}.  

We can use the equality of the Wilson coefficient $C_i$ for all four operators $Q^{qq'}_i$ for $i=1$ or 2 to focus on the normalization of the six operators $\{Q_j\}_{j=1,2,\ldots,6}$ whose Wilson coefficients we denote by $\{C_i\}_{i=1,2,\ldots,6}$. 

Using this basis of six operators, we first transform the operators to the RI-SMOM scheme using the $6\times 6$ mixing matrix $Z^{\Lat\rightarrow \RI}$:
\begin{equation}
	Q^\RI_k = Z^{\Lat\rightarrow \RI}_{kj} Q^\Lat_j.
	\label{eq:Lat-to-RI}
\end{equation}
Then we transform the RI/SMOM operators to the $\MS$ scheme using the $6\times 6$ matrix $\Delta r$ obtained from Ref.~\cite{Lehner_note}:
\begin{equation}
	Q_i^{\MS} = (1 + \Delta r)^{\RI\rightarrow \MS}_{ik} Q^{\RI}_k.
	\label{eq:RI-to-MSbar}
\end{equation}
Finally, we can substitute Eq.~\eqref{eq:Lat-to-RI} into Eq.~\eqref{eq:RI-to-MSbar} to obtain:
\begin{equation}
	\sum_{i=1}^6 C_i^{\MS} Q_i^{\MS} = \sum_{i=1}^6 C_i^{\Lat} Q_i^{\Lat}.
\end{equation}
where
\begin{equation}
	C_j^{\Lat} = \sum_{i,k=1}^6 C_i^{\MS} (1 + \Delta r)^{\RI\rightarrow \MS}_{ik} Z^{\Lat\rightarrow \RI}_{kj}.
\end{equation}

We have performed the needed NPR calculations on 100 configurations from a $16^3\times 32$ Iwasaki ensemble which has the same lattice spacing as the ensemble we used in the $\epsilon_K$ calculation.  In the $(\gamma_\mu, \gamma_\mu)$ scheme with $Z_q^{\gamma_\mu} = 0.7404(4)$ we find the mixing matrix:
	\small{
	\begin{equation}
    Z^{\Lat\rightarrow \RI} = \begin{pmatrix} 
0.505(0.000)& -0.050(0.000)& 0.004(0.002)& -0.003(0.002)& 0.001(0.002)& -0.003(0.001) \\
 -0.050(0.000)& 0.505(0.000)& -0.003(0.001)& 0.010(0.001)& -0.003(0.001)& 0.008(0.001) \\
 0 & 0 & 0.514(0.008)& -0.043(0.007)& -0.001(0.009)& 0.006(0.005) \\
 0 & 0 & -0.056(0.006)& 0.540(0.005)& -0.008(0.006)& 0.027(0.004) \\
 0 & 0 & 0.002(0.007)& -0.006(0.006)& 0.537(0.008)& -0.089(0.005) \\
 0 & 0 & -0.012(0.003)& 0.033(0.003)& -0.040(0.003)& 0.410(0.002) \\
		\end{pmatrix}
		\label{Eq:Z_gg}
	\end{equation}
}

Using this mixing matrix and $\Delta r^{\RI\rightarrow \MS}$ for the $(\gamma_\mu, \gamma_\mu)$ scheme, we find the following lattice Wilson coefficients:
\begin{equation}
	C^{\Lat} =\begin{pmatrix} 
	-0.202(0.000)& 0.588(0.000)& 0.012(0.001)& -0.024(0.001)& 0.009(0.001)& -0.027(0.001)\\
		\end{pmatrix}
		\label{Eq:C_lat_gg}
\end{equation}

In the $(\gamma_\mu,\slashed{q})$ scheme, we use $Z_q^{\slashed{q}} = 0.8016(3)$.  The mixing matrix $Z^{\Lat\rightarrow \RI}$ differs from that given in Eq.~\eqref{Eq:Z_gg} by an over-all factor of $(Z_q^{\gamma_\mu}/Z_q^{\slashed{q}})^2$.  Using this rescaled mixing matrix and $\Delta r^{\RI\rightarrow\MS}$ for the $(\gamma_\mu, \slashed{q})$ scheme, we find the following lattice Wilson coefficients: 
\begin{equation}
	C^{\Lat} =\begin{pmatrix} 
  -0.222(0.000)& 0.645(0.000)& 0.013(0.001)& -0.027(0.001)& 0.010(0.001)& -0.030(0.001)\\
		\end{pmatrix}
		\label{Eq:C_lat_gq}
\end{equation}

This second determination of the lattice Wilson coefficients differ by an overall-factor of 1.09 from those in Eq.~\eqref{Eq:C_lat_gg}.  This discrepancy is a useful indicator of the size of the systematic errors in the lattice Wilson coefficients arising from two sources: i) The QCD perturbation theory truncation errors associated with the perturbative calculation of the coefficients $\Delta r_{ik}$ that appear in Eq.~\eqref{eq:RI-to-MSbar} and ii) the discretization errors that enter the lattice QCD calculation of the coefficients $Z^{\Lat\to\RI}$ in Eqs.~\eqref{Eq:Z_gg}.  The discrepancy arising from QCD perturbation theory can be reduced by performing the NPR calculation at a higher energy scale or to a higher order in $\alpha_s$.  The lattice discretization error can be made smaller by using a finer lattice for these $\epsilon_K$ measurements or by using step-scaling in the NPR calculation so that the large energy scale needed to reduce the perturbative errors need not be used at the lattice scale adopted for the $\epsilon_K$ calculation.

\bibliography{paper}

\end{document}